\documentclass[prd,aps,showpacs,preprintnumbers,nofootinbib,superscriptaddress]{revtex4}

\usepackage{graphicx}
\usepackage{amsmath}
\usepackage{amssymb}
\usepackage{color}
\usepackage[dvipsnames]{xcolor}
\usepackage{hyperref}
\usepackage{url}
\usepackage{float}
\usepackage{epstopdf}
\usepackage{extarrows}
\usepackage{enumerate}
\usepackage{enumitem}
\newcommand{\beq}{\begin{equation}}
\newcommand{\eeq}{\end{equation}}
\newcommand{\be}{\begin{equation}}
\newcommand{\ee}{\end{equation}}
\newcommand{\bea}{\begin{eqnarray}}
\newcommand{\eea}{\end{eqnarray}}
\newcommand{\nn}{\nonumber}

\newcommand{\tf}{\texorpdfstring}
\newcommand{\gev}{~\text{GeV}}
\newcommand{\sech}{~\text{sech}}
\newcommand{\arcsinh}{~\text{arcsinh}}

\begin{document}


\title{Double Higgs production at the 14 TeV LHC and the 100 TeV $pp$-collider}
\author{Qing-Hong Cao}
\email{qinghongcao@pku.edu.cn}
\affiliation{Department of Physics and State Key Laboratory
of Nuclear Physics and Technology, Peking University, Beijing 100871, China}
\affiliation{Collaborative Innovation Center of Quantum Matter, Beijing, China}
\affiliation{Center for High Energy Physics, Peking University, Beijing 100871, China}

\author{Gang Li}
\email{ligang@pku.edu.cn}
\affiliation{Department of Physics and State Key Laboratory of
Nuclear Physics and Technology, Peking University, Beijing 100871, China}

\author{Bin Yan}
\email{binyan@pku.edu.cn}
\affiliation{Department of Physics and State Key Laboratory of
Nuclear Physics and Technology, Peking University, Beijing 100871, China}

\author{Dong-Ming Zhang}
\email{zhangdongming@pku.edu.cn}
\affiliation{Department of Physics and State Key Laboratory of
Nuclear Physics and Technology, Peking University, Beijing 100871, China}

\author{Hao Zhang}
\email{zhanghao@ihep.ac.cn}
\affiliation{Institute of High Energy Physics, Chinese Academy of Sciences,
Beijing 100049, China}
\date{\today}
\begin{abstract}

We consider effective Higgs boson couplings, including both the CP-even and
CP-odd couplings, that affect Higgs boson pair production in this study. Through the partial wave analysis, we find that the process $gg\to hh$ is dominated by the $s$-wave component even at a 100~TeV $pp$-collider. Making use of the $s$-wave kinematics, we propose a cut efficiency function to mimic the collider simulation and obtain the potential of measuring Higgs effective couplings at the 14~TeV LHC with an integrated luminosity of $3000~{\rm fb}^{-1}$ and at a 100 TeV $pp$-collider. Analytical expressions of the $2\sigma$ exclusion limits at the LHC and the $5\sigma$ discovery bands at the 100 TeV machine are given.

\end{abstract}
\maketitle

\section{Introduction}
\label{sec:1}

Double Higgs boson production is important to measure
the trilinear Higgs coupling in order to determine the structure of the Higgs potential~\cite{Glover:1987nx, Baur:2003gpa,Baur:2003gp,Dolan:2012rv,
Baglio:2012np,Papaefstathiou:2012qe,
Barger:2013jfa,Barr:2013tda,Yao:2013ika,deLima:2014dta,Li:2015yia}.  In addition to the trilinear Higgs coupling, the gluon-initiated process $gg\to hh$ also involves the coupling of Higgs boson to top quarks. Besides, in composite Higgs models~\cite{Agashe:2004rs,Contino:2006qr} and Little Higgs models~\cite{ArkaniHamed:2001nc,ArkaniHamed:2002qy},  the contact interactions $hh\bar{t}_Lt_R+h.c.$ and $h(h)gg$ are naturally predicted.
So far no new particle beyond the standard model (SM) is observed yet.  It is natural to adopt the effective field theory (EFT)~\cite{Buchmuller:1985jz,
Grzadkowski:2010es,Contino:2013kra} approach to study the double Higgs production. 
In this paper, we extend the previous studies~\cite{Pierce:2006dh,Contino:2012xk,
Chen:2014xra,Goertz:2014qta,Azatov:2015oxa,Dawson:2015oha}, which focus on the CP-even Higgs effective couplings, and include all the possible CP-odd Higgs effective couplings. The general effective Lagrangian of interest to us is~\cite{Chen:2014xra,
Goertz:2014qta,Azatov:2015oxa,Dawson:2015oha}
\begin{align}
\label{eq:effLag}
\mathcal{L}_{\rm{eff}}&=-\frac{m_{t}}{v}\bar{t}(c_{t} +i\tilde{c}_{t}\gamma_{5})
th-\frac{m_t}{2v^2}\bar{t}(c_{2t}+i\tilde{c}_{2t}\gamma_{5})th^2+
\frac{\alpha_s h}{12\pi v}(c_gG_{\mu\nu}^{A}G^{A,\mu\nu}
+\tilde{c}_gG_{\mu\nu}^{A}\tilde{G}^{A,\mu\nu})\nn\\
&+\frac{\alpha_s h^2}{24\pi v^2}(c_{2g}G_{\mu\nu}^{A}G^{A,\mu\nu}
+\tilde{c}_{2g}G_{\mu\nu}^{A}\tilde{G}^{{A,}\mu\nu})-c_{3h}\frac{m_h^2}{2v}h^3,
\end{align}
where $m_t$ is the top quark mass, $v$ is the vacuum expectation value, $\alpha_s$ is the strong coupling constant and
$G_{\mu\nu}^A(\equiv\partial_{\mu}G_{\nu}^A-\partial_{\nu}
G_{\mu}^A-g_sf^{ABC}G^{{A}}_{\mu}G^{{A}}_{{\nu}})$ is
the field strength of gluon and its dual is defined as $\tilde{G}^{{A,}\mu\nu}
=\frac{1}{2}\varepsilon^{\mu\nu\rho\sigma}G^{{A}}_{\rho\sigma}$
with $\varepsilon^{0123}=1$.
The terms of $c_t$, $c_{2t}$, $c_g$, $c_{2g}$ and $c_{3h}$ describe the
CP-even interactions while the terms of $\tilde{c}_{t}$, $\tilde{c}_{2t}$,
$\tilde{c}_{g}$ and $\tilde{c}_{2g}$ represent the CP-odd interactions.
In the SM $c_t=1$ and $c_{3h}=1$ while all other coefficients vanish at the
tree level. It is worth mentioning that the $c_{g}$ and $c_{2g}$ terms in
the above interaction might be correlated in a given NP model. For example,
$c_{g}=c_{2g}$ when both terms arise from the same dimension-6
operator $\mathcal{O}_{H G}=H^{\dag}HG_{\mu\nu}^{A}G^{{A,}\mu\nu}$
where $H$ denotes the Higgs doublet. Similarly, $\tilde{c}_g
=\tilde{c}_{2g}$ when they are from the operator $\tilde{\mathcal{O}}_{H G}
=H^{\dag}HG_{\mu\nu}^{A}\tilde{G}^{{A,}\mu\nu}$~\cite{Buchmuller:1985jz,
Grzadkowski:2010es,Contino:2013kra}~\footnote{The operator
$\tilde{\mathcal{O}}_{H G}=H^{\dag}HG_{\mu\nu}^{A}\tilde{G}^{{A,}\mu\nu}$
can also induce a QCD $\bar{\theta}$-term. However, it can be removed
through invoking Peccei-Quinn mechanism~\cite{Peccei:1977hh}}.

The remainder of this paper is organized as follows.
In Sec.~\ref{sec:2}, we present expressions of the single Higgs production
amplitude under the Lagrangian in Eq.~\eqref{eq:effLag} and obtain constraints from current Higgs signal strength measurements and electric dipole moments (EDMs).
In Sec.~\ref{sec:31} and \ref{sec:32}, we give the expression of the amplitude in the double
Higgs production and perform partial wave analysis to show this process is dominated by the $s$-wave component, respectively.
We obtain a cut efficiency function based on the $s$-wave dominant feature of the amplitude in Sec.~\ref{sec:41}.
Then we use the cut efficiency function to mimic the collider simulation to get the
$m_{hh}$ distribution in Sec.~\ref{sec:42} and the cross section before and after the selection cuts at the HL-LHC and a 100 TeV $pp$-collider in Sec.~\ref{sec:43}. The correlation and sensitivity of the Higgs effective couplings at the HL-LHC and a 100 TeV $pp$-collider is investigated in Sec.~\ref{sec:44x} and Sec.~\ref{sec:44y}, respectively.
Finally, we conclude in Sec.~\ref{sec:5}.

\section{Constraints from Single Higgs measurements and Electric Dipole Moments}
\label{sec:2}
The effective couplings $c_t,\tilde{c}_t$, $c_g,\tilde{c}_g$ in Eq.~\eqref{eq:effLag}
which are related to the double Higgs production also contribute to the single
Higgs production and decay processes.
Therefore, we consider the current constraints from the single Higgs
measurements at the 7 TeV, 8 TeV and 13 TeV LHC as well as the low energy experiments.

The partonic amplitude of the single Higgs procution $g^{{a,\mu}}
(p_{{1}})g^{{b,\nu}}(p_{{2}})\to h$  at the leading order (LO) is
\beq
\label{eq:1h}
\mathcal{M}_{h}=-\frac{\alpha_s\hat{s}\delta^{ab}}{4\pi v}
[(c_t F_{\triangle}+\frac{2}{3}c_g)A^{\mu\nu}-(\tilde{c}_t F_\triangle^{(1)}
+\frac{2}{3}\tilde{c}_g)C^{\mu\nu}]\epsilon^{{a}}_{\mu}(p_{{1}})
\epsilon^{{b}}_{\nu}(p_{{2}}),
\eeq
where $\hat{s}=(p_{{1}}+p_{{2}})^2={m_h}^2$, $\alpha_{s}=
g_{s}^{2}/(4\pi)$ and $\text{Tr}(t^at^b)=\delta^{ab}{{/2}}$. The form
factors $F_{\triangle}$ and $F_\triangle^{(1)}$ can be expressed in terms
of piecewise function~\cite{Djouadi:2005gi,Djouadi:2005gj}
\beq
F_{\triangle}=\tau_t[1+(1-\tau_t)f(\tau_t)], \quad F_\triangle^{(1)}=-\tau_t f(\tau_t),
\eeq
where $ \tau_t=4m_t^2/{m_h}^2$ and
\beq
\label{ftau}
f(\tau)=
\begin{cases}
  \arcsin ^2\displaystyle{\left(\frac{1}{\sqrt{\tau}}\right)} & \tau\geq 1, \\
  \displaystyle{-\frac{1}{4}\left[\log\left(\frac{1+\sqrt{1-\tau}}{1
  -\sqrt{1-\tau}}\right)-i\pi\right]^2} & \tau<1.
\end{cases}
\eeq
In the large $m_t$ limit~\cite{Plehn:1996wb}, we have
\be
F_\triangle=\frac{2}{3}+\mathcal{O}(\hat{s}/m_t^2), \quad F_\triangle^{(1)}
=-1+\mathcal{O}(\hat{s}/m_t^2).
\ee
The Lorentz structures $A^{\mu\nu}$ {and} $C^{\mu\nu}$ are defined as
\bea
\label{eq:lor}
A^{\mu\nu}=g^{\mu\nu}-\frac{p_{{1}}^{\nu}p_{{2}}^{\mu}}
{p_{{1}}\cdot p_{{2}}},\quad C^{\mu\nu}=\frac{p_{{1}\rho}p_{{2}\sigma}}
{p_{{1}}\cdot p_{{2}}}\varepsilon^{\mu\nu\rho\sigma}.
\eea

The single Higgs production cross section and partial decay width in the NP model, normalized to the SM values, are  
\beq
\label{eq:kappag}
\kappa^{2}_{g}=\frac{\sigma(gg\to h)}{\sigma(gg\to h)_{\rm{SM}}}=
\frac{|c_tF_{\triangle}+\frac{2}{3}c_g|^2+|\tilde{c}_tF_\triangle^{(1)}
+\frac{2}{3}\tilde{c}_g|^2}{|F_{\triangle}|^2},
\eeq
and 
\beq
\label{eq:kappaa}
\kappa^{2}_{\gamma}=\frac{\Gamma(h\to \gamma\gamma)}
{\Gamma(h\to \gamma\gamma)_{\rm{SM}}}=\frac{|F_1(\tau_W)
+\frac{4}{3}c_tF_{\triangle}|^2+|\frac{4}{3}\tilde{c}_t
F_\triangle^{(1)}|^2}{|F_1(\tau_W)+\frac{4}{3}F_{\triangle}|^2},
\eeq
respectively known as the $\kappa$-framework~\cite{Heinemeyer:2013tqa}, where
the $hWW$ coupling is assumed to be the SM value. The
form factor $F_1(\tau_W)$ is defined as~\cite{Djouadi:2005gi}
\beq
F_1(\tau_W)=-\frac{1}{2}[2+3\tau_W+3\tau_W(2-\tau_W)f(\tau_W)],\
\tau_W\equiv 4m_W^2/{m_h}^2.
\eeq
The current bound on $(\kappa_{\gamma},\kappa_{g})$ from the combination of the ATLAS and CMS results at the 7 and 8 TeV LHC is shown in Refs.~\cite{ATLAS:2015comb,CMS:2015kwa,Khachatryan:2016vau}. Note that the 13 TeV LHC data
does not give a stronger constraint~\cite{CMS:2016ixj,ATLAS:2016nke}
\begin{figure}[!htb]
\centering
\includegraphics[width=0.25\textwidth]{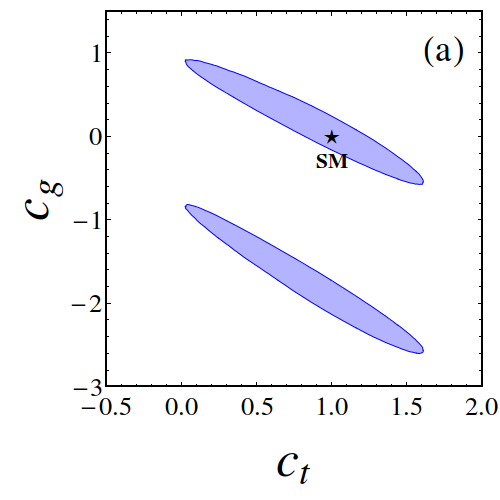}
\includegraphics[width=0.25\textwidth]{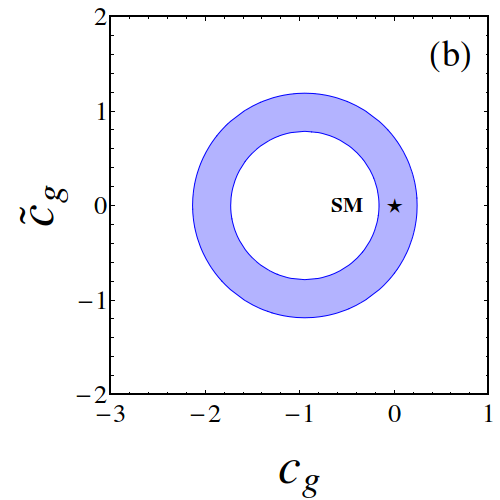}
\includegraphics[width=0.25\textwidth]{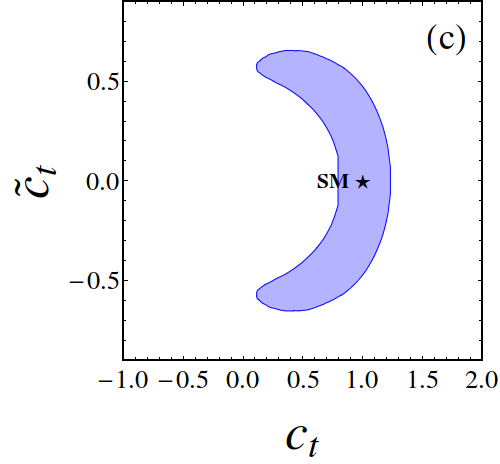}\\
\includegraphics[width=0.25\textwidth]{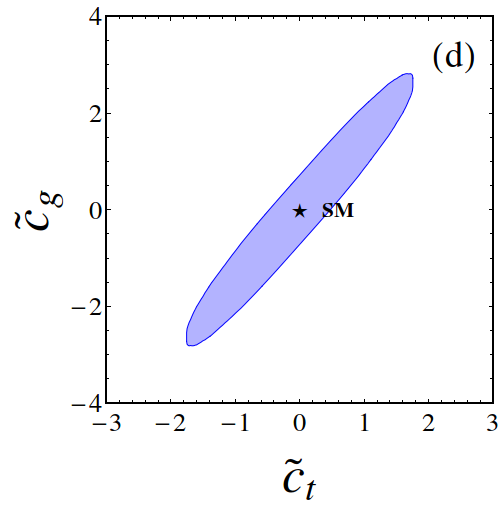}
\includegraphics[width=0.25\textwidth]{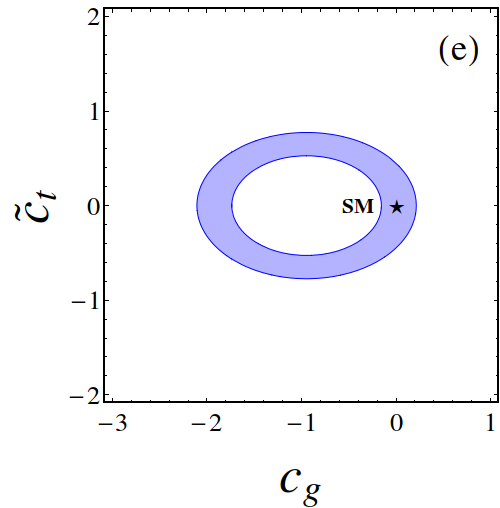}
\includegraphics[width=0.25\textwidth]{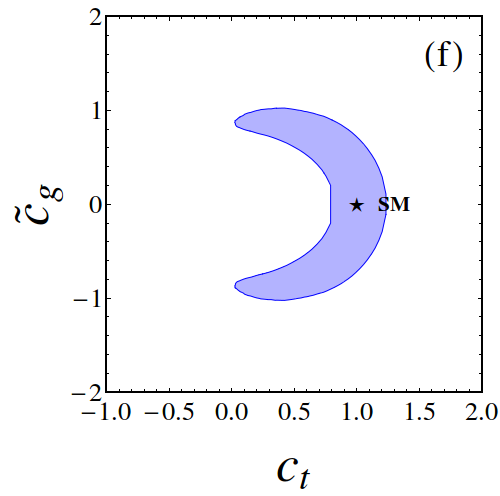}
\caption{The current constraints at $95\%$ CL on $(c_t,c_g)$,
$(c_g,\tilde{c}_g)$, $(c_t,\tilde{c}_t)$, $(\tilde{c}_t,\tilde{c}_g)$, $(c_g,\tilde{c}_t)$ and $(c_t,\tilde{c}_g)$. In each case, we consider
only two couplings and set others to be the SM values.}
\label{fig:constraint}
\end{figure}

In Fig.~\ref{fig:constraint}, we display the allowed regions of the effective couplings by the current single Higgs measurements\footnote{For the ``single Higgs measurements", we mean that there is only one Higgs boson that contributes to the process measured. } at the LHC at $95\%$ confidence level (CL), which are shown in blue bands. Only two effective couplings vary in each plot while other couplings are set to be the SM values.
Scenario $(c_t,c_g)$ shows two isolated bands, since the constraint from $\kappa_g^2$ is proportional to
$|c_tF_{\triangle}+\frac{2}{3}c_g|^2$, which can be simplified as $|c_g+c_t|^2$
in the infinite $m_t$ limit. Therefore, there are two degenerate regions
satisfying the constraints, $c_g+c_t\sim\pm 1$. Besides, the lower and upper limits of $c_t$ come from the constraint of
$\kappa_\gamma^2$. So $c_t$ cannot be too small or large,
otherwise $\Gamma(h\to\gamma\gamma)$ will be too large.
The constraint on scenario $(c_g,\tilde{c}_g)$ comes only from $\kappa_g^2$,
which is $|c_g+1|^2+|\tilde{c}_g|^2$ in the infinite $m_t$ limit. It's obvious that
in order to satisfy the single Higgs measurements, the allowed parameter
space must be a ring.
In scenario $(c_t,\tilde{c}_t)$, we have to consider both constraints on
$\kappa_g^2$ and $\kappa_\gamma^2$. In the infinite $m_t$ limit, $\kappa_g^2$
is approximated to be $|c_t|^2+\frac{9}{4}|\tilde{c}_t^2|$. The allowed region
from this constraint is an elliptical ring. The constraint on $\kappa_\gamma^2$ will further reject the $c_t\lesssim 0$ region. In scenario $(\tilde{c}_t,\tilde{c}_g)$, $\kappa_g^2\sim 1+|\tilde{c}_g
-\frac{3}{2}\tilde{c}_t|^2$ in the infinite $m_t$ limit. The
constraint on $\kappa_g^2$
will lead to $\tilde{c}_g\sim \frac{3}{2}\tilde{c}_t$, and $\kappa_\gamma^2$ will
give lower and upper limits on $\tilde{c}_t$. The correlation in this scenario is different
from scenario $(c_t,c_g)$ due to the relative minus sign between $F_\triangle$
and $F_\triangle^{(1)}$.
Similar to scenario $(c_g,\tilde{c}_g)$ the allowed region is a ring in scenario $(c_g,\tilde{c}_t)$ as a fact of $\kappa_g^2\sim |c_g+1|^2+\frac{9}{4}|\tilde{c}_t|^2$.
In principle, $\kappa_\gamma^2$ will further give a constraint on $\tilde{c}_t$. But
it turns out that the constraint from $\kappa_\gamma^2$ is not so strong such that the region
allowed by $\kappa_g^2$ satisfies the $\kappa_\gamma^2$ constraint.
The situation of scenario $(c_t,\tilde{c}_g)$ is also similar to scenario $(c_t,\tilde{c}_t)$.
In the infinite $m_t$ limit, we have $\kappa_g^2\sim |c_t|^2+|\tilde{c}_g|^2$.

On the other hand, the individual constraints on $c_t$, $\tilde{c}_t$, $c_g$ and $\tilde{c}_g$ at $95\%$ CL from the single Higgs measurements are
\begin{align}
\label{eq:1para_constr}
c_t \in [0.839, 1.24],\ \tilde{c}_t \in [-0.345, 0.186],\
c_g \in {[-2.134,-1.731]\cup[-0.161, 0.242]},\ \tilde{c}_g \in [-0.279, 0.517].
\end{align}
{Apart} from the LHC measurements, the CP-odd couplings are also constrained by the electric dipole moments of electron, neutron and
mercury atom (Hg)~\cite{Brod:2013cka,Chien:2015xha},
\begin{align}
|\tilde{c}_t|<0.01,\ |\tilde{c}_g|<0.05.
\end{align}

\section{Double Higgs production}\label{sec:3}
\subsection{Amplitude}\label{sec:31}
In this section, we will discuss the amplitude of double Higgs production
via gluon fusion $gg\to hh$.
The LO partonic amplitude of $g^{{a,\mu}}(p_{{1}})g^{{b,\nu}}
(p_{{2}})\to h(p_{{3}})h(p_{{4}})$ with the effective Lagrangian in Eq.~\eqref{eq:effLag} is
\bea
\label{eq:2hamp}
\mathcal{M}_{hh}&=&\biggl[c_{t}^2 \mathcal{M}_{\Box}^{SM}+\tilde{c}_{t}^2
\mathcal{M}_{\Box}^{(1)}+c_{t}\tilde{c}_{t}\mathcal{M}_{\Box}^{(2)}
+c_{t}c_{3h}\mathcal{M}_{\triangle}^{SM}
+\tilde{c}_{t}c_{3h}\mathcal{M}_{\triangle}^{(1)}
+c_{2t}\mathcal{M}_{\triangle}^{(2)} + \tilde{c}_{2t}\mathcal{M}_{\triangle}^{(3)}\nn\\
&& ~~ +c_gc_{3h}\mathcal{M}_{g}^{(1)}
+c_g\mathcal{M}_{g}^{(2)}
+\tilde{c}_{g}c_{3h}\mathcal{M}_{g}^{(3)}
+\tilde{c}_{g}\mathcal{M}_{g}^{(4)}\biggr]\epsilon^{{a}}_{\mu}
(p_{{1}})\epsilon^{{b}}_{\nu}(p_{{2}}) \delta^{ab},
\eea
where $a$ and $b$ in the superscript denote the color index of gluons, 
\begin{align}
&\mathcal{M}_{\Box}^{SM}= -\frac{\alpha_s\hat{s}}{4\pi v^2}
(F_{\Box}A^{\mu\nu}+G_{\Box}B^{\mu\nu}),\nn\\
&\mathcal{M}_{\Box}^{(1)}= -\frac{\alpha_s\hat{s}}{4\pi v^2}
(F_{\Box}^{(1)}A^{\mu\nu}+G_{\Box}^{(1)}B^{\mu\nu}),
&& \mathcal{M}_{\Box}^{(2)}=\frac{\alpha_s\hat{s}}{4\pi v^2}
F_{\Box}^{(2)}C^{\mu\nu},\nn\\
&\mathcal{M}_{\triangle}^{SM}= -\frac{\alpha_s\hat{s}}
{4\pi v^2}\frac{{m_h}^2}{\hat{s}-{m_h}^2}F_{\triangle}A^{\mu\nu},
&& \mathcal{M}_{\triangle}^{(1)}=\frac{\alpha_s\hat{s}}{4\pi v^2}
\frac{3m_H^2}{\hat{s}-{m_h}f^2}F_\triangle^{(1)}C^{\mu\nu},\nn\\
&\mathcal{M}_{\triangle}^{(2)}= -\frac{\alpha_s\hat{s}}{4\pi v^2}
F_{\triangle}A^{\mu\nu},
&& \mathcal{M}_{\triangle}^{(3)}=\frac{\alpha_s\hat{s}}
{4\pi v^2}F_\triangle^{(1)}C^{\mu\nu},\nn\\
&\mathcal{M}_{g}^{(1)}= -\frac{\alpha_s\hat{s}}{4\pi v^2}
\frac{3{m_h}^2}{\hat{s}-{m_h}^2}\frac{2}{3} A^{\mu\nu},
&&\mathcal{M}_{g}^{(2)} = -\frac{\alpha_s\hat{s}}{4\pi v^2}
\frac{2}{3} A^{\mu\nu},\nn\\
& \mathcal{M}_{g}^{(3)}= \frac{\alpha_s\hat{s}}{4\pi v^2}
\frac{3{m_h}^2}{\hat{s}-{m_h}^2}\frac{2}{3} C^{\mu\nu},
&& \mathcal{M}_{g}^{(4)}=\frac{\alpha_s\hat{s}}{4\pi v^2}
\frac{2}{3} C^{\mu\nu}.
\end{align}
Here the Mandelstam variables of the partonic process are defined as
\beq
\hat{s}=(p_{{1}}+p_{{2}})^2,\ \hat{t}=(p_{{1}}-p_{{3}})^{2},\
\hat{u}=(p_{{2}}-p_{{3}})^{2}.
\eeq
The Lorentz structures $A^{\mu\nu}$ and $C^{\mu\nu}$ are defined in Eqs.
\eqref{eq:lor} while $B^{\mu\nu}$ is~\cite{Glover:1987nx,Plehn:1996wb}
\bea
B^{\mu\nu}&=&g^{\mu\nu}+\frac{p_{{3}}^{2}p_{{1}}^{\nu}
p_{{2}}^{\mu}}{p_{T}^2 p_{{1}}\cdot p_{{2}}}-\frac{2p_{{2}}
\cdot p_{{3}}p_{{1}}^{\nu}p_{{3}}^{\mu}}{p_{T}^2 p_{{1}}
\cdot p_{{2}}}-\frac{2p_{{1}}\cdot p_{{3}}p_{{2}}^{\mu}
p_{{3}}^{\nu}}{p_{T}^2 p_{{1}}\cdot p_{{2}}}
+\frac{2p_{{3}}^{\mu}p_{{3}}^{\nu}}{p_{T}^{2}},
\eea
with $p_{T}^{2}=(\hat{u}\hat{t}-{m_h}^4)/\hat{s}$. It can be easily verified
that $A^{\mu\nu}A_{\mu\nu}=B^{\mu\nu} B_{\mu\nu}=C^{\mu\nu} C_{\mu\nu}=2$
and $A^{\mu\nu} B_{\mu\nu}=B^{\mu\nu} C_{\mu\nu}=A^{\mu\nu} C_{\mu\nu}=0$.
Therefore we can expand the amplitude in terms of those tensor structures as follows
\bea
\label{eq:amp}
\mathcal{M}_{hh}&=&-\frac{\alpha_s \hat{s}\delta^{ab}}{4\pi v^2}\epsilon_{\mu}(p_{a})
\epsilon_{\nu}(p_{b})\{[c_t^2F_{\Box}+\tilde{c}_t^2F_{\Box}^{(1)}+\frac{3{m_h}^2}
{\hat{s}-{m_h}^2}c_{3h}(c_t F_{\triangle}+\frac{2}{3}c_g)+\frac{2}{3}c_g+c_{2t}
F_{\triangle}]A^{\mu\nu}\nn\\
&+&(c_t^2G_{\Box}+\tilde{c}_t^2G_{\Box}^{(1)})B^{\mu\nu}
-[c_t\tilde{c}_tF_{\Box}^{(2)}+\frac{3{m_h}^2}{\hat{s}-{m_h}^2}c_{3h}(\tilde{c}_t
F_\triangle^{(1)}+\frac{2}{3}\tilde{c}_g)+\frac{2}{3}\tilde{c}_g+\tilde{c}_{2t}
F_\triangle^{(1)}]C^{\mu\nu}\}.
\eea
The expressions of the form factors $F_{\Box}$, $G_{\Box}$, $F_{\Box}^{(1)}$,
$G_{\Box}^{(1)}$, $F_{\Box}^{(2)}$, $F_{\triangle}$ and $F_\triangle^{(1)}$ can
be found in Appendix \ref{formfactors}. In the large $m_t$ limit
\cite{Plehn:1996wb}, we have
\begin{equation}
\label{eq:ffLET}
\begin{split}
F_{\Box}&=-\frac{2}{3}+\mathcal{O}(\hat{s}/m_t^2),\qquad
G_{\Box}=\mathcal{O}(\hat{s}/m_t^2),\qquad
F_{\triangle}=\frac{2}{3}+\mathcal{O}(\hat{s}/m_t^2),\\
F_{\Box}^{(1)}&=\frac{2}{3}+\mathcal{O}(\hat{s}/m_t^2),\qquad
F_{\Box}^{(2)}=2+\mathcal{O}(\hat{s}/m_t^2),\qquad
G_{\Box}^{(1)}=\mathcal{O}(\hat{s}/m_t^2),\qquad
F_\triangle^{(1)}= -1+\mathcal{O}(\hat{s}/m_t^2),\\
\end{split}
\end{equation}
which can be obtained from the low energy theorem (LET)
\cite{Shifman:1979eb,Kniehl:1995tn}.

\begin{figure}[!htb]
\centering
\includegraphics[width=0.245\textwidth]{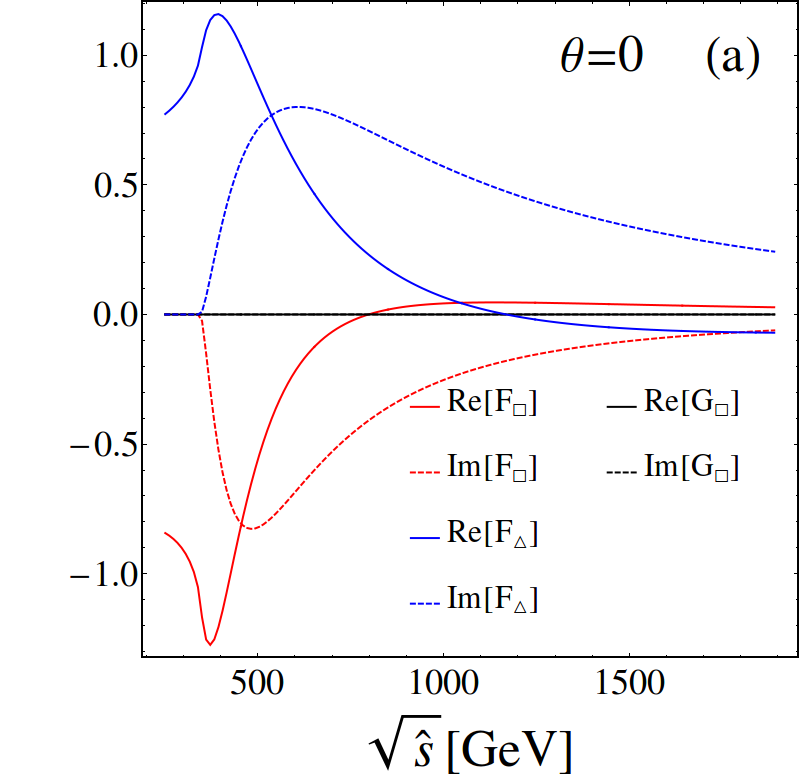}
\includegraphics[width=0.235\textwidth]{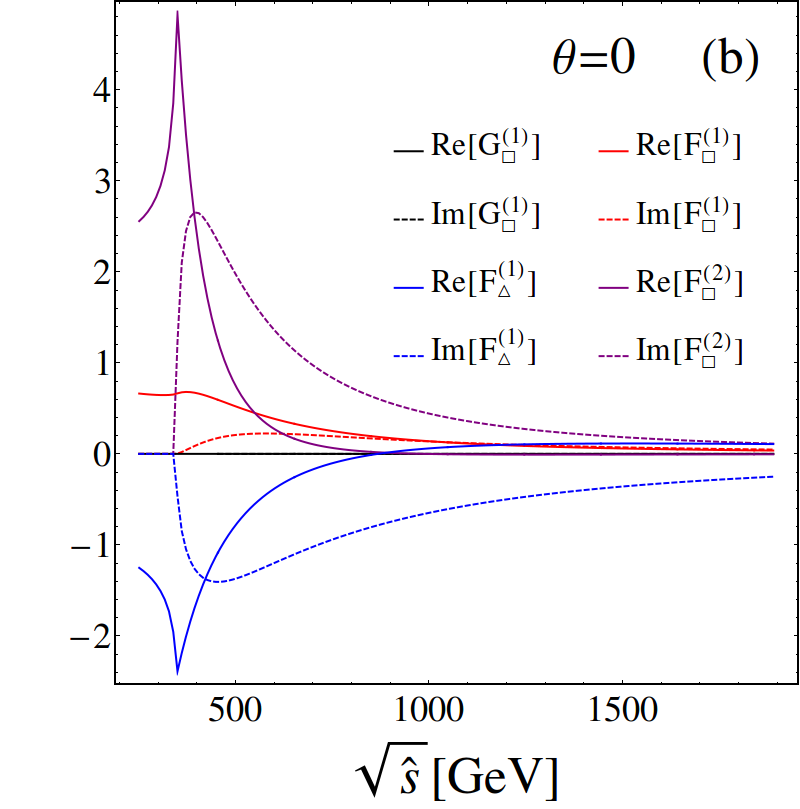}
\includegraphics[width=0.245\textwidth]{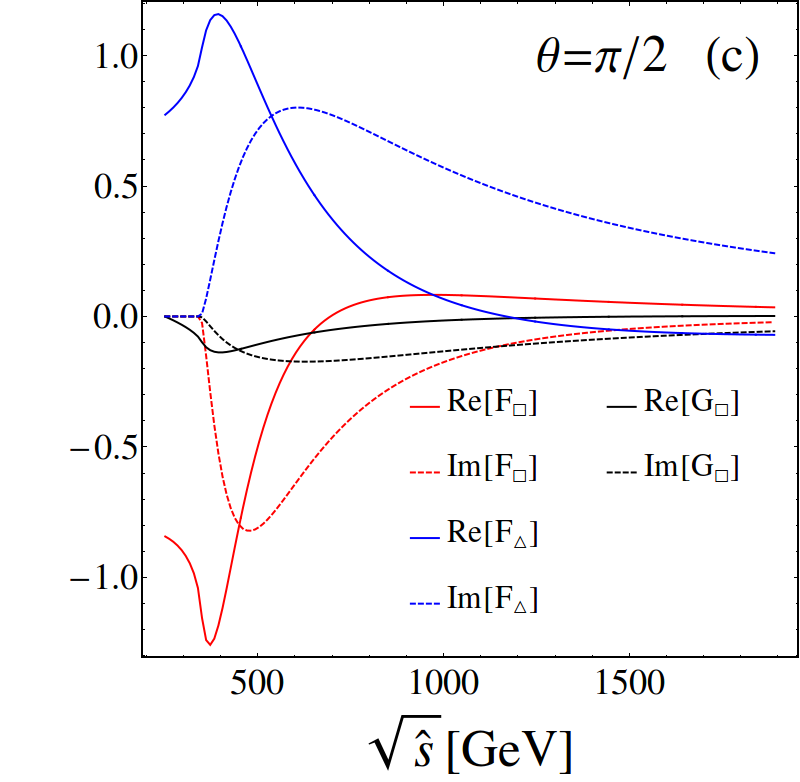}
\includegraphics[width=0.235\textwidth]{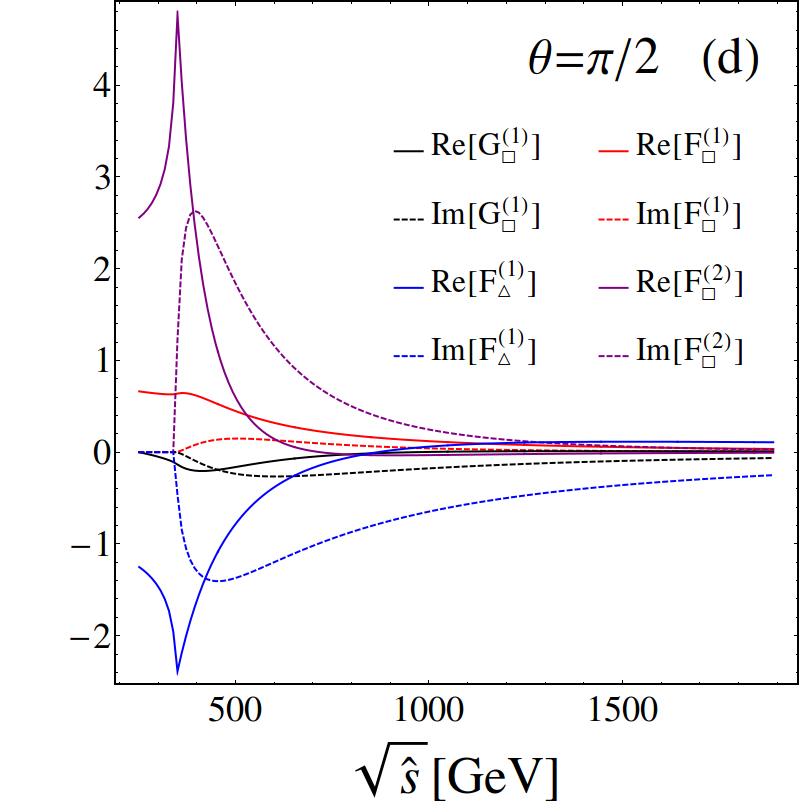}
\caption{The form factors dependence on the partonic c.m. energy $\sqrt{\hat{s}}$,
where we have chosen the scattering angle $\theta=0$ and $\pi/2$.}
\label{fig:formfactors}
\end{figure}

Figure~\ref{fig:formfactors} shows the {$\sqrt{\hat{s}}$-}dependence of
each form factor, where we have chosen two specific values of
$\theta$ which is defined as the scattering angle of the initial gluon
and final Higgs boson.
Numerically, {the} $F$ form factors are always larger than {the} $G$ form factors around the threshold region 
$\sqrt{\hat{s}}\sim 400$ GeV, where the dominant cross section arises.
Unlike the $G$ form factors, the $F$ form factors is insensitive to $\theta$.
Thus the partonic cross section is dominated by the $s$-wave around the threshold region.
To evaluate the $s$-wave and $d$-wave contributions at a large $\sqrt{\hat{s}}$, it is necessary to perform a partial wave analysis.

\subsection{Partial wave analysis}\label{sec:32}
\begin{figure}[!htb]
\centering
\includegraphics[width=0.252\textwidth]{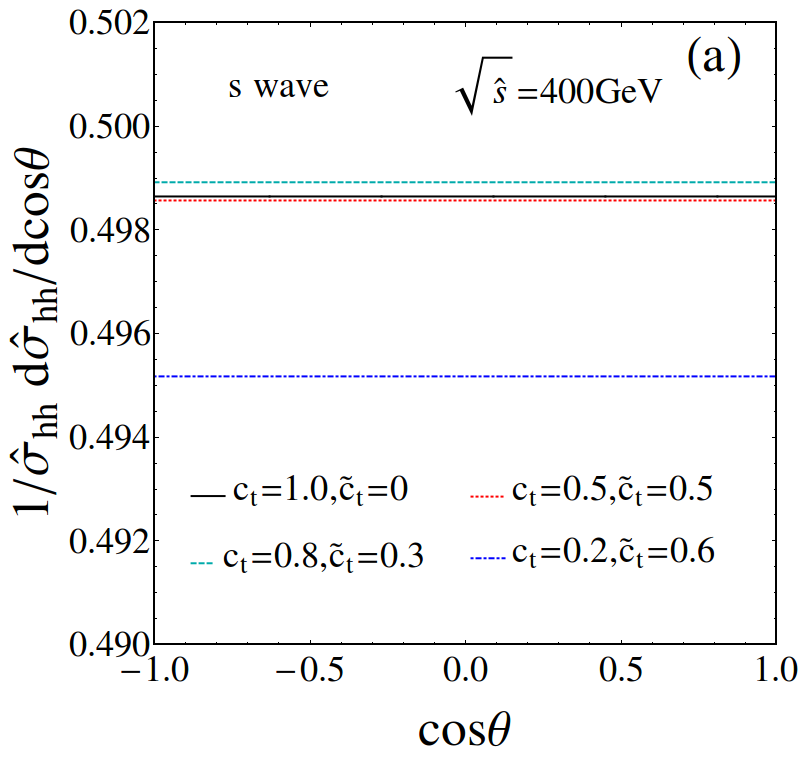}
\includegraphics[width=0.238\textwidth]{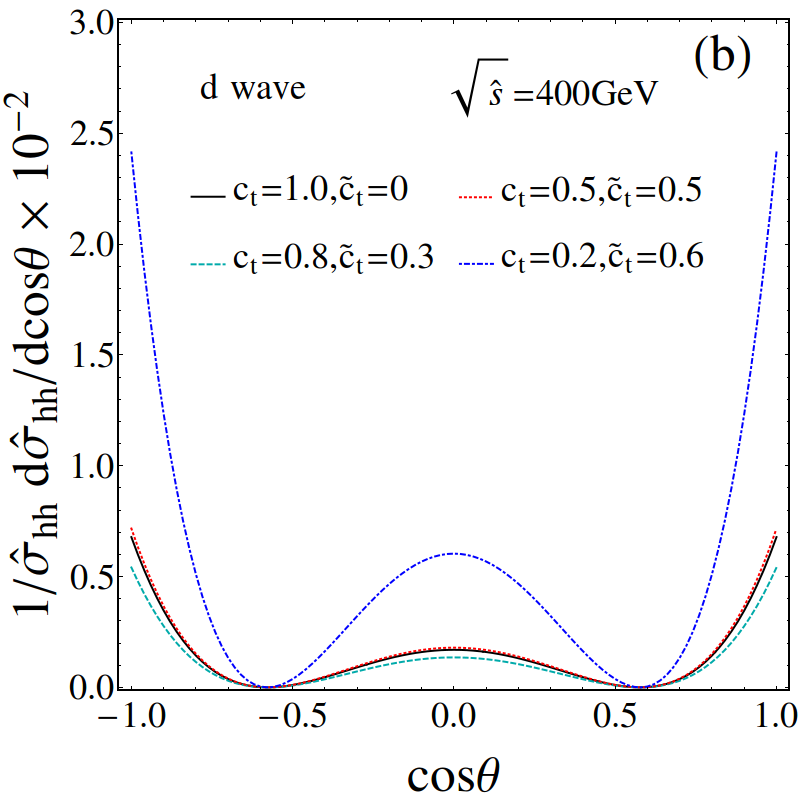}
\includegraphics[width=0.248\textwidth]{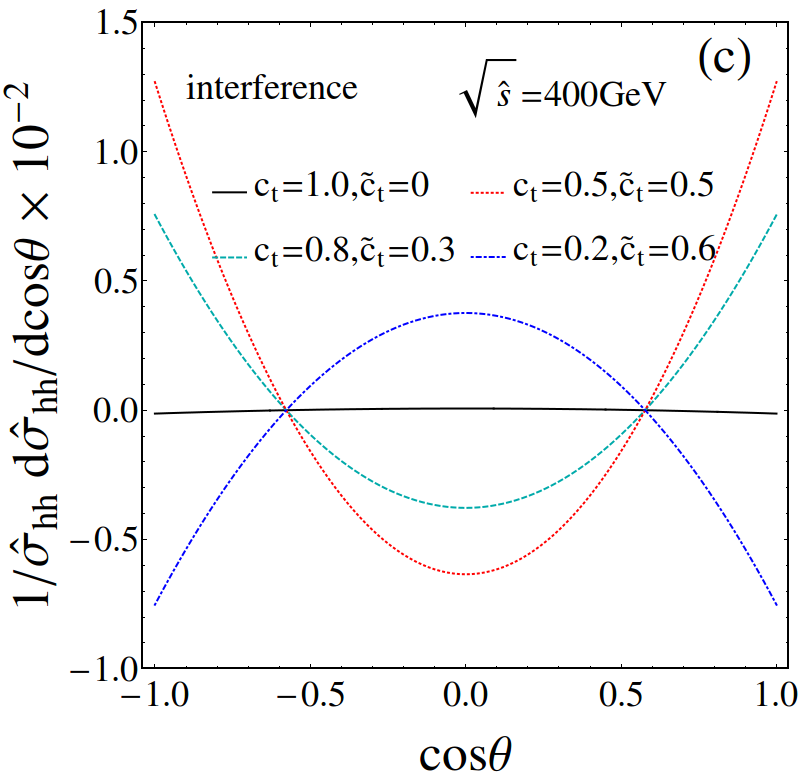}\\
\includegraphics[width=0.248\textwidth]{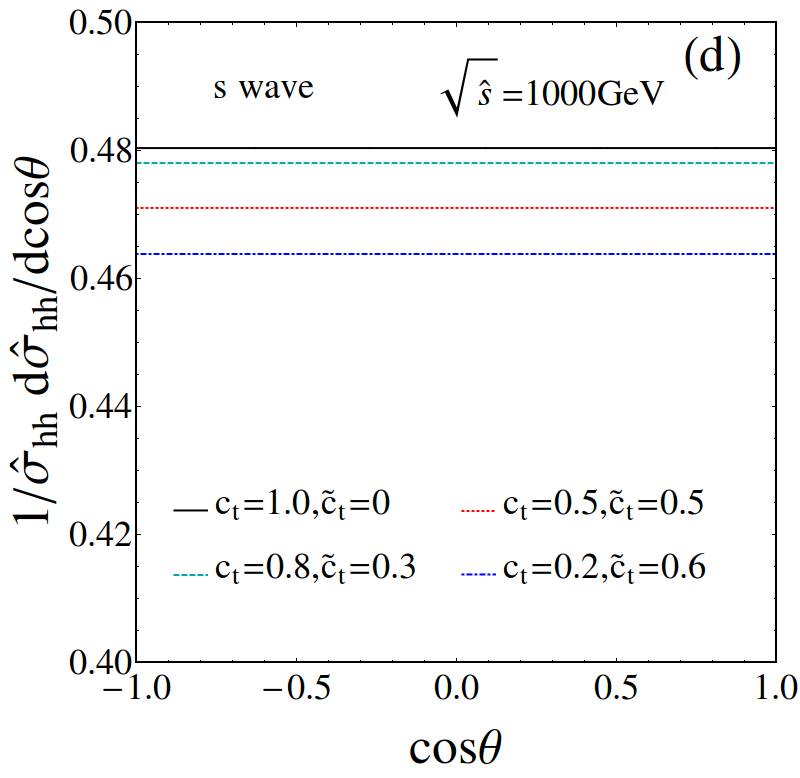}
\includegraphics[width=0.243\textwidth]{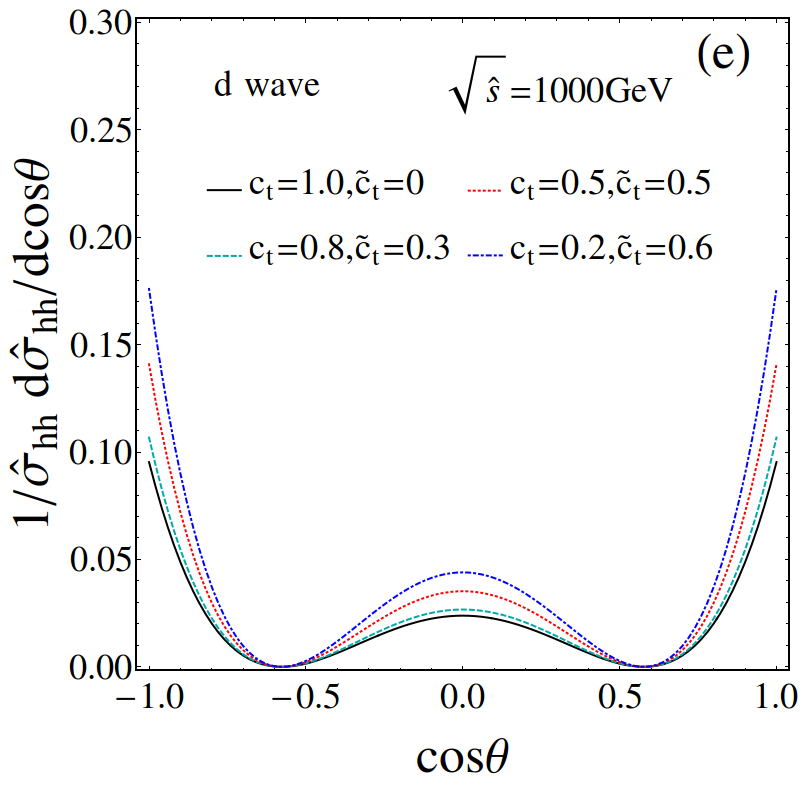}
\includegraphics[width=0.248\textwidth]{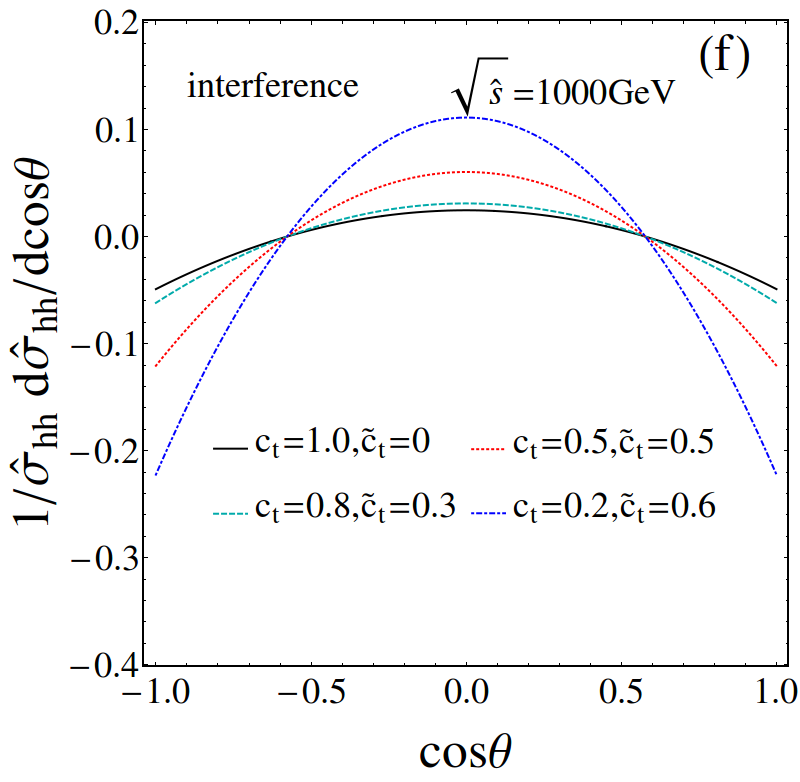}
\caption{$\cos\theta$ distributions from the $s$-wave  and $d$-wave contributions
at $\sqrt{s}=400\gev,1000\gev$. The black (solid), cyan (dashed), red
(dot-dashed) and blue (dotted) lines correspond to $(c_t,\tilde{c}_t)=(1,0)$,
$(0.8,0.3)$, $(0.5,0.5)$, $(0.2,0.6)$, respectively.}
\label{fig:dx400}
\end{figure}

\begin{figure}[!htb]
\centering
\includegraphics[width=0.3\textwidth]{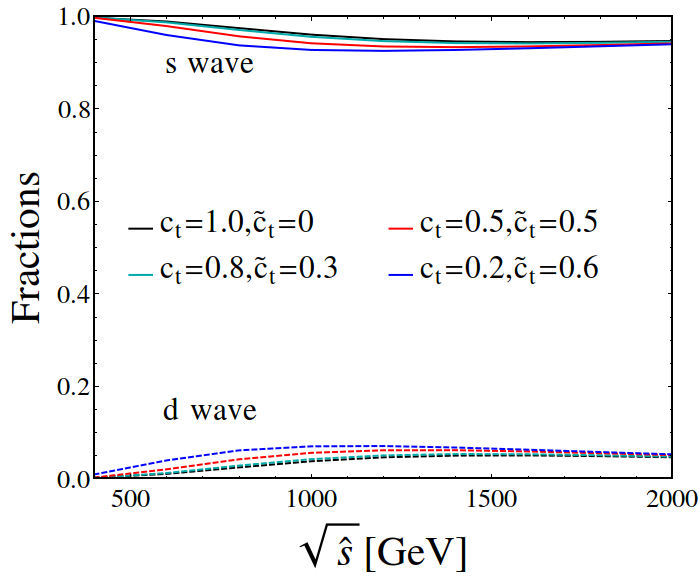}
\caption{Fractions of the $s$-wave and $d$-wave contributions to the partonic cross section as a function of $\sqrt{\hat{s}}$ for different $c_t$ and $\tilde{c}_t$.}
\label{fig:PWX}
\end{figure}

The amplitude in the partial wave {expansion} is given by
\be
\mathcal{M}_{hh}(\hat{s},\theta)=\sum_{\ell=0,2} (2\ell +1) \mathcal{M}_{\ell}
(\hat{s}) P_{\ell}(\cos\theta),
\ee
where $P_{\ell}(x)$ are the Legendre polynomials satisfy ing the orthogonal
relation $\displaystyle{\int_{-1}^{1}dxP_\ell(x)P_{\ell^{\prime}}(x)=\frac{2}{(2\ell+1)}
\delta_{\ell\ell^{\prime}}}$.
The $s$-wave and $d$-wave components of the amplitude {are proportional}
to $P_0(x)$ and $P_2(x)$ respectively, which are
\bea
\mathcal{M}_0(\hat{s})&=&\frac{1}{2}\int_{-1}^1 d\cos\theta \mathcal{M}_{hh}
(\hat{s},\theta) P_0(\cos\theta),\\
\mathcal{M}_2(\hat{s})&=&\frac{1}{2}\int_{-1}^1 d\cos\theta \mathcal{M}_{hh}
(\hat{s},\theta) P_2(\cos\theta).
\eea
Now the {partonic level} differential cross section with respect to
$\cos\theta$ can be expanded into three terms
\be
\label{eq:pw_0}
\frac{d\hat{\sigma}_{hh}(\hat{s})}{d\cos\theta}=\hat{\sigma}_0(\hat{s})
+\hat{\sigma}_2(\hat{s})P_2(\cos\theta)^2
+\hat{\sigma}_{\text{int}}{(\hat{s})}P_2(\cos\theta),
\ee
where the first and the second terms denote the $s$-wave and $d$-wave contributions, respectively. The third term, which arises from the interference of the $s$-wave and $d$-wave components of the amplitude, vanishes after integrating over $\cos\theta$. $\hat{\sigma}_0(\hat{s})$, $\hat{\sigma}_2(\hat{s})$ and $\hat{\sigma}_{\text{int}}{(\hat{s})}$ can be obtained numerically.
{The angular dependence of the form factors can be clearly revealed by choosing different combinations of ($c_t,\tilde{c}_t$).} 
In Fig.~\ref{fig:dx400}, we show the {three} terms in Eq.~\eqref{eq:pw_0} with $\sqrt{\hat{s}}=400~\rm GeV,1000~\rm GeV$ for
different $(c_t,\tilde{c}_t)$, {where we fix other Higgs effective couplings to be the SM values and normalize the three terms to the total cross sections.} 
Since the $s$-wave has no angular dependence, the distributions
in Figs.~\ref{fig:dx400}(a) and (d) are flat. {The} $d$-wave {and the interference contributions have nontrival angular dependences}, which are reflected in Figs.~\ref{fig:dx400}(b, e) and (c, f), respectively. From the distributions, the {$d$-wave} and the {interference contributions} are comparable for $\sqrt{\hat{s}}= 400~\rm GeV$ and $1000~\rm GeV$. This is because the imaginary (real) parts of the form factors at $\sqrt{\hat{s}}= 400~(1000)~\rm GeV$ are small {such that} the interference {contribution is} suppressed. 
While increasing the $\sqrt{\hat{s}}$ from 400 GeV to 1000 GeV, the fractions of the $d$-wave and the interference contributions grow almost one order of magnitude. However, their contributions are still overwhelmed by the $s$-wave.
So the $d$-wave and the interference contributions to $\cos\theta$ distributions at the hadron level, such as transverse momentum or rapidity distributions, are small.
Figure~\ref{fig:PWX} shows the $s$ and $d$-wave contributions to the total cross sections after integrating over $\cos\theta$. It is clear that the $d$-wave contributions are at most {of} $10\%$ of the total cross sections. As a result, the invariant mass distributions at the hadron level are dominated by the $s$-wave.

To be concrete, we show the $s$-wave and $d$-wave contributions to the total cross section at the hadron level,
\begin{align}
\sigma_0&=\int_{\tau_0}^{1}d\tau \int_{\tau}^{1}\frac{dx}{x}f_g(x;\mu_F^{2})
f_g(\frac{\tau}{x};\mu_F^{2})\int_{-1}^{1} d\cos\theta \hat{\sigma}_0({\tau s})
P_0(\cos\theta)^2,\\
\sigma_2&=\int_{\tau_0}^{1}d\tau \int_{\tau}^{1}\frac{dx}{x}f_g(x;\mu_F^{2})f_g
(\frac{\tau}{x};\mu_F^{2})\int_{-1}^{1} d\cos\theta \hat{\sigma}_2({\tau s})
P_2(\cos\theta)^2,
\end{align}
{where $\tau_0=4{m_h}^2/s$, }$f_g$'s are the PDF functions of the initial gluons,
and $\mu_F$ is the factorization scale. The hadronic cross sections can be expanded as follows,
\begin{align}
 \frac{\sigma_0}{\sigma^{SM}_{hh}}&=a_1 c_t^4+a_2 c_t^3+a_3 c_t^2\tilde{c}_t^2 +a_4 c_t^2 +a_5 c_t
 \tilde{c}_t^2 + a_6 \tilde{c}_t^4 + a_7 \tilde{c}_t^2{,}\\
 \frac{\sigma_2}{\sigma^{SM}_{hh}}&=b_1 c_t^4+b_2 c_t^3+b_3 c_t^2\tilde{c}_t^2 +b_4 c_t^2 +b_5 c_t
\tilde{c}_t^2 + b_6 \tilde{c}_t^4 + b_7 \tilde{c}_t^2,
\end{align}
where $\sigma^{SM}_{hh}$ denotes the hadronic cross section of $gg\to hh$ in the SM, which has been calculated at the LO
\cite{Eboli:1987dy,Glover:1987nx,Plehn:1996wb}, NLO
\cite{Dawson:1998py,Grigo:2013rya,Frederix:2014hta,Maltoni:2014eza,
Borowka:2016ehy,Borowka:2016ypz,Kerner:2016msj}, NLL
\cite{Ferrera:2016prr}, NNLO~\cite{deFlorian:2013uza,deFlorian:2013jea,
Grigo:2014jma,Grigo:2015dia,Degrassi:2016vss,deFlorian:2016uhr,
Hoff:2016bdo} and NNLL~\cite{Shao:2013bz,deFlorian:2015moa}. The coefficients of the expansions are displayed in {Table}~\ref{tbl:coeff1}. {Total} cross sections at the 14 TeV LHC and the 100 TeV $pp$-collider are dominated by the $s$-wave component. Besides, {the fractions of the $s$-wave contribution at the 100 TeV $pp$-collider are smaller than the fractions at the 14 TeV LHC, while the fractions of the $d$-wave contribution at the 100 TeV $pp$-collider are larger than the fractions at the 14 TeV LHC.}

\begin{table}[!htb]
\tabcolsep=10pt
\caption{{The coefficients $a_i$ and $b_i$ at $\sqrt{s}=14~\rm TeV$ and 100 TeV.} }
\begin{tabular}{cccccccccccccccc}
\hline
 $\sqrt{s}$ & $a_1$ & $a_2$ & $a_3$ & $a_{4}$ & $a_{5}$ & $a_{6}$& $a_{7}$ \\
\hline
$14~\rm TeV$ & 2.069 & -1.351 & 13.858 & 0.276 & -6.219 & 0.706 & 0.861 \\
$100~\rm TeV$ &  1.891 & -1.108 & 11.280 & 0.208 & -4.795 & 0.663 & 0.634 \\
\hline
 $\sqrt{s}$& $b_1$ & $b_2$ & $b_3$ & $b_{4}$ & $b_{5}$ & $b_{6}$& $b_{7}$\\
\hline
$14~\rm TeV$ & 0.006 & 0 & 0.020 & 0 & -0.136 & 0.013 & 0 \\
$100~\rm TeV$ & 0.009 & 0 & 0.027 & 0 & -0.137 & 0.017 & 0\\
\hline
\end{tabular}
\label{tbl:coeff1}
\end{table}
From the above partial wave analysis, we draw a few conclusions, which do not rely on the Higgs effective couplings.
\begin{enumerate}
\item[(1)] The $d$-wave contributions to the $\cos\theta$ distributions at the hadron level, such as transverse momentum or rapidity distributions, are always small;
\item[(2)] The $d$-wave contributions to the invariant mass distributions at the hadron level are small;
\item[(3)] {The} $d$-wave contributions to the total cross sections are small.
\end{enumerate}
Therefore, it is justified that the double Higgs production is dominated by the $s$-wave.

\section{Collider simulation}\label{sec:4}
In {Table}~\ref{tab:hig-pair} we collect references of the searches of double Higgs production at the 8 TeV
and the 13 TeV LHC as well as the {projected} analyses at HL-LHC and 100 TeV
$pp$-collider in the final states $b\bar{b}\gamma\gamma$, $b\bar{b}b\bar{b}$,
$b\bar{b}\tau^+\tau^-$, $b\bar{b}W^+W^-$, $\gamma\gamma W^+W^-$ and $b\bar{b}VV(V=W,Z)$. {Although $b\bar{b}b\bar{b}$ channel has the largest cross section}, the QCD background
is hard to control. On the other hand, the $b\bar{b}\gamma\gamma$ channel, despite of its small decay branching ratio, exhibits clear collider signature. Therefore, it has been studied extensively in the literature~\cite{Baur:2003gp,Contino:2012xk,
Baglio:2012np,
Chen:2014xra,Azatov:2015oxa,Cao:2015oaa}. In this study, we focus our attention on the $b\bar{b}\gamma\gamma$ channel and use the cut efficiency function $\mathcal{A}(m_{hh})$ to mimic the detector effects in different NP scenarios\footnote{Hereafter, NP in this paper denotes those which can be described by the effective Lagrangian in Eq.~\eqref{eq:effLag}.}. At the 14 TeV HL-LHC, we follow the analysis done by the ATLAS Collaboration~\cite{Aad:2015} and adopt the cut efficiency function in Ref.~\cite{Cao:2015oaa}. At the 100 TeV $pp$-collider, we follow the projected analysis done by the 100 TeV group~\cite{Contino:2016spe}. In the rest of this section, we will first derive cut efficiency function of $gg\to hh \to b\bar{b}\gamma\gamma$ at the 100 TeV $pp$-collider, then we will discuss the correlations and sensitivities of Higgs effective couplings.
\begin{table}[!htb]
\tabcolsep=5pt
\caption{Searches of double Higgs production at the $8~\rm{TeV}$ and $13~\rm TeV$ by the ATLAS and
CMS Collaborations, as well as the projected analysis at the $14~\rm{TeV}$
High-Luminosity LHC (HL-LHC) and the $100~\rm TeV$ $pp$-collider.}
\begin{tabular}{cccccccc}
\hline
 &  & $b\bar{b}\gamma\gamma$ & $b\bar{b}b\bar{b}$ & $b\bar{b}\tau^+\tau^-$
 & $\gamma\gamma W^+W^-$ & $b\bar{b}VV$ \\
\hline
$8~\rm{TeV}$&ATLAS&\cite{Aad:2014yja,Aad:2015xja}&\cite{Aad:2015uka,Aad:2015xja}&
~\cite{Aad:2015xja}&\cite{Aad:2015xja} & -\\
\hline
$8~\rm{TeV}$&CMS&\cite{CMS:2014ipa}&\cite{Khachatryan:2015yea}
&\cite{CMS:2016zxv}\cite{Khachatryan:2015tha}&- & -\\
\hline
$13~\rm{TeV}$&ATLAS&\cite{ATLAS:2016bbaa}&\cite{Aaboud:2016xco}&-&- & -\\
\hline
$13~\rm{TeV}$&CMS&\cite{CMS:2016vpz}&\cite{CMS:2016foy}\cite{CMS:2016tlj}
&\cite{CMS:2016ugf}\cite{CMS:2016guv}\cite{CMS:2016knm}\cite{CMS:2016ymn}&
-&\cite{CMS:2016cdj}\cite{CMS:2016rec}\\
\hline
$14~\rm{TeV}$ HL-LHC&ATLAS&\cite{Aad:2015}&-&\cite{Aad:2013il}&- & -\\
\hline
$14~\rm{TeV}$ HL-LHC&CMS&\cite{CMS:2015nat}&-&\cite{CMS:2015nat}&
\cite{CMS:2015nat} & -\\
\hline
$100~\rm{TeV}$ $pp$-collider& Contino, et. al. &\cite{Contino:2016spe}
&\cite{Contino:2016spe}&\cite{Contino:2016spe}&\cite{Contino:2016spe}
& \cite{Contino:2016spe}\\
\hline
\end{tabular}
\label{tab:hig-pair}
\end{table}

\subsection{Cut efficiency function}\label{sec:41}
The Born level differential cross section of double Higgs production can be written as
\be
\frac{d^2\sigma}{dm_{hh}d\eta}=\frac{1}{2S}\int dx_1dx_2
H\left(\hat s,\hat\eta,\mu_r,\{\theta_{NP}\}\right)f_{g/p}\left(x_1,\mu_f\right)
f_{g/p}\left(x_2,\mu_f\right)\delta\left(x_1x_2S-\hat s\right)
\det\left[\frac{\partial\left(\hat s,\hat\eta\right)}
{\partial\left(m_{hh},\eta\right)}\right],
\ee
where $H$ is the hard scattering cross section
depending on the center of mass energy (c.m.) square
$\hat s$ and Higgs boson pseudo-rapidity $\hat\eta$ in the
c.m. frame,
$S$ is the collision energy of the hadron collider, $\mu_f$
is the factorization scale, $f_{i/j}$ is the parton distribution
function (PDF) of parton $i$
from $j$, $m_{hh}$ is the invariant mass of Higgs boson pair in the
lab frame, $\eta$ ($\hat\eta$) is the pseudo-rapidity of Higgs boson
in the lab frame (c.m. frame),  $\{\theta_{NP}\}$ are the parameters
from the new physics model. We do not write the parameters
${m_h}^2$ and $m_t^2$ explicitly, the azimuthal angle has been
integrated out.

We know that $\hat s=m_{hh}^2$ at the Born level.
{The} Jacobian determinant is $2m_{hh}|\partial\hat\eta/\partial\eta|$, and
\bea
\frac{d^2\sigma}{dm_{hh}d\eta}&=&\frac{m_{hh}}{S}\int dx_1
dx_2\left|\frac{\partial\hat\eta}
{\partial\eta}\right|
H\left(m_{hh},\hat\eta\left(m_{hh},\eta,\frac{x_1}{x_2}\right)
,\mu_r\right)f_{g/p}\left(x_1,\mu_f\right)
f_{g/p}\left(x_2,\mu_f\right)
\delta\left(x_1x_2S-m_{hh}^2\right)
\nonumber\\
&=&\int^1_{m_{hh}^2/S} \frac{m_{hh}dx_1}{x_1S^2}
H\left(m_{hh},\hat\eta\left(m_{hh},\eta,\frac{x_1^2S}{m_{hh}^2}\right)
,\mu_r\right)f_{g/p}\left(x_1,\mu_f\right)f_{g/p}\left(\frac{m_{hh}^2}{x_1 S},\mu_f\right)
\left|\frac{\partial\hat\eta}
{\partial\eta}\right|.
\eea
{The $\eta$ and $\hat{\eta}$ are related by
\be
\eta=-\frac{1}{2}\log\left[\frac{\sqrt{\Delta^2+
4x_1x_2\beta^2{\text{sech}}^2\hat\eta}-\Delta}{\sqrt{\Delta^2+
4x_1x_2\beta^2{\text{sech}}^2\hat\eta}+\Delta}\right],
\ee
where
\bea
\beta\equiv&\displaystyle\sqrt{1-\frac{4m_h^2}{m_{hh}^2}},\
\Delta\equiv&\left(x_1-x_2\right)+\left(x_1+x_2\right)\beta\tanh\hat\eta.
\eea
Thus}
\be
\frac{\partial \hat \eta}{\partial \eta}
=\frac{\sqrt{\Delta^2+4x_1x_2\beta^2{\text{sech}}^2\hat\eta}}
{\left(x_1+x_2\right)\beta+\left(x_1-x_2\right)\tanh\hat\eta},
\ee
For gluon-fusion initial state, the main contribution comes
from the small-$x$ region with $x_1\sim x_2$. In that limit,
it is a good approximation that $\hat\eta=\eta$.

Owing to the scalar feature of Higgs boson, the kinematics of Higgs boson decay products is mainly controlled by the Higgs kinematics, e.g. $p_T$ and $\eta$ of the Higgs boson.
Thus the cut efficiency depends on the $p_{\text{T}}$ and
$\eta$ distributions of Higgs bosons. 
The transverse momentum of Higgs boson is 
\be
p_{\text{T}}=\frac{\sqrt{\hat s}}{2}\beta{\text{sech}}\hat\eta
=\frac{m_{hh}}{2}\beta{\text{sech}}
\left[\hat\eta\left(m_{hh},\eta_1,\frac{x_1^2S}{m_{hh}^2}\right)\right].
\ee
{Denote
$\epsilon\left(p_{\text{T}},\eta_1,\eta_2\right)$ to be the
differential cut efficiency function.} The pseudo-rapidity
$\eta_2$ is determined by $\hat\eta$, $m_{hh}$ and $x_1$.
{Therefore, $\epsilon$ is} a function of $m_{hh}$, $x_1$ and $\eta$
(which is just $\eta_1$).

The hard scattering function, $H$, is generically $\eta$-dependent.
Fortunately, for the SM-like double Higgs production induced by the
effective Lagrangian {given in Eq.~\eqref{eq:effLag}}, higher partial wave components
are highly suppressed.
{Therefore, we can treat $\{\theta_{NP}\}$ as $\hat \eta$-independent.} Then the amplitude
square will be $\hat \eta$-independent. Under {such} assumptions,
the differential cross section can be factorized as {following,}
\bea
\frac{d^2\sigma}{dm_{hh}d\eta}&=&\frac{m_{hh}}{S^2}H\left(m_{hh}
,\mu_r\right)\int^1_{m_{hh}^2/S} \frac{dx_1}{x_1}f_{g/p}\left(x_1,\mu_f\right)f_{g/p}\left(\frac{m_{hh}^2}{x_1 S},\mu_f\right)
\left|\frac{\partial\hat\eta}
{\partial\eta}\right|_{m_{hh},\eta,x_1}.
\eea
Integrating the pseudo-rapidity out, we have
\bea
\frac{d\sigma}{dm_{hh}}&=&\frac{m_{hh}}{S^2}H\left(m_{hh}
,\mu_r\right)\int^1_{m_{hh}^2/S} \frac{dx_1}{x_1}
f_{g/p}\left(\frac{m_{hh}^2}{x_1 S},\mu_f\right)f_{g/p}\left(x_1,\mu_f\right)
\int d\eta \left|\frac{\partial\hat\eta}
{\partial\eta}\right|_{m_{hh},\eta,x_1}\nonumber\\
&\equiv&\frac{m_{hh}}{S^2}H\left(m_{hh}
,\mu_r\right)\Sigma\left(m_{hh},S,\mu_f\right).
\eea

We can also write down the differential cross section after {kinematic cuts used by experimental groups},
\bea
\frac{d\sigma_{\text{after~cuts}}}{dm_{hh}}&=&\int d\tilde{m}_{hh}\frac{\tilde{m}_{hh}}
{S^2}H\left(\tilde{m}_{hh}
,\mu_r\right)\int^1_{\tilde{m}_{hh}^2/S} \frac{dx_1}{x_1}
f_{g/p}\left(\frac{\tilde{m}_{hh}^2}{x_1 S},\mu_f\right)f_{g/p}\left(x_1,\mu_f\right)
\nonumber\\&&
\times \int d\eta \left|\frac{\partial\hat\eta}
{\partial\eta}\right|_{\tilde{m}_{hh},\eta,x_1}
\epsilon\left(m_{hh},\tilde{m}_{hh},x_1,\eta\right),
\label{eq:full}
\eea
where $m_{hh}$ is the invariant mass of the Higgs-pair system
measured in the experiment, $\tilde m_{hh}$ is the real invariant mass of the
Higgs-pair system of the same event, which is introduced to describe the finite
energy smearing effect. For an ideal detector, we have
\be
\epsilon\left(m_{hh},\tilde{m}_{hh},x_1,\eta\right)
=\epsilon\left(m_{hh},x_1,\eta\right)
\delta\left(\tilde{m}_{hh}-m_{hh}\right),
\ee
which will be broken by finite energy smearing effect.
Due to the cut effect, {in general}
\be
\epsilon\left(m_{hh},\tilde{m}_{hh},x_1,\eta\right)
\neq\epsilon\left(\tilde{m}_{hh},m_{hh},x_1,\eta\right).
\ee
{To} investigate the inclusive result, {one} can integrate
$m_{hh}$ and have
\bea
\sigma_{\text{after~cuts}}&=&\int dm_{hh}d\tilde{m}_{hh}\frac{\tilde{m}_{hh}}
{S^2}H\left(\tilde{m}_{hh}
,\mu_r\right)\int^1_{\tilde{m}_{hh}^2/S} \frac{dx_1}{x_1}
f_{g/p}\left(\frac{\tilde{m}_{hh}^2}{x_1 S},\mu_f\right)f_{g/p}\left(x_1,\mu_f\right)
\nonumber\\&&
\times \int d\eta \left|\frac{\partial\hat\eta}
{\partial\eta}\right|_{\tilde{m}_{hh},\eta,x_1}
\epsilon\left(m_{hh},\tilde{m}_{hh},x_1,\eta\right).
\eea
{Define}
\bea
\tilde\Sigma\left(\tilde{m}_{hh},S,\mu_f\right)&\equiv&
\int dm_{hh}\int^1_{\tilde{m}_{hh}^2/S} \frac{dx_1}{x_1}
f_{g/p}\left(\frac{\tilde{m}_{hh}^2}{x_1 S},\mu_f\right)f_{g/p}\left(x_1,\mu_f\right)
\nonumber\\&&
\times \int d\eta \left|\frac{\partial\hat\eta}
{\partial\eta}\right|_{\tilde{m}_{hh},\eta,x_1}
\epsilon\left(m_{hh},\tilde{m}_{hh},x_1,\eta\right),
\eea
{we obtain}
\be
\sigma_{\text{after~cuts}}=\int d\tilde{m}_{hh}\frac{\tilde{m}_{hh}}{S^2}H\left(\tilde{m}_{hh}
,\mu_r\right)\tilde\Sigma\left(\tilde{m}_{hh},S,\mu_f\right).
\ee
Then it is natural to define a differential cut efficiency as
\be
{\mathcal{A}}\left(m_{hh},S,\mu_f\right)
=\frac{\tilde\Sigma\left(m_{hh},S,\mu_f\right)}
{\Sigma\left(m_{hh},S,\mu_f\right)}.
\label{eq:defA}
\ee
Such a differential cut acceptance function only depends
on the collision energy and the detail of the PDF.
When the new physics contribution is dominated by the $s$-wave, 
one can calculate the total cross section after cuts
by a convolution of the differential cross section of {of $m_{hh}$} and the differential cut acceptance function ${\mathcal{A}}(m_{hh},S,\mu_f)$. Hence we obtain the \textit{master formula} for our study as following
\bea
\sigma_{\text{after~cuts}}&=&\int d\tilde{m}_{hh}\frac{\tilde{m}_{hh}}
{S^2}H\left(\tilde{m}_{hh}
,\mu_r\right)\tilde\Sigma\left(\tilde{m}_{hh},S,\mu_f\right)
\nonumber\\&=&\int \frac{m_{hh}dm_{hh}}{S^2}
H\left(m_{hh}
,\mu_r\right)\tilde\Sigma\left(m_{hh},S,\mu_f\right)
\nonumber\\&=&\int \frac{m_{hh}dm_{hh}}{S^2}
H\left(m_{hh}
,\mu_r\right)\Sigma\left(m_{hh},S,\mu_f\right){\mathcal{A}}\left(m_{hh},S,\mu_f\right)\nonumber\\
&=&\int dm_{hh}\frac{d\sigma}{dm_{hh}}
{\mathcal{A}}\left(m_{hh},S,\mu_f\right).
\label{eq:convolution}
\eea
Equation \eqref{eq:defA} also tells us how to
calculate the integral kernel ${\mathcal{A}}$
practically. It can be calculated by generating $s$-wave
events with fixed $m_{hh}$ and counting the fraction
of the events which pass the cuts.
It is worth {emphasizing} that without the integration {of} $m_{hh}$, the
result is not exactly the differential distribution due to the finite
invariant mass smearing effect. However, when the smearing
effect is not too large, it is a good approximation to mimic the
differential distribution after cut as following
\be
\left(\frac{d\sigma}{dm_{hh}}\right)_{\text{after cuts}}=\frac{d\sigma}{dm_{hh}}
{\mathcal{A}}\left(m_{hh},S,\mu_f\right).
\label{eq:cut_dist}
\ee
As to be shown soon, this approximation works well for the Higgs boson pair production. Thus we will use this approximation to illustrate the differential
cross section in our work.

At the 100 TeV $pp$-collider, we can also use this analytical function to include
all the detector effects {as we did for 14 TeV LHC in \cite{Cao:2015oaa}}. We
follow the strategy in the 100 TeV report~\cite{Contino:2016spe}.  The main
backgrounds {consist of} $b\bar{b}\gamma\gamma$, $b\bar{b}j\gamma$,
$jj\gamma\gamma$, $b\bar{b}h(\gamma\gamma)$ and $t\bar{t}h(\gamma\gamma)$.
The cuts used are
\bea
&&\gamma~\text{isolation}~R=0.4,\
\text{jets: anti-kt, parameter}~R=0.4,\
\Delta R_{b b}<3.5~,~\Delta R_{\gamma\gamma}<3.5,\nonumber\\
&&p_{T}^{b_1}>60~{\rm{GeV}},~
p_{T}^{b_2}>35~{\rm{GeV}},~|\eta^{b}|<4.5,\
p_{T}^{\gamma_1}>60~{\rm GeV},~p_{T}^{\gamma_2}>35~{\rm GeV},~|\eta^{\gamma}|<4.5,\nonumber\\
&&p_T(bb)>100~{\rm GeV}, p_T(\gamma\gamma)>100~{\rm GeV},\
100{~\text{GeV}}<m_{b\bar b}<150{~\text{GeV}},\
{120.5}{~\text{GeV}}<m_{\gamma\gamma}<{129.5}{~\text{GeV}},
\label{eq:cuts2}
\eea
where $b_1$ ($b_2$) and $\gamma_1$ ($\gamma_2$) represent the leading (subleading) ${b}${-}jet and photon, respectively.
The ${b}$-tagging {probability and faking rates are}
\beq
p_{b\to b}=0.75,~p_{c\to b}=0.1,~p_{j\to b}=0.01.
\eeq
The light-jet-to-photon {faking} probability is parametrized via
\beq
p_{j\to\gamma}=\alpha \exp(-p_{T,j}/\beta)~,\alpha=0.01~, \beta=30~\rm{GeV}.
\eeq
The photon {identification} efficiency {is}
\beq
\epsilon_{\gamma}(p_T)=
\begin{cases}
95\%,& ~\text{for}~ |\eta|<1.5,\\
90\%,& ~\text{for}~ 1.4<|\eta|<4,\\
80\%,& ~\text{for}~ 4<|\eta|<6.
\end{cases}
\eeq
To get the cut efficiency function, we generate partonic level
$pp\to hh\to b\bar b\gamma\gamma$
events with MadGraph5\_aMC@NLO event generator~\cite{Alwall:2014hca}
with CT14 PDF \cite{Dulat:2015mca}.
As we are interested only in the $s$-wave component, the
default SM with trilinear Higgs coupling is enough. The events are generated with fixed $m_{hh}$
for each 10 GeV interval. The detector effects are mimicked with
Gaussian smearing effects with the parameters given in
\cite{Contino:2016spe}.

\begin{figure}[!htb]
 \centering
 \includegraphics[width=0.3\textwidth]{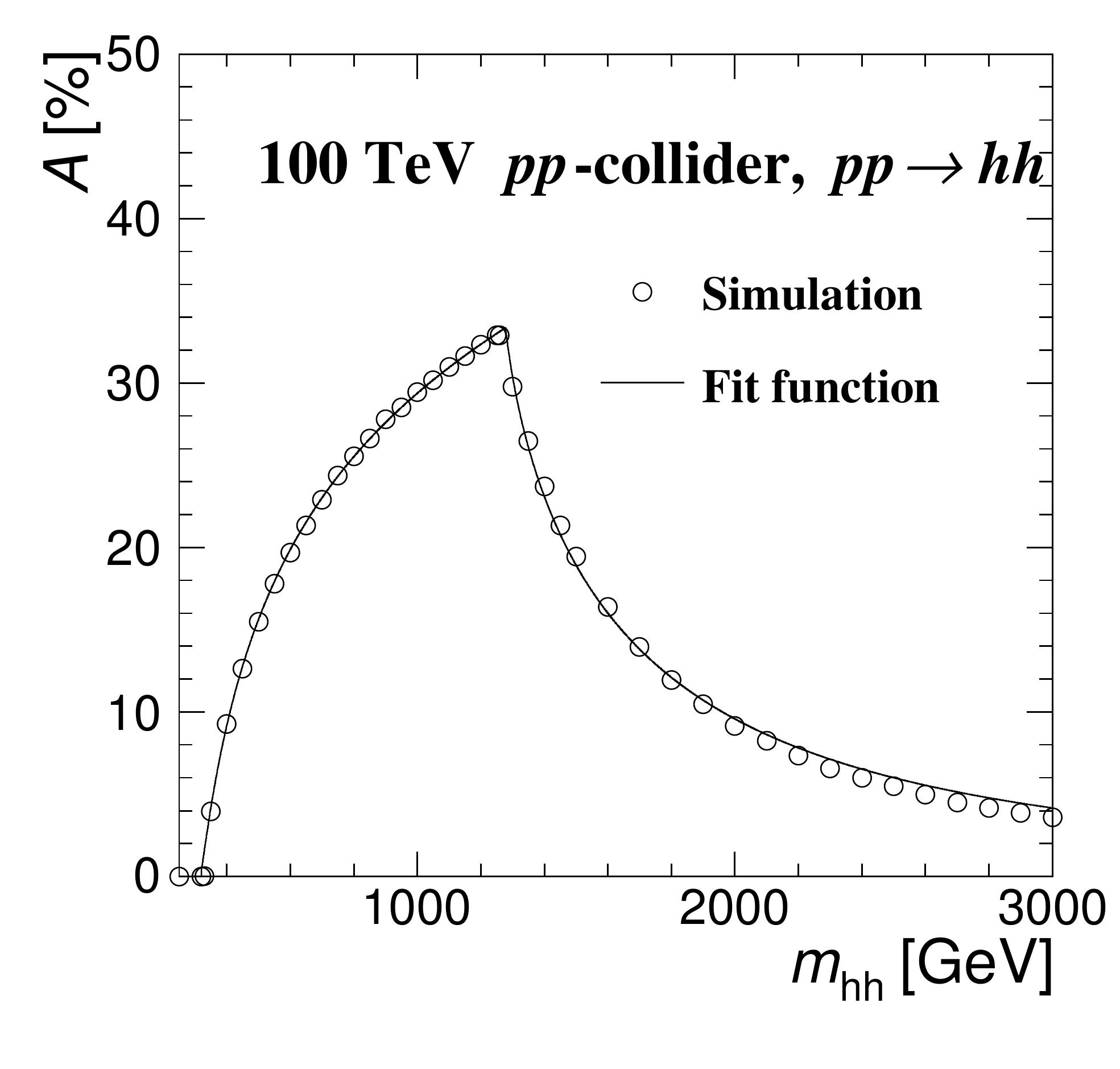}
  \caption{ The cut efficiency function at the 100 TeV $pp$-collider.}
 \label{fig:100TeV_01}
 \end{figure}

We show the cut efficiency  function for  the 100 TeV $pp$-collider in
Fig.~\ref{fig:100TeV_01}.
The structures in {the figure can be easily understood as follows.} For the ``peak''
structure, 
the boost factor of the Higgs boson is $\gamma\simeq 5$ around the crossing point. The angular distance between the Higgs decay products, i.e. $\Delta R\equiv \sqrt{(\Delta\eta)^2+(\Delta \phi)^2}$ is approximated by $\Delta R\simeq2/
\gamma\simeq0.4$. So {crossing} this point, the typical $\Delta R$
of the $b\bar b$ {system and the} $\gamma\gamma$ system will become
smaller. {The signal events are likely to fail the
$\Delta R$ cuts to yield a smaller cut efficiency}. We would like to estimate the result analytically
with some approximations. Let us define the {4-momenta} of the
partons (photons) in the Higgs rest frame with the $z$-direction
defined by the Higgs 3-momentum in the lab frame. Then the exact
result of the $\Delta R$ is
\begin{align}
\Delta R^2 &= 
\arccos^2\biggl\{\biggl[4r^2\beta_T^2+\left(1-z^2\right)
\left(\beta_T^2-\cos^2\phi-\sin^2\phi\tanh^2\eta\right)-z^2\sech^2\eta \nn\\
&+2z\sin\phi\tanh\eta
\times\sqrt{\left(1-z^2\right)\left(\beta_T^2+\sech^2\eta\right)}\biggr]\nonumber\\
&\times\biggl[\left(1-z^2\right)\cos^2\phi+\biggl(\sqrt{1+4r^2}\beta_T
-\sqrt{1-z^2}\sin\phi\tanh\eta
+z\sqrt{\beta_T^2
+\sech^2\eta}\biggr)^2
\biggr]^{-\frac{1}{2}}\nonumber\\
&\times\biggl[\left(1-z^2\right)\cos^2\phi+\biggl(\sqrt{1+4r^2}\beta_T
+\sqrt{1-z^2}\sin\phi\tanh\eta
-z\sqrt{\beta_T^2+\sech^2\eta}\biggr)^2
\biggr]^{-\frac{1}{2}}\biggr\}\nonumber\\
&+\biggl\{\arcsinh\biggl\{\biggl[
\sinh\eta\left(\sqrt{1+4r^2}\beta_T+z\sqrt{\beta^2_T+\sech^2\eta}\right)
+\sin\phi\sech\eta\sqrt{1-z^2}\biggr]\nonumber\\
&\times\biggl[
\left(1-z^2\right)\cos^2\phi+\biggl(\sqrt{1+4r^2}\beta_T
-\sqrt{1-z^2}\sin\phi\tanh\eta
+z\sqrt{\beta_T^2+\sech^2\eta}\biggr)^2
\biggr]^{-\frac{1}{2}}\biggr\}\nonumber\\
&+\arcsinh\biggl\{\biggl[
\sinh\eta\left(-\sqrt{1+4r^2}\beta_T+z\sqrt{\beta^2_T+\sech^2\eta}\right)
+\sin\phi\sech\eta\sqrt{1-z^2}\biggr]\nonumber\\
&\times\biggl[
\left(1-z^2\right)\cos^2\phi+\biggl(\sqrt{1+4r^2}\beta_T
+\sqrt{1-z^2}\sin\phi\tanh\eta
-z\sqrt{\beta_T^2+\sech^2\eta}\biggr)^2
\biggr]^{-\frac{1}{2}}\biggr\}
\biggl\}^2,
\end{align}
where  $\beta_T\equiv p_{\text{T}}/m_h$ is the ratio between the
transverse momentum and the mass of the Higgs boson in the
lab frame, $r=m/m_h$ is the mass ratio between the final state
particle and the Higgs boson, which is 0 for photon and $m_b/m_h$ for
bottom quark, $\eta$ is the pseudo-rapidity of the Higgs boson in the
lab frame, $z\equiv\cos\theta$ is the cosine value of the polar
angle of one parton in the Higgs rest frame, $\phi$ is the azimuthal 
angle of one parton in the Higgs rest frame. In the highly boost
region, $\beta_T\gg 1$, and
\be
\Delta R=\arccos\left[1-\frac{2}{1+\left(1-z^2\right)\beta_T^2}\right]
\ee
is a good approximation for the massless final state particle.
To pass the $\Delta R$ cuts, we need
to solve this equation. The solutions are
\be
z_\pm=\pm\sqrt{\frac{1-\beta_T^{-2}-\left(1+\beta_T^{-2}\right)\cos\Delta R}{1-\cos\Delta R}}.
\ee
The region allowed by the cut is $[-1,z_-]\cup[z_+,1]$. This is a hint that
we can fit the high invariant mass tail with the function
\be
\mathcal{A}\left(m_{hh}\right)=c_1\left[1-\sqrt{\frac{m_{hh}^2\left(1-\cos\Delta R\right)
-8\left(m_h-\delta m_h\right)^2}{\left(1-\cos\Delta R\right)\left(m_{hh}^2
-4\left(m_h-\delta m_h\right)^2\right)}}\right]^{\gamma_c},
\ee
where $\Delta R=0.4$ is the angular distance cut,
the parameter $\gamma_c$ and $\delta m$ reflect the energy
resolution effect and the invariant mass cut effect, $c$ is a normalization
constant.

For massive final state particle, we have
\be
\Delta R=\arccos\left[\frac{\left(1-z^2+r^2\right)\beta_T^2-1}
{\sqrt{\left[\left(1-z^2+r^2\right)\beta_T^2-1\right]^2+4\left(1-z^2\right)
\left(1+4r^2\right)\beta_T^2}}\right].
\ee
The moving direction of a massive particle can be flipped by a
Lorentz boost. For a very large $\beta_T>1/r$, $z\to\pm1$,
we have $\Delta R=0$. In this case, the region allowed by the cut
is $[z_{-2},z_{-1}]\cup[z_{+1},z_{+2}]$ and could be $\varnothing$.
However, because $m_b\ll m_h$, this will be only a tiny correction
at very high $m_{hh}$ region ($p_{\text{T,cut}}^h>3.3$ TeV) and could be neglected.

The behavior of the cut efficiency function in the low invariant mass region can be understood as follows.
The $p_{\text{T}}$ cuts on the Higgs bosons require
the Higgs bosons have a large $p_{\text{T}}$, which means that
the energy of the Higgs bosons must be larger than
$\sqrt{m_H^2+p_{\text{T,cut}}^2}\simeq 160$ GeV. This is the reason why
the events with $m_{hh}<320$ GeV have a very tiny (close to 0) cut
acceptance.
It is easy to know that the integration region of the polar angle in the
c.m. frame is
\be
\left[-\sqrt{1-\frac{4(p_{{\text{T,cut}}}^{h})^2}{m_{hh}^2-4m_h^2}},\sqrt{1-\frac{4(p_{{\text{T,cut}}}^h)^2}{m_{hh}^2-4m_h^2}}\right].
\ee
This is a hint that we could fit the low invariant mass region
with
\be
\mathcal{A}\left(m_{hh}\right)=a_1\left[1-\frac{4(p_{{\text{T,cut}}}^h)^2}{m_{hh}^2-4\left(m_h
-\delta m_2\right)^2}\right]^{\beta_a}\left(\frac{2m_h}{m_{hh}}\right)^{\beta_b}
\left[1+a_2\left(\frac{2m_h}{m_{hh}}\right)\log\left(\frac{2m_h}{m_{hh}}\right)\right],
\ee
where $p_{{\text{T,cut}}}^h=100$ GeV.

Finally, we obtain the analytic function $\mathcal{A}({m}_{hh},s,\mu_F)$ at the 100 TeV $pp$-collider in the following form
\beq
\mathcal{A}\left({m}_{hh}\right)=
\begin{cases}
c_1\left[1-\sqrt{\dfrac{{m}_{hh}^2\left(1-\cos\Delta R_0\right)
-8\left({m_h}-\delta m_1\right)^2}{\left(1-\cos\Delta R_0\right)\left({m}_{hh}^2
-4\left({m_h}-\delta m_1\right)^2\right)}}\right]^{\gamma_c}, & {m}_{hh}>M^{(t)}_{hh},\\
c_2 \left[1-\displaystyle \frac{4(p_{{\text{T,cut}}}^{h})^2}{{m}_{hh}^2-4\left({m_h}
-\delta m_2\right)^2}\right]^{\beta_a}\left(\dfrac{{2m_h}}{{m}_{hh}}\right)^{\beta_b}
\left[1+\beta_c\left(\dfrac{{2m_h}}{{m}_{hh}}\right)\log
\left(\dfrac{{2m_h}}{{m}_{hh}}\right)\right],& 319.9{~\text{GeV}}<{m}_{hh}<M^{(t)}_{hh},\\
0,& {m}_{hh}<319.9{~\text{GeV}}.
\end{cases}
\label{eq:cuteff2}
\eeq
where the fitting parameters $\delta m_1=\delta m_2=0.15~\rm GeV$, $\Delta R_0=0.4$,  $c_1={40.30}$, $\gamma_c={0.938}$, $c_2=8.269$, $\beta_a=1.241$, $\beta_b=-0.565$, $\beta_c=-2.057$, {and $M_{hh}^{(t)}=1277.5$GeV},  in the low detector performance scenario.

For completeness, we also show the cut efficiency function at the HL-LHC below. The selection cuts used by the ATLAS Collaboration~\cite{Aad:2015} are
\bea
&&p_{T}^{b_1}>40~{\rm{GeV}},~
p_{T}^{b_2}>25~{\rm{GeV}},~|\eta^{b}|<2.5,\nonumber\\
&&p_{T}^\gamma>30~{\rm GeV},~|\eta^{\gamma}|<1.37~{\rm or}~
1.52<|\eta^{\gamma}|<2.37,\nonumber\\
&&\Delta R_0 <\Delta R_{b b,\gamma\gamma}<2.0~, ~
\Delta R_{b\gamma}>\Delta R_0,~~ \Delta R_0=0.4~,\nonumber\\
&&100{~\text{GeV}}<m_{b b}<150{~\text{GeV}},~
p_{T}^{b b}>110{~\text{GeV}},\nonumber\\
&&123{~\text{GeV}}<m_{\gamma\gamma}<128{~\text{GeV}},~
p_{T}^{\gamma\gamma}>110{~\text{GeV}},
\label{eq:14cuts}
\eea
To mimic the detector effects, the final state parton momenta are smeared by a Gaussion distribution. The $b$-tagging efficiency is~\cite{Aad:2013hl,Cao:2015oaa}
\bea
\epsilon_b\left(p_{T},\eta\right)&=&0.135\tanh\left(\frac{p_{T}+50}
{75}\right)\tanh\left(\frac{450}
{p_{T}+80}\right)\times \big[3+e^{-\left(|\eta|-\sqrt{p_{T}/
1000}\right)^2/1.6}\big] e^{-|\eta|^3 p_{T}/1000}.
\label{eq:btag}
\eea
and the photon energy resolution and identification efficiency are~\cite{Aad:2013gl}
 \bea
\sigma\left({\text{GeV}}\right)&=&0.3\oplus0.10\times\sqrt{E({\text{GeV}})}
\oplus0.010\times E({\text{GeV}}),~~{\text{for}}~~|\eta|<1.37,\nonumber\\
\sigma\left({\text{GeV}}\right)&=&0.3\oplus0.15\times\sqrt{E({\text{GeV}})}
\oplus0.015\times E({\text{GeV}}),~~{\text{for}}~~1.52<|\eta|<2.37,
\eea
and
\begin{equation}
\epsilon_\gamma\left(p_{T}\right)=0.76-1.98\exp\left(-\frac{p_{T}}
{16.1{\text{GeV}}}\right),
\end{equation}
respectively.

After fitting the Monte Carlos simulation results with all the detector effects, we obtain the following cut efficiency function with $p_{\rm T,cut}^h=110~{\rm GeV}$, which is slightly different from the function of the 100~TeV machine,
\beq
\mathcal{A}\left(M_{hh}\right)=
\begin{cases}
c_1\left[1-\sqrt{\dfrac{M_{hh}^2\left(1-\cos\Delta R_0\right)
-8\left(m_H-\delta m_1\right)^2}{\left(1-\cos\Delta R_0\right)\left(M_{hh}^2
-4\left(m_H-\delta m_1\right)^2\right)}}\right]^{\gamma_c}, & M_{hh}>M^{(t)}_{hh},\\
c_2 \left[1-\displaystyle \frac{4(p_{\rm T, cut}^{h})^2}{M_{hh}^2-4\left(m_H
-\delta m_2\right)^2}\right]^{\beta_a}\left(\dfrac{M_{hh}}{\sqrt{s}}\right)^{\beta_b}
\left[1+\beta_c\left(\dfrac{M_{hh}}{\sqrt{s}}\right)\log
\left(\dfrac{2M_{hh}}{\sqrt{s}}\right)\right],& 329.3{~\text{GeV}}<M_{hh}<M^{(t)}_{hh},\\
0,& M_{hh}<329.3{~\text{GeV}}.
\end{cases}
\label{eq:cuteff}
\eeq
The fitting parameters are $c_1=1.1378$, $c_2=11.02$, $\delta m_1=50~\rm GeV$, $\gamma_c=1.675$, $\delta m_2=2.5~\rm GeV$, $\beta_a=1.13$, $\beta_b=1.48$, $\beta_c=4.88$, $\Delta R_0=0.4$ and $M_{hh}^{t}=1260~{\rm GeV}$~\cite{Cao:2015oaa}.

\subsection{\tf{The $m_{hh}$ distribution}{The mhh distribution}}\label{sec:42}

Once knowing the cut efficiency function $\mathcal{A}(m_{hh})$, one can calculate numbers of events of Higgs boson pair production after a series of kinematic cuts listed in Eq.~\eqref{eq:cuts2} or Eq.~\eqref{eq:14cuts} using the master formula shown in Eq.~\eqref{eq:convolution}. That requires knowledge of the inclusive $m_{hh}$ distribution. Below we examine the impact of various effective couplings on the $m_{hh}$ distribution before and after imposing experimental cuts.  

\begin{figure}[!htb]
\centering
\includegraphics[width=0.2\textwidth]{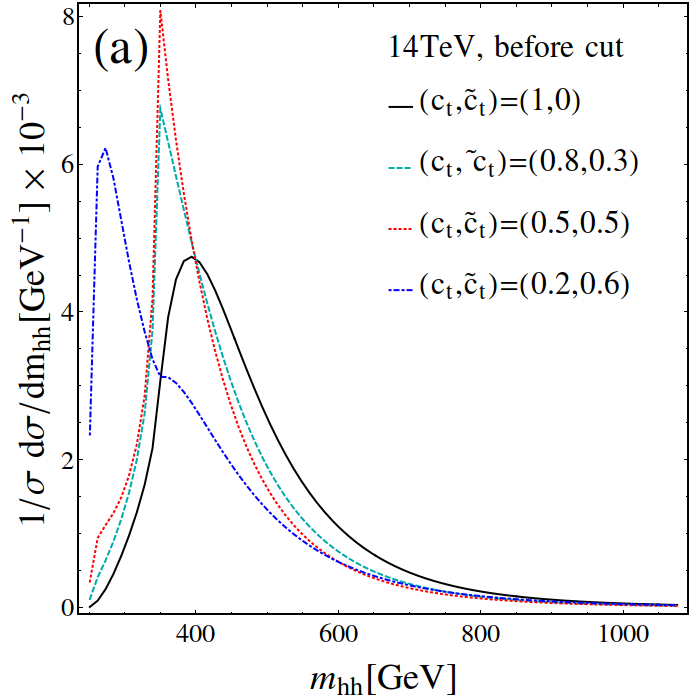}
\includegraphics[width=0.2\textwidth]{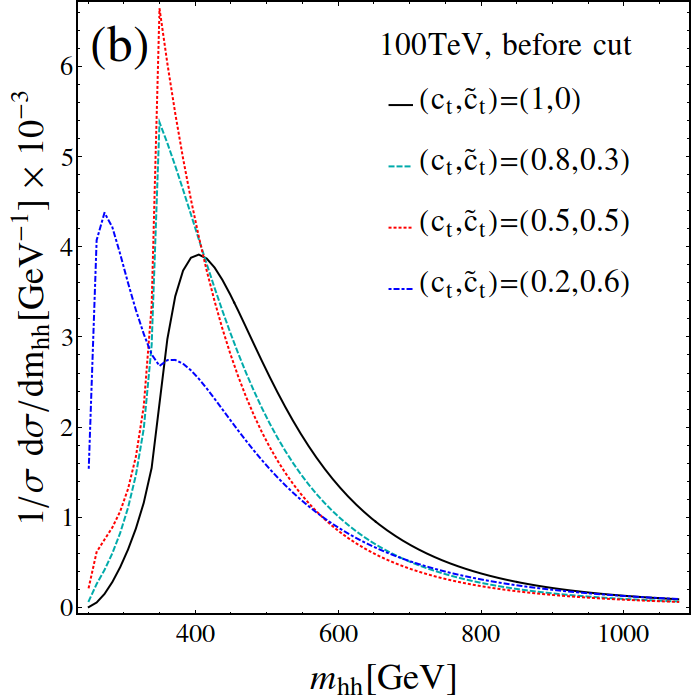}
\includegraphics[width=0.2\textwidth]{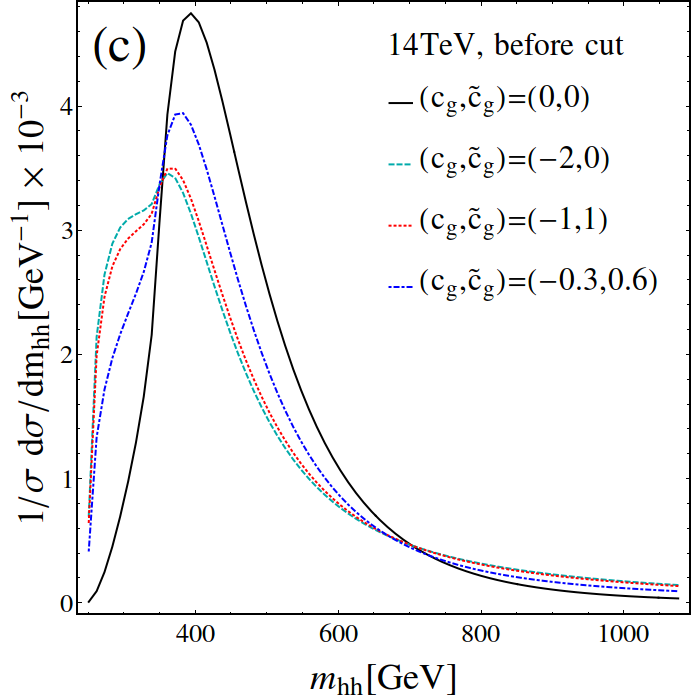}
\includegraphics[width=0.2\textwidth]{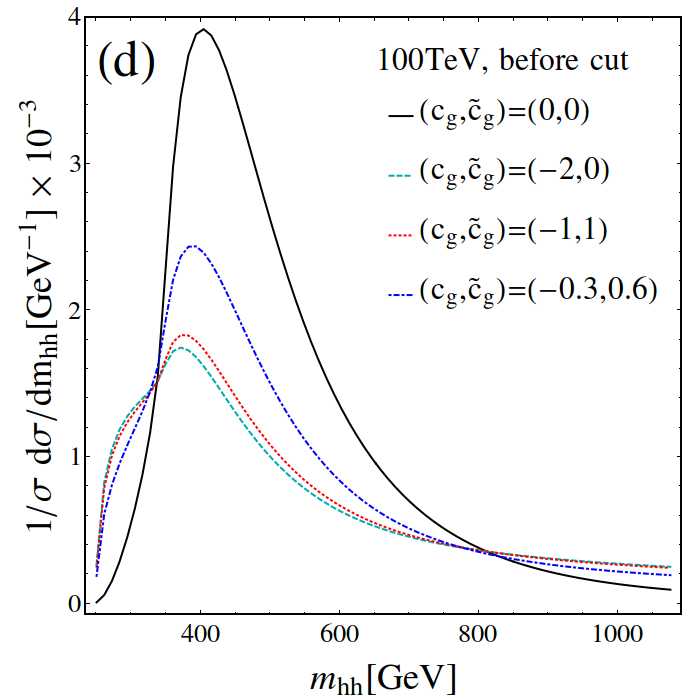}\\
\includegraphics[width=0.2\textwidth]{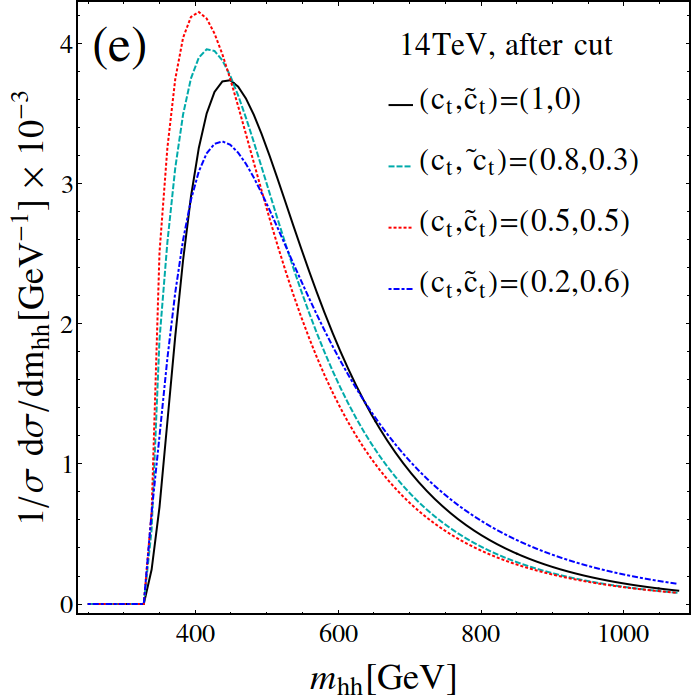}
\includegraphics[width=0.2\textwidth]{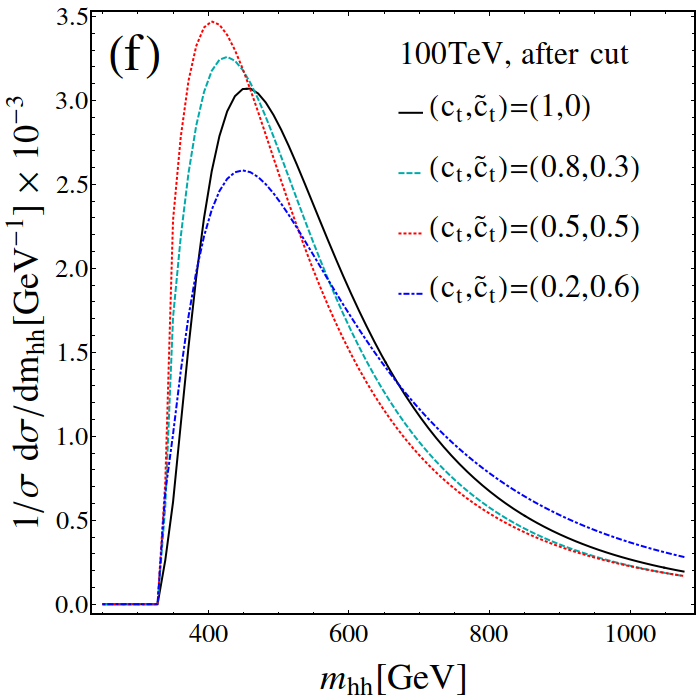} 
\includegraphics[width=0.2\textwidth]{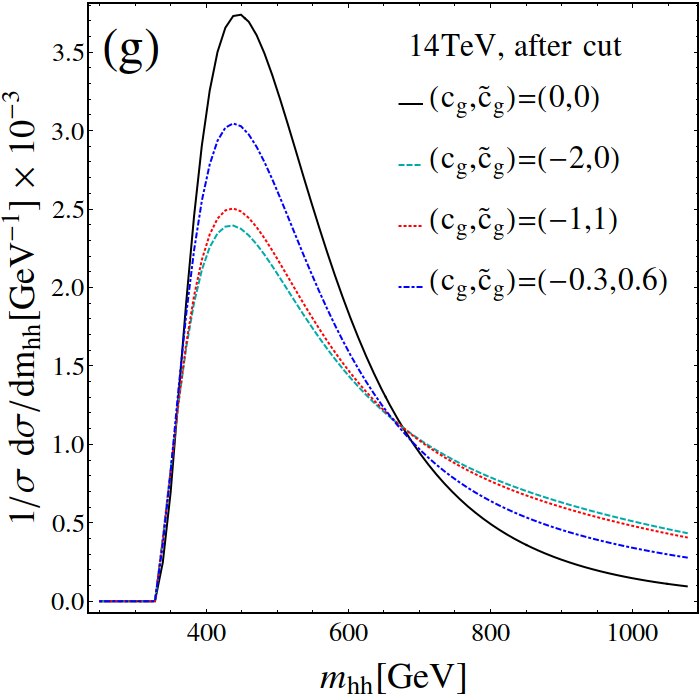}
\includegraphics[width=0.2\textwidth]{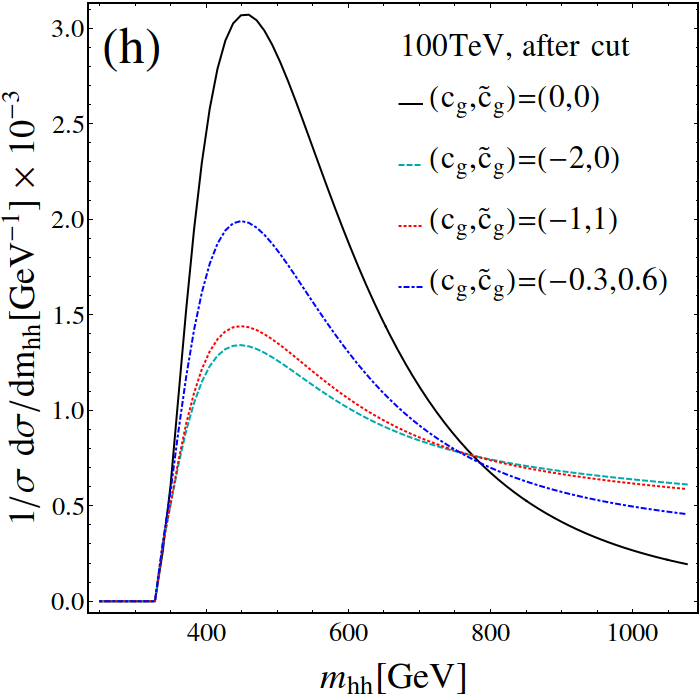}
\caption{$m_{hh}$ distributions in the double Higgs production with CP-violating $htt$ and $h(h)gg$ couplings at the $14~\rm TeV$ LHC and the $100~\rm TeV$ $pp$-collider before (top row) and after the kinematic cuts (bottom row). The black (solid), green (dashed), red (dotdashed) and blue (dotted) lines in the upper panel correspond to $(c_t,\tilde{c}_t)=(1,0)$, $(0.8,0.3)$, $(0.5,0.5)$, $(0.2,0.6)$, respectively. While the black (solid), green blue (dashed), red (dotdashed) and blue (dotted) lines in the lower panel correspond to $(c_g,\tilde{c}_g)=(0,0),(-2,0),(-1,1),(-0.3,0.6)$, respectively.}
\label{fig:mhhnocut}
\end{figure}

Figure~\ref{fig:mhhnocut} displays the $m_{hh}$ distributions in the double Higgs production with CP-violating $htt$ and $h(h)gg$ couplings before and after the selection cuts at the 14~TeV LHC and at a future 100~TeV $pp$-collider, respectively.
We derive the $m_{hh}$ distribution after cuts by convoluting the inclusive distribution with the cut efficiency function as stated in Eq.~\eqref{eq:cut_dist}. Two combinations of Higgs effective couplings, $(c_t,\tilde{c}_t)$ and $(c_g,\tilde{c}_g)$, are considered. We fix all the other effective couplings as the SM values while varying the two effective couplings in each combination. We choose a few benchmark couplings listed as follows:
\bea
(c_t,\tilde{c}_t)&=&(1,~0),~~(~0.8,~0.3), ~~(~~0.5,~0.5), ~~~~(~0.2,0.6),\nn \\
(c_g,\tilde{c}_g)&=&(0,~0),~~(-2.0,~0), ~~~(-1,~1), ~~~~~~~~(-0.3,0.6),
\eea
which are well consistent with the measurements of single Higgs production at the LHC Run-I.

For the case of $(c_t,\tilde{c}_t)$, the invariant mass distribution of Higgs boson pairs peaks around 400~GeV in the SM, i.e. $(c_t,\tilde{c}_t)=(1,0)$; see the black-solid curves in Figs.~\ref{fig:mhhnocut}(a) and (b). Other values of $c_t$ and $\tilde{c}_t$ shift the peak to small $m_{hh}$ regions both at the 14~TeV and at the 100~TeV. It can be understood as follows. In the SM, a large cancellation between $F_\Box$ and $F_\triangle\times 3m_h^2/(\hat{s}-m_h^2)$ occurs near the threshold $m_{hh}\sim 2m_h\approx 250~{\rm GeV}$~\cite{Shifman:1979eb,Kniehl:1995tn}. However, the cancellation is spoiled when the $c_t$ coupling deviates sizably from the SM value $c_t=1$. That shifts the peak position. In addition, the contribution from $\tilde{c}_t(c_t F_\Box^{(2)}+F_\triangle^{(1)}\times 3m_h^2/(\hat{s}-m_h^2) )$ increases dramatically with $\tilde{c}_t$. Therefore, a large $\tilde{c}_t$, e.g. $(c_t, \tilde{c}_t)=(0.2, 0.6)$, distorts the smooth $m_{hh}$ distribution; see the blue curves in Figs~\ref{fig:mhhnocut}(a) and (b). We notice that the $m_{hh}$ distributions do not change much when increasing the collider energy from 14 TeV to 100 TeV. The $m_{hh}$ distributions in the small $m_{hh}$ region is sensitive to $c_t$ and $\tilde{c}_t$ before imposing any cuts. Different choices of $c_t$ and $\tilde{c}_t$ couplings yield distinct distributions. Unfortunately, 
the differences in low $m_{hh}$ region are washed out once imposing a hard $p_T$ cut on the Higgs boson in order to disentangle the signal out of huge SM background at the 14 TeV LHC and the 100 TeV $pp$-collider. Figures~\ref{fig:mhhnocut}(e) and (f) show the $m_{hh}$ distributions after the selection cuts given in Eq.~\eqref{eq:14cuts}. After cuts all the curves are quite similar. If NP models only modify the $c_t$ and $\tilde{c}_t$ coupling, then it is difficult to discriminate the NP models through the $m_{hh}$ distributions.

We also show the $m_{hh}$ distributions for various combinations of $(c_g,\tilde{c}_g)$ in Fig.~\ref{fig:mhhnocut}. The $c_g$ and $\tilde{c}_g$ couplings introduce a momentum dependence to the double Higgs production, and they are expected to play an important role in large $m_{hh}$ region. In the small $m_{hh}$ region, the invariant mass distributions are distorted at the 14 TeV and 100 TeV colliders, owing to the weaker cancellation when $c_g<0$. In the high $m_{hh}$ regions, say $m_{hh}\gtrsim 400\gev$, the distributions are distinctly different, especially at the 100 TeV collider. See Figs.~\ref{fig:mhhnocut}(c) and (d). It is because, unlike the $F$ form factors, the contributions from $h(h)gg$ interaction, which are proportional to $c_{g}$ and $\tilde{c}_{g}$, do not decrease in the large $m_{hh}$ region. More importantly, the differences of the $m_{hh}$ distributions remain even after imposing the selection cuts; see Figs.~\ref{fig:mhhnocut}(g) and (h). As a consequence, it is possible to discriminate different NP models that modify $c_g$ and $\tilde{c}_g$ through the $m_{hh}$ distributions. 

Next, we will discuss the correlation and sensitivity of the Higgs effective couplings in the scattering of $gg\to hh \to b\bar{b}\gamma\gamma$ at the 14 TeV LHC and the 100 TeV $pp$-collider.

\subsection{Signal strength and Higgs effective couplings}\label{sec:43}
{With the help of the cut efficiency function}, we can easily obtain the total cross section of any NP described by the Higgs effective couplings after the selection cuts. Making use of the narrow width of the Higgs boson, the signal strength of the signal process, $pp\to hh \to b\bar{b}\gamma\gamma$, can be factorized as follows
\begin{align}
\label{eq:ratio}
{\frac{\sigma(pp\to hh\to b\bar{b}\gamma\gamma)}{\sigma_{SM}(pp\to hh\to b\bar{b}\gamma\gamma)}}&=\frac{\sigma_{hh}(pp\to hh)}{\sigma_{hh}^{SM}(pp\to hh)}\times
\frac{\text{Br}(h\to b\bar{b})}{\text{Br}(h\to b\bar{b})_{SM}}\times
\frac{\text{Br}(h\to\gamma\gamma)}{\text{Br}(h\to\gamma\gamma)_{SM}}\equiv \mu_{hh}\times\mu_{b\bar{b}} \times \mu_{\gamma\gamma},
\end{align}
where $\mu_{hh,b\bar{b},\gamma\gamma}$ denote the signal strength of the cross section of double Higgs production, of the branching ratio of $h\to b\bar{b}$ decay, of the branching ratio of $h\to \gamma\gamma$ decay, defined as follows:
\beq
\mu_{hh}\equiv \frac{\sigma_{hh}}{\sigma_{hh}^{SM}},\quad
\mu_{b\bar{b}}\equiv \frac{\text{Br}(h\to b\bar{b})}{\text{Br}(h\to b\bar{b})_{SM}},\quad \mu_{\gamma\gamma}\equiv \frac{\text{Br}(h\to\gamma\gamma)}{\text{Br}(h\to\gamma\gamma)_{SM}}.
\eeq
The dependence of $\mu_{hh}$ on the effective couplings is
\begin{align}
\mu_{hh}
&=A_1c_{3h}^2c_g^2+A_2c_{3h}^2c_gc_t+A_3c_{3h}^2c_t^2+A_4c_{3h}c_g^2
+A_5c_{3h}c_gc_t^2+A_6c_{3h}c_gc_t+A_7c_{3h}c_g\tilde{c}_t^2\nn\\
&+A_8c_{3h}c_t^3+A_9c_{3h}c_t\tilde{c}_t^2+A_{10}c_g^2+A_{11}c_gc_t^2
+A_{12}c_g\tilde{c}_t^2+A_{13}c_t^4
+A_{14}c_t^2\tilde{c}_t^2+A_{15}\tilde{c}_t^4\nn\\
&+A_{16}c_{3h}^2\tilde{c}_g^2+A_{17}c_{3h}^2\tilde{c}_g\tilde{c}_t
+A_{18}c_{3h}^2\tilde{c}_t^2+A_{19}c_{3h}\tilde{c}_g^2+A_{20}c_{3h}\tilde{c}_gc_t\tilde{c}_t
+A_{21}c_{3h}\tilde{c}_g\tilde{c}_t\nn\\
&+A_{22}\tilde{c}_g^2+A_{23}\tilde{c}_gc_t\tilde{c}_t+A_{24}c_{2t}^2
+A_{25}c_{2t}c_{3h}c_{g}+A_{26}c_{2t}c_{3h}c_{t}
+A_{27}c_{2t}c_g+A_{28}c_{2t}c_t^2\nn\\
&+A_{29}c_{2t}\tilde{c}_t^2+A_{30}c_t\tilde{c}_t\tilde{c}_{2t}
+A_{31}c_{3h}\tilde{c}_t\tilde{c}_{2t}+A_{32}c_{3h}\tilde{c}_g\tilde{c}_{2t}
+A_{33}\tilde{c}_{2t}^2+A_{34}\tilde{c}_g\tilde{c}_{2t}.
\end{align}
The product of $\mu_{b\bar{b}}$ and $\mu_{\gamma\gamma}$ is 
\begin{align}
\mu_{b\bar{b}}\times \mu_{\gamma\gamma}
&=\frac{\Gamma(h\to\gamma\gamma)}{\Gamma(h\to\gamma\gamma)_{SM}}
\left(\frac{\Gamma_{tot}^{SM}}{\Gamma_{tot}}\right)^2
=\frac{\kappa_{\gamma}^2}{\left[1+(\kappa_{g}^2-1)\text{BR}_{g}^{SM}
+(\kappa_{\gamma}^2-1)\text{BR}_{\gamma}^{SM}\right]^2}
\label{eq:brs}
\end{align}
where we assume the Yukawa coupling of bottom quarks is not altered by NP effects. 
The $\kappa_{g}$ and $\kappa_{\gamma}$ couplings are defined in Eqs.~\eqref{eq:kappag}
and \eqref{eq:kappaa}. The SM branching ratios are $\text{BR}_{g}^{SM}\equiv
\text{BR}(h\to gg)_{SM}=8.187\%$ and $\text{BR}_{\gamma}^{SM}\equiv \text{BR}
(h\to \gamma\gamma)_{SM}=0.227\%$~\cite{Olive:2016xmw}. 

The values of the coefficients $A$'s are listed in Table~\ref{tbl:As0} at the 14 TeV LHC and the 100 TeV $pp$-collider, before imposing any cuts (top panel) and after the series of cuts defined in Eq.~\ref{eq:cuts2} (bottom panel). The values of those coefficients at the 14~TeV LHC without any cut agree exactly with those values given in Ref.~\cite{Azatov:2015oxa}. We notice that the $A_{10,12,22,24}$ coefficients are larger at the 100 TeV $pp$-collider than at the 14 TeV LHC. Those coefficients correspond to the couplings of $c_g^2$, $c_g\tilde{c}_t^2$, $\tilde{c}_g^2$ and $c_{2t}^2$, which modify the $h(h)gg$ and $hhtt$ interactions and contribute significantly to the double Higgs production at the large $m_{hh}$ region.

\begin{table}
\tabcolsep=5pt
\caption{The cross sections of $gg\to hh\to b\bar{b}\gamma\gamma$ in terms of the Higgs effective couplings at the 14 TeV LHC and 100 TeV $pp$-collider before (top panel) and after (bottom panel) the selection cuts.}
\begin{tabular}{ccccccccccccccccccccc}
\hline
$\sqrt{s}$ & $A_1$ & $A_2$ & $A_3$ & $A_4$ & $A_5$ & $A_6$ & $A_7$
& $A_8$ & $A_9$ & $A_{10}$ & $A_{11}$ & $A_{12}$ \\
\hline
$14~\rm TeV$  & 0.138 & 0.370 & 0.276 & 0.640 & -0.766 & 0.821 & 0.535
& -1.35 & -6.22 & 1.37 & -1.82 & 1.58\\
$100~\rm TeV$ & 0.101 & 0.267 & 0.208 & 0.592 & -0.569 & 0.658 & 0.425
& -1.11 & -4.79 & 3.32 & -1.30 & 1.67\\
\hline
$\sqrt{s}$ & $A_{13}$ & $A_{14}$ & $A_{15}$ & $A_{16}$ & $A_{17}$ & $A_{18}$
& $A_{19}$ & $A_{20}$ & $A_{21}$ & $A_{22}$ & $A_{23}$ & $A_{24}$\\
\hline
$14~\rm TeV$ & 2.07 & 13.9 & 0.719 & 0.138 & -0.611 & 0.861 & 0.640
& 2.13 & -1.24 & 1.37 & 4.64 & 2.55\\
$100~\rm TeV$ & 1.90 & 11.3 & 0.680 & 0.101 & -0.428 & 0.634 & 0.592
& 1.53 & -0.928 & 3.32 & 3.51 & 2.90\\
\hline
$\sqrt{s}$ & $A_{25}$ & $A_{26}$ & $A_{27}$ & $A_{28}$ & $A_{29}$ &
$A_{30}$ & $A_{31}$ & $A_{32}$ & $A_{33}$ & $A_{34}$ \\
\hline
$14~\rm TeV$ & 0.821 & 1.39 & 2.44 & -4.24 & 2.30 & -18.8 & 4.04
& -1.24 & 6.19 & -3.02 \\
$100~\rm TeV$  & 0.658 & 1.21 & 2.06 & -4.13 & 2.16 & -16.3 & 3.28
& -0.928 & 6.10 & -2.08 \\
\hline\\
\hline
$\sqrt{s}$ & $A_1$ & $A_2$ & $A_3$ & $A_4$ & $A_5$ & $A_6$ & $A_7$
& $A_8$ & $A_9$ & $A_{10}$ & $A_{11}$ & $A_{12}$ \\
\hline
$14~\rm TeV$ & 0.0369 & 0.0975 & 0.0993 & 0.406 & -0.264 & 0.410
& 0.239 & -0.739 & -2.46 & 1.80 & -0.888 & 1.42\\
$100~\rm TeV$ & 0.0347 & 0.0846 & 0.0880 & 0.465 & -0.215 & 0.372
& 0.229 & -0.671 & -2.14 & 3.20 & -0.531 & 1.67\\
\hline
$\sqrt{s}$ & $A_{13}$ & $A_{14}$ & $A_{15}$ & $A_{16}$ & $A_{17}$
& $A_{18}$ & $A_{19}$ & $A_{20}$ & $A_{21}$ & $A_{22}$ & $A_{23}$ & $A_{24}$ \\
\hline
$14~\rm TeV$ & 1.64 & 7.18 & 0.517 & 0.0369 & -0.120 & 0.257 & 0.406 & 0.517
& -0.428 & 1.80 & 1.76 & 2.85\\
$100~\rm TeV$ & 1.58 & 6.46 & 0.806 & 0.0347 & -0.102 & 0.222 & 0.465
& 0.435 & -0.361 & 3.20 & 1.43 & 3.33 \\
\hline
$\sqrt{s}$ & $A_{25}$ & $A_{26}$ & $A_{27}$ & $A_{28}$ & $A_{29}$
& $A_{30}$ & $A_{31}$ & $A_{32}$ & $A_{33}$ & $A_{34}$ \\
$14~\rm TeV$  & 0.410 & 0.920 & 2.11 & -3.79 & 1.91 & -12.2 & 2.04
& -0.428 & 5.28 & -1.64 \\
$100~\rm TeV$ & 0.372 & 0.889 & 1.96 & -3.87 & 1.92 & -11.6 & 1.88
& -0.361 & 5.68 & -1.10  \\
\hline
\end{tabular}

\label{tbl:As0}
\end{table}

Equipped with the inclusive $m_{hh}$ distributions and cut efficiency function, we are 
ready to explore the sensitivity of the HL-LHC and 100~TeV $pp$-collider on the Higgs effective couplings. The expected discovery significance and the exclusion limit can be evaluated with~\cite{Cowan:2010js}
\bea
Z_0&=&\sqrt{2\left[(n_s+n_b)\log\frac{n_s+n_b}{n_b}-n_s\right]},\\
Z_\mu&=&\sqrt{-2\left(n_b\log\frac{n_s+n_b}{n_b}-n_s\right)},
\eea
respectively, where $n_s$ and $n_b$ denote the numbers of the signal
and background events. The signal and background events in the SM at the 14 TeV HL-LHC with an integrated luminosity $\mathcal{L}=3000~\rm fb^{-1}$ ~\cite{Aad:2015} and the 100 TeV $pp$-collider with $\mathcal{L}=30~\rm ab^{-1}$~\cite{Contino:2016spe} are
\bea
{\rm 14~TeV} &:& n_s=8.4, \qquad n_b=47,\nn\\
{\rm 100~TeV}&:& n_s=12061,~~n_b=27118.
\eea

\subsection{Sensitivity to Higgs effective couplings at the HL-LHC}\label{sec:44x}

\begin{figure}
 \centering
 \includegraphics[width=0.23\textwidth]{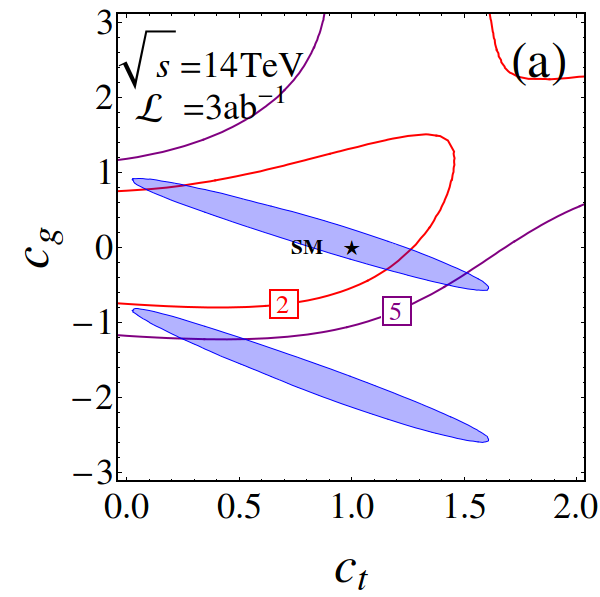}
 \includegraphics[width=0.23\textwidth]{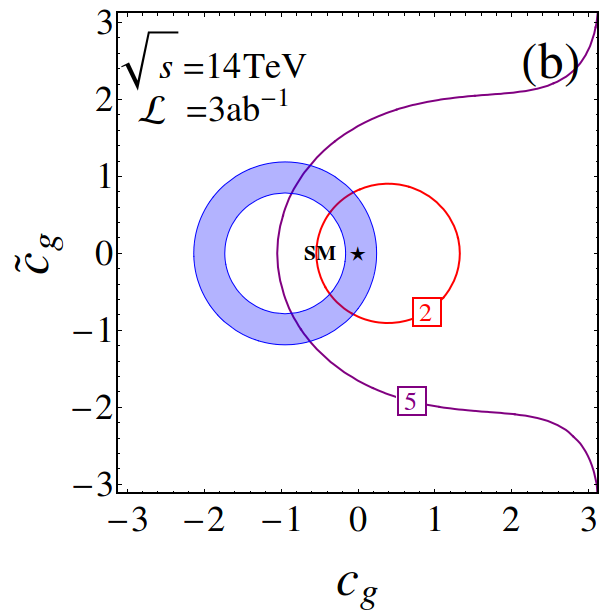}
 \includegraphics[width=0.23\textwidth]{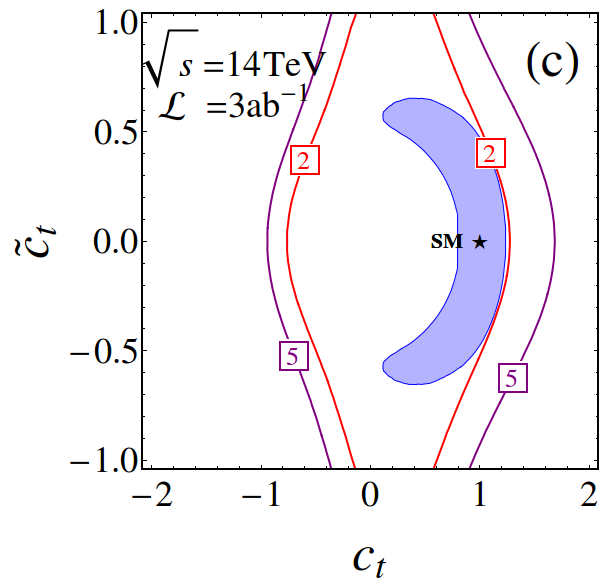} \\
 \includegraphics[width=0.23\textwidth]{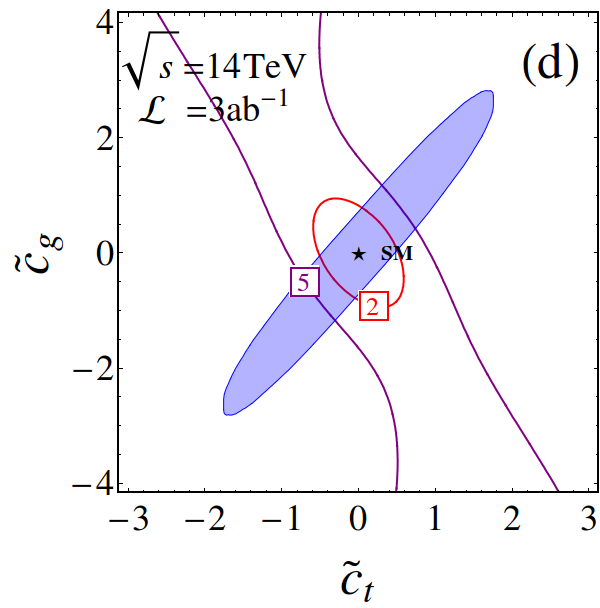}
 \includegraphics[width=0.23\textwidth]{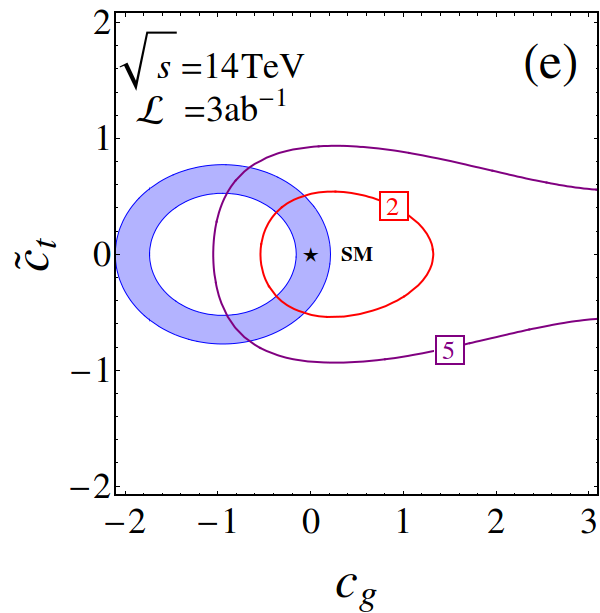}
 \includegraphics[width=0.23\textwidth]{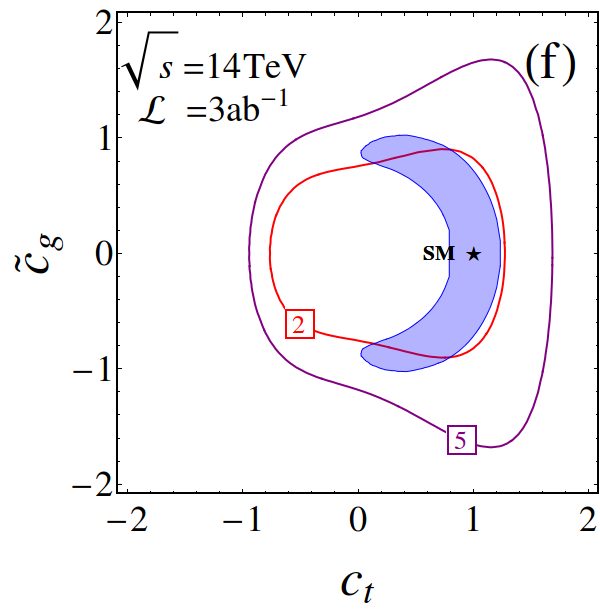}
\caption{Contours of $2\sigma$ exclusion (red curves) and $5\sigma$ discovery (purple curves)
for different combinations of Higgs effective couplings at the HL-LHC. The blue
regions represent the $95\%$ CL constraints at the 7 and 8 TeV LHC.}
 \label{fig:14TeV_0}
 \end{figure}

\begin{figure}[!htb]
 \centering
 \includegraphics[width=0.23\textwidth]{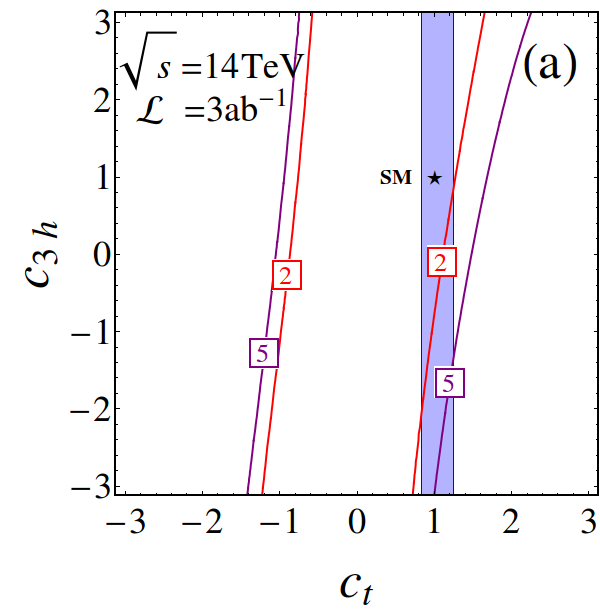}
 \includegraphics[width=0.23\textwidth]{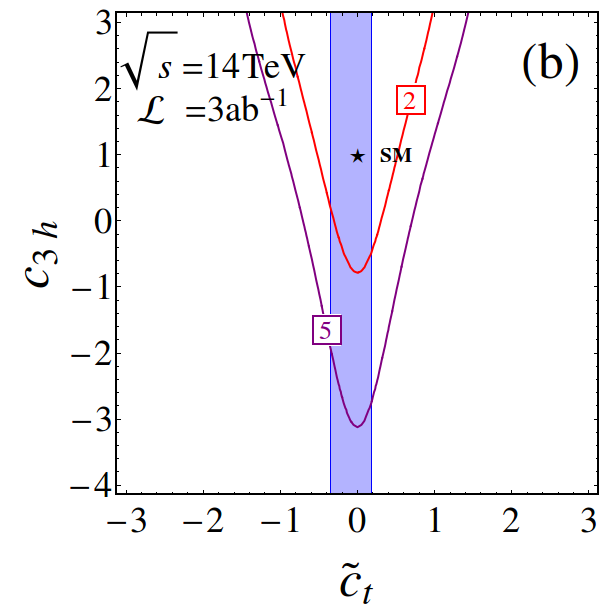}
 \includegraphics[width=0.23\textwidth]{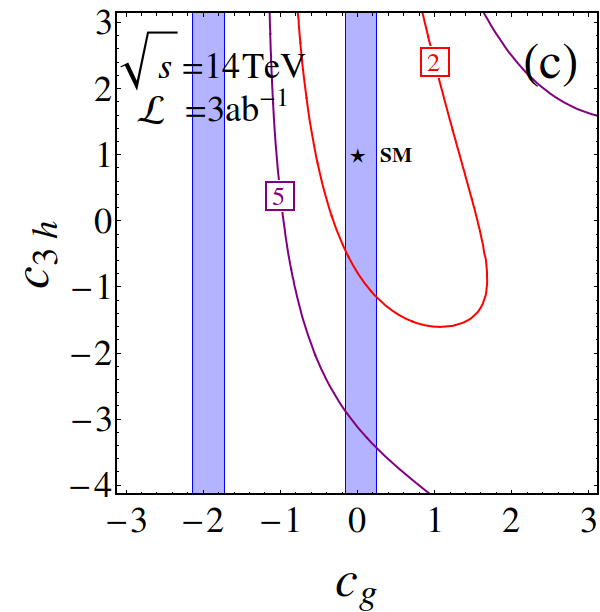}
 \includegraphics[width=0.23\textwidth]{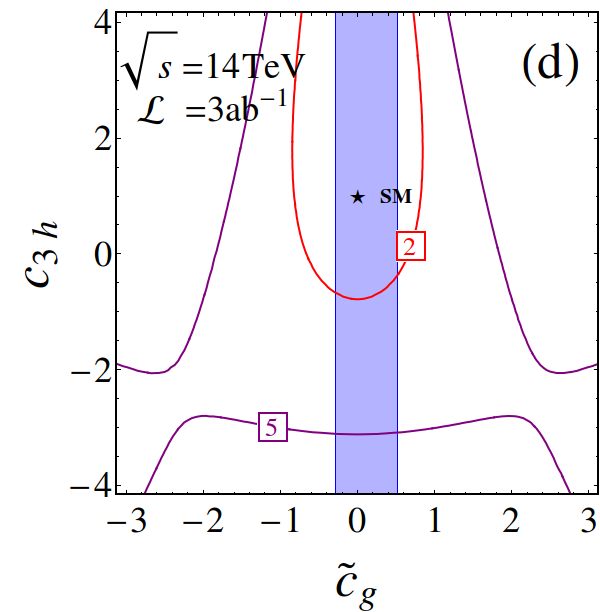}
 \includegraphics[width=0.23\textwidth]{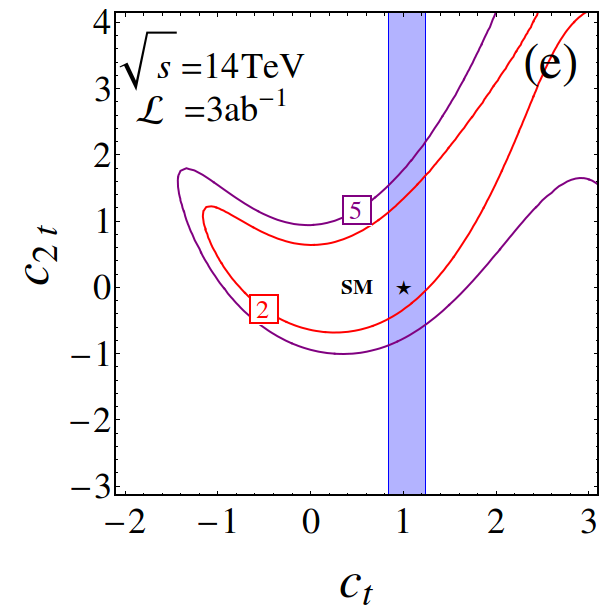}
 \includegraphics[width=0.23\textwidth]{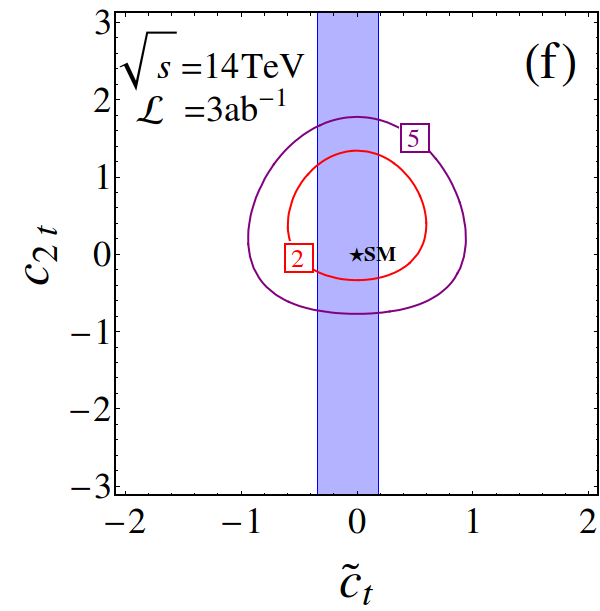}
 \includegraphics[width=0.23\textwidth]{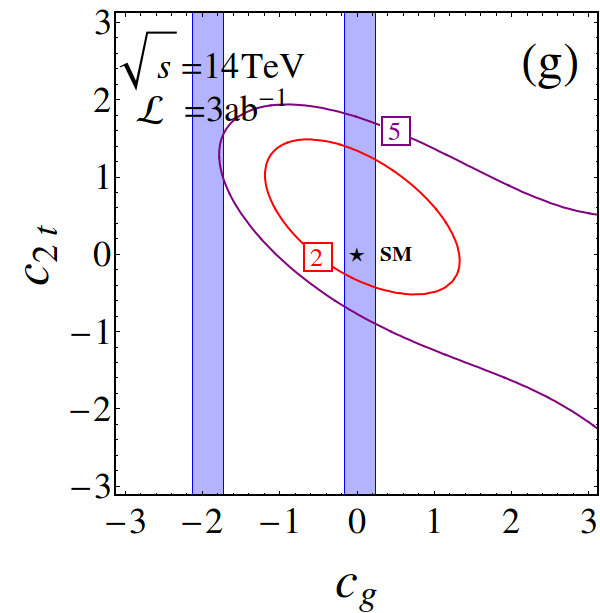}
 \includegraphics[width=0.23\textwidth]{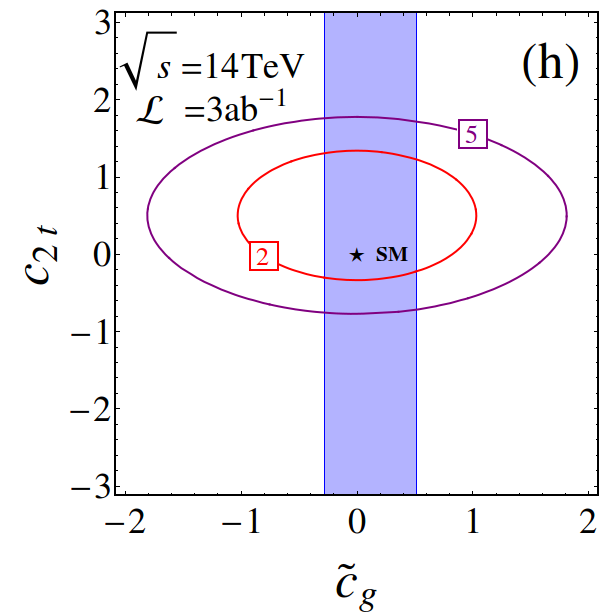}
 \includegraphics[width=0.23\textwidth]{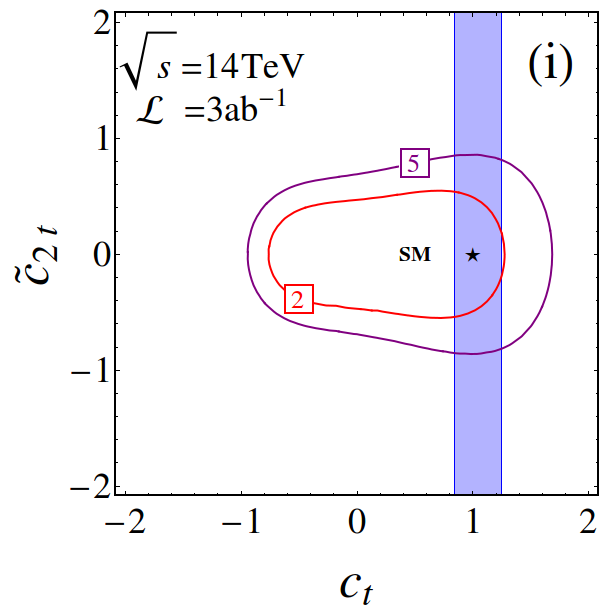}
 \includegraphics[width=0.23\textwidth]{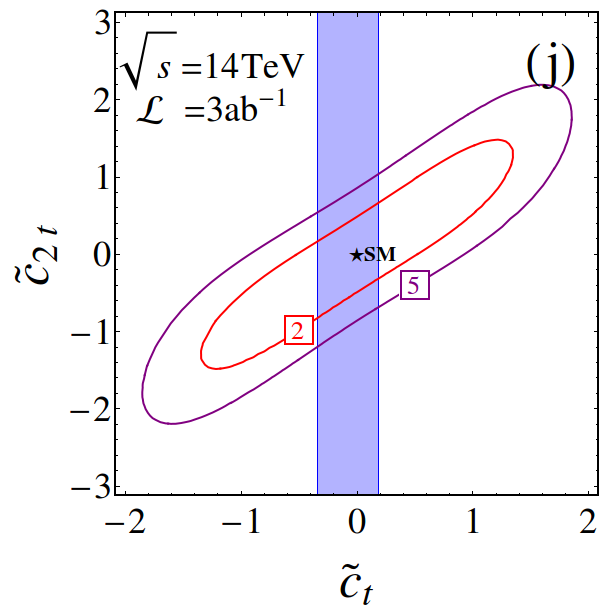}
 \includegraphics[width=0.23\textwidth]{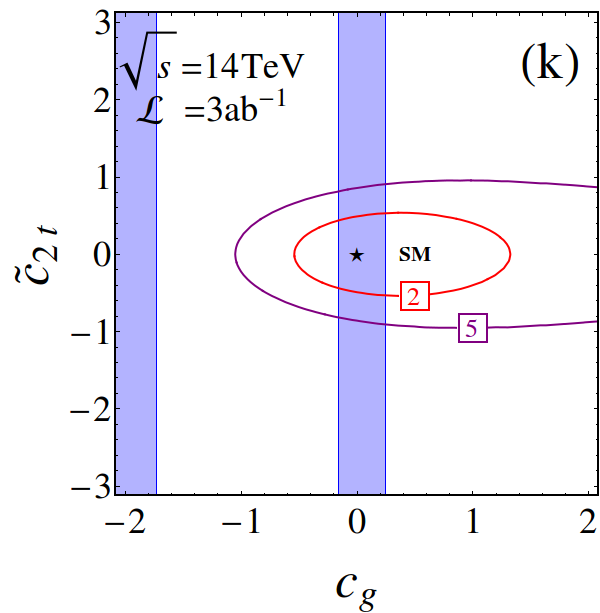}
 \includegraphics[width=0.23\textwidth]{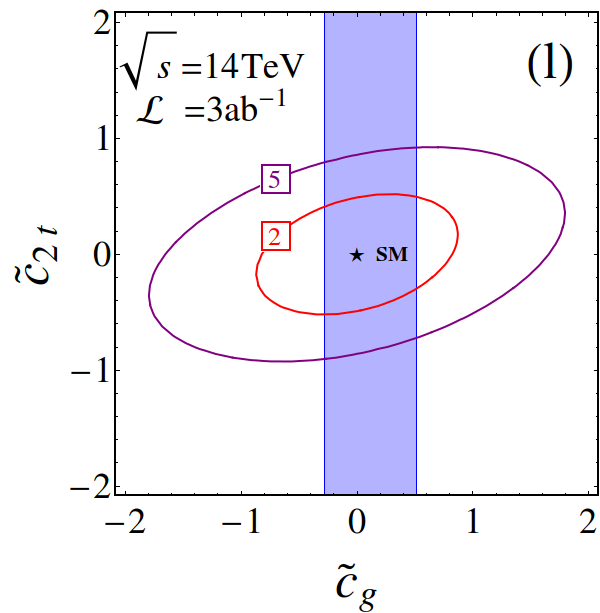}
  \caption{Contours of $2\sigma$ exclusion (red curves) and $5\sigma$ discovery (purple curves)
for different combinations of Higgs effective couplings at the HL-LHC. The blue
regions represent the $95\%$ CL constraints at the 7 and 8 TeV LHC.
  }
 \label{fig:14TeV_1}
 \end{figure}

\begin{figure}
 \includegraphics[width=0.23\textwidth]{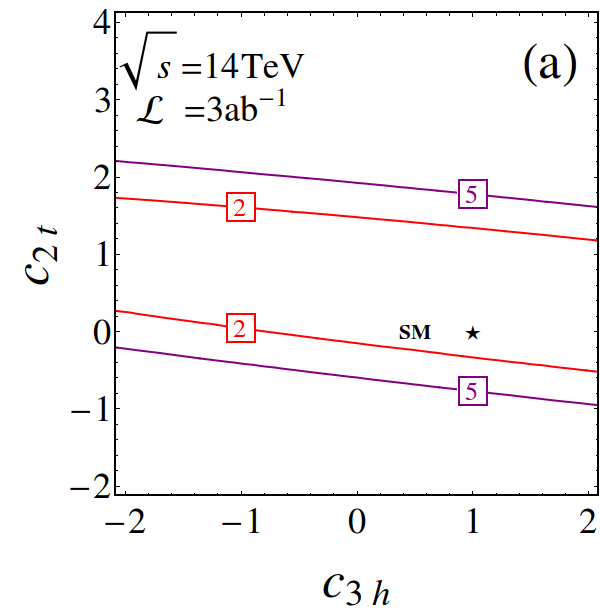}
 \includegraphics[width=0.23\textwidth]{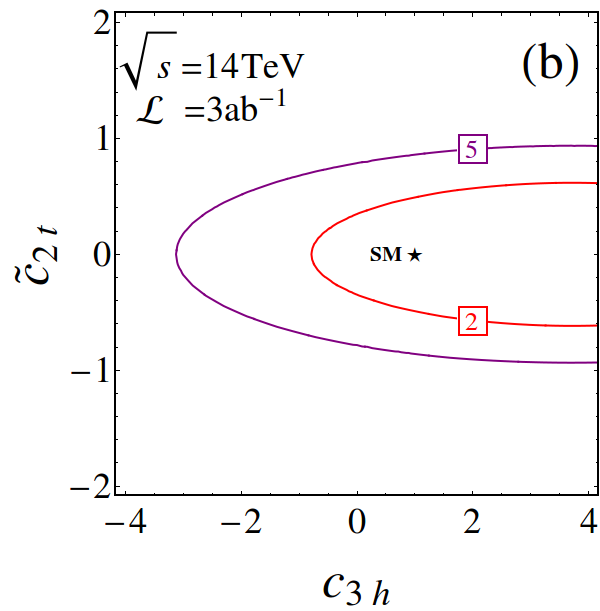}
 \includegraphics[width=0.23\textwidth]{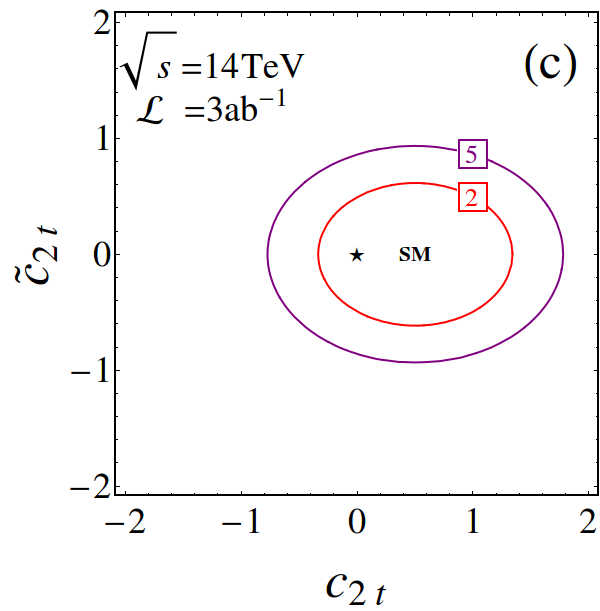}
 \caption{Contours of $2\sigma$ exclusion (red curves) and $5\sigma$ discovery (purple curves)
for different combinations of Higgs effective couplings at the HL-LHC. 
 }
 \label{fig:14TeV_2}
 \end{figure}

Figures~\ref{fig:14TeV_0}, \ref{fig:14TeV_1} and \ref{fig:14TeV_2} show the $2\sigma$ exclusion
(red curves) and $5\sigma$ discovery (purple curves) contours for the double
Higgs production at the 14~TeV LHC with an integrated luminosity $\mathcal{L}=3000~\rm fb^{-1}$, named as high luminosity LHC (HL-LHC). Throughout this study we vary only two effective couplings at a time. The blue regions denote the parameter space that is allowed by the current single Higgs measurements.
The pair production of the SM Higgs bosons is expected to be observed at the HL-LHC at only $1.3\sigma$ confidence level~\cite{Aad:2015}. Even though it is less promising to detect the double Higgs event, one can set an $2\sigma$ exclusion limit on the NP. On the other hand, if this process is discovered at the $5\sigma$ confidence level, it is a clear evidence of NP. We also show the $5\sigma$ discovery contours below.

In general, the shapes of the $2\sigma$ and $5\sigma$ boundary are similar. The large distortion occurs around the corners of the correlation contour of $c_g$ and $\tilde{c}_g$. The large $c_g$ and $\tilde{c}_g$ couplings could increase the total width of Higgs boson sizably\footnote{The current bound on the Higgs boson total width is about $\Gamma_{tot}\lesssim 6~\Gamma_{tot}^{SM}$~\cite{Aad:2015xua,Khachatryan:2015mma}, which is still too weak to constrain the Higgs effective couplings. };  see Eq.~\eqref{eq:brs}. The enlarged width inevitably reduces the branching ratio of Higgs boson decaying into a pair of bottom quarks or photons and then reduces the discovery potential of Higgs pair events, especially in the region of $c_g\gtrsim 2$ or $|\tilde{c}_g|\gtrsim 2$. In order to compensate the reduction of branching ratio, the double Higgs production rate has to be dramatically enhanced to reach a $5\sigma$ discovery.  

Figure~\ref{fig:14TeV_0} shows the sensitivity of the HL-LHC to a few combinations of effective couplings that can affect the single Higgs signal strength simultaneously. 
In the scenario $(c_t,c_g)$, one can use the double Higgs production to exclude
the degenerate parameter space in the lower band allowed by the single Higgs
measurements \cite{Cao:2015oaa}, but only a portion of the upper band consisting of the SM is 
excluded; see the red curve. 
In the scenarios of $(c_g,\tilde{c}_g)$, $(\tilde{c}_t,\tilde{c}_g)$ and $(c_g,\tilde{c}_t)$,  the parameter space away from the SM can be excluded; see Figs.~\ref{fig:14TeV_0}(b), (d) and (e).  
That is mainly owing to the different correlations of effective couplings in single Higgs productions and double Higgs productions. For example, consider $(\tilde{c}_g, \tilde{c}_t)$. The double Higgs production rate is proportional to $(\tilde{c}_t + \tfrac{4}{3}\tilde{c}_g)^2$ while the single Higgs production rate proportional to $(-\tilde{c}_t +\tfrac{2}{3}\tilde{c}_g)^2$; see Eqs.~\eqref{eq:amp} and \eqref{eq:kappag}. That yields the different slopes of the blue band and red (purple) curves. 
Unfortunately, the double Higgs process has less sensitivity to the parameter space in the scenarios of 
$(c_t,\tilde{c}_t)$ and $(c_t,\tilde{c}_g)$; see  Figs.~\ref{fig:14TeV_0}(c) and (f). 

Not all the effective couplings affect the single Higgs production and Higgs boson decay. We separate the effective couplings into two categories: couplings sensitive to single Higgs production, say $c_{t,g}$ and $\tilde{c}_{t,g}$, and others. Figure~\ref{fig:14TeV_1} shows the correlation among $c_{t,g}$ ($\tilde{c}_{t,g}$) and others effective couplings. Plots in the first row in Fig.~\ref{fig:14TeV_1} show the correlations
between $c_{3h}$ and $c_t, \tilde{c}_t, c_g, \tilde{c}_g$, respectively. The sign of $c_{3h}$ coupling is important as it could alter the cancellation between the triangle diagram and the box diagram in the SM. A negative $c_{3h}$ leads to an enhancement of the double Higgs production, easily yielding a $5\sigma$ discovery. On the other hand, the $2\sigma$ exclusion limit demands the $c_{3h}$ being not too negatively large when $c_t=1$; see Figs.~\ref{fig:14TeV_1}(b), (c) and (d). The tension is slightly alleviated in $(c_{3h}, c_t)$; it requires $c_{3h}>-2$ if the double Higgs event is not observed at the HL-LHC; see Fig.~\ref{fig:14TeV_1}(a). There is no stringent bound on $c_{3h}$ from top, indicating that the double Higgs production is not sensitive to the quartic term in the Higgs potential if the coefficient is positive. It has been pointed out in the comprehensive study in Ref.~\cite{Azatov:2015oxa} which considers the CP-conserving operators. Our study shows the conclusion also holds for a CP-violating model.     
 
Plots in the second (third) row of Fig.~\ref{fig:14TeV_1} show the correlations
between $c_{2t}(\tilde{c}_{2t})$ and $c_t, \tilde{c}_t, c_g, \tilde{c}_g$, respectively. 
If the NP model generates a sizable $c_{2t}$, then it is very promising to see its effects in the Higgs boson pair productions in both CP-conserving and CP-violating models; see the purple curves. 
Similar to the case of $c_{3h}$, the cancellation between $F_{\triangle}$
and $F_{\Box}$ also imposes a bound on $c_{2t}$ from bottom. Unlike the $c_{3h}$, the $c_t$ coupling is also bounded from top. If no deviation is observed in the double Higgs production, then one can impose a bound on $c_{2t}$ ($\tilde{c}_{2t}$), together with constraints obtained from the single Higgs production, as follows: 
 \begin{align}
& (c_t,c_{2t}): -0.469\lesssim c_{2t} \lesssim 1.69, 
&&  (c_t,\tilde{c}_{2t}):-0.545\lesssim \tilde{c}_{2t} \lesssim 0.542,
\nn\\
& (\tilde{c}_t,c_{2t}):-0.339\lesssim c_{2t} \lesssim 1.34,
 &&  (\tilde{c}_t,\tilde{c}_{2t}):-0.817\lesssim \tilde{c}_{2t} \lesssim 0.671,
 \nn\\
 &(c_g,c_{2t}):-0.429\lesssim c_{2t} \lesssim 1.40, 
  &&(c_g,\tilde{c}_{2t}):-0.546\lesssim \tilde{c}_{2t} \lesssim 0.529,
  \nn\\
 &(\tilde{c}_g,c_{2t}):-0.339\lesssim c_{2t} \lesssim 1.34, 
 &&  (\tilde{c}_g,\tilde{c}_{2t}):-0.519\lesssim \tilde{c}_{2t} \lesssim 0.516~.
 \end{align}
It is worth mentioning that the degenerate parameter spaces in $c_g$, i.e. the two blue bands in Figs.~\ref{fig:14TeV_1} (c), (g) and (k),  can be fully resolved at the HL-LHC.

Figure~\ref{fig:14TeV_2} shows the correlations among effective couplings ($c_{2t}$, $\tilde{c}_{2t}$ and $c_{3h}$) that do not affect the single Higgs production. The three couplings are completely free. They are constrained only by double Higgs production at the HL-LHC. If the NP effects are hidden in the three couplings, then one is not able to probe the NP effects no matter how accurately one measures the single Higgs boson production. The double Higgs production is sensitive to both magnitude and sign of the $c_{3h}$ coupling. If $c_{3h}$ is the only non-zero effective coupling, then null results of Higgs pair searches will require $c_{3h}>-1$. Including $c_{2t}$ completely relax the constraint on $c_{3h}$; see Fig.~\ref{fig:14TeV_2}(a). It is owing to the interference between $c_{3h}F_\triangle$ and $c_{2t}F_\triangle$ terms in Eq.~\eqref{eq:amp}. As a result, a large negative $c_{3h}$ is still allowed. 
The $\tilde{c}_{2t}$ coupling, which does not interfere with $c_{3h}$, has no strong impact on $c_{3h}$. The $c_{2t}$ and $\tilde{c}_{2t}$ do not interfere and result in the symmetric eclipse bound. 

Finally, we list analytical expressions of all the $2\sigma$ exclusion limits below: 
\begin{align}
&(c_t,c_g):\ 2.25 c_g^2+c_g (0.508-1.15 c_t) c_t+1.64 c_t^4-0.739 c_t^3+0.0993 c_t^2-1.00<1.28~,\nn\\
&(c_g,\tilde{c}_g):\ 2.25 c_g^2-0.645 c_g+2.25 \tilde{c}_g^2<1.28~,\nn\\
&(c_t,\tilde{c}_t):1.64 c_t^4-0.739 c_t^3+c_t^2 (7.18 \tilde{c}_t^2+0.0993)-2.46 c_t \tilde{c}_t^2+0.571 \tilde{c}_t^4+0.257 \tilde{c}_t^2-1.00<1.28~,\nn\\
&(\tilde{c}_t,\tilde{c}_g):\ 2.25 \tilde{c}_g^2 + 1.72 \tilde{c}_g \tilde{c}_t + 0.571 \tilde{c}_t^4 + 4.97 \tilde{c}_t^2<1.28~,\nn\\
&(c_g,\tilde{c}_t):\ 2.25 c_g^2+c_g (1.65 \tilde{c}_t^2-0.645)+0.571 \tilde{c}_t^4+4.97 \tilde{c}_t^2<1.28~,\nn\\
&(c_t,\tilde{c}_g):\ 2.25 \tilde{c}_g^2+1.64 c_t^4-0.739 c_t^3+0.0993 c_t^2-1.00<1.28~,\nn\\
&(c_t,c_{3h}):\ 0.0993 c_{3h}^2 c_t^2-0.739 c_{3h} c_t^3+1.64 c_t^4-1.00<1.28~,\nn\\
&(\tilde{c}_t,c_{3h}):\ c_{3h}^2 (0.257 \tilde{c}_t^2+0.0993)+c_{3h} (-2.46 \tilde{c}_t^2-0.739)+0.571 \tilde{c}_t^4+7.18 \tilde{c}_t^2+0.639<1.28~,\nn\\
&(c_g,c_{3h}):\ c_{3h}^2 (0.0369 c_g^2+0.0975 c_g+0.0993)+c_{3h} (0.406 c_g^2+0.146 c_g-0.739)+1.80 c_g^2-0.888 c_g+0.639<1.28~,\nn\\
&(\tilde{c}_g,c_{3h}):\ c_{3h}^2 (0.0369 \tilde{c}_g^2+0.0993)+c_{3h} (0.406 \tilde{c}_g^2-0.739)+1.80 \tilde{c}_g^2+0.639<1.28~,\nn\\
&(c_{t},c_{2t}):\ 2.85 c_{2t}^2+c_{2t} (0.920-3.79 c_{t}) c_{t}+1.64 c_{t}^4-0.739 c_{t}^3+0.0993 c_{t}^2-1.00<1.28~,\nn\\
&(\tilde{c}_t,c_{2t}):\ 2.85 c_{2t}^2 + c_{2t} (1.91 \tilde{c}_t^2 - 2.87) + 0.571 \tilde{c}_t^4 + 4.97 \tilde{c}_t^2<1.28~,\nn\\
&(c_g,c_{2t}):\ 2.85 c_{2t}^2+c_{2t} (2.52 c_g-2.87)+2.25 c_g^2-0.645 c_g<1.28~,\nn\\
&(\tilde{c}_g,c_{2t}):\ 2.85 c_{2t}^2 - 2.87 c_{2t} + 2.25 \tilde{c}_g^2<1.28~,\nn\\
&(c_t,\tilde{c}_{2t}):\ 5.28 \tilde{c}_{2t}^2+1.64 c_t^4-0.739 c_t^3+0.0993 c_t^2-1.00<1.28~,\nn\\
&(\tilde{c}_t,\tilde{c}_{2t}):\ 5.28 \tilde{c}_{2t}^2 - 10.2 \tilde{c}_{2t} \tilde{c}_t + 0.571 \tilde{c}_t^4 + 4.97 \tilde{c}_t^2<1.28~,\nn\\
&(c_g,\tilde{c}_{2t}):\ 5.28 \tilde{c}_{2t}^2+2.25 c_g^2-0.645 c_g<1.28~,\nn\\
&(\tilde{c}_g,\tilde{c}_{2t}):\ 5.28 \tilde{c}_{2t}^2 - 2.07 \tilde{c}_{2t} \tilde{c}_g + 2.25 \tilde{c}_g^2<1.28~,\nn\\
&(c_{3h},c_{2t}):\ 2.85 c_{2t}^2+c_{2t} (0.920 c_{3h}-3.79)+0.0993 c_{3h}^2-0.739 c_{3h}+0.639<1.28~,\nn\\
&(c_{3h},\tilde{c}_{2t}):\ 5.28 \tilde{c}_{2t}^2+0.0993 c_{3h}^2-0.739 c_{3h}+0.639<1.28~,\nn\\
&(c_{2t},\tilde{c}_{2t}):\ 2.85 c_{2t}^2-2.87 c_{2t}+5.28 \tilde{c}_{2t}^2<1.28~.
\end{align}
Effective couplings violating the above inequalities can be excluded at the HL-LHC. 

\subsection{\tf{Sensitivity to Higgs effective couplings at a future 100~TeV $pp$-collider}{Sensitivity to Higgs effective couplings at a future 100~TeV $pp$-collider}}\label{sec:44y}

Now we study the potential of a future 100 TeV $pp$-collider on Higgs effective couplings.
It is shown that increasing the collider energy improves the sensitivity significantly~\cite{Azatov:2015oxa, Contino:2016spe}.  
Our simulation shows that the performance at the 100~TeV machine with an integrated luminosity of $10~\rm fb^{-1}$ is comparable to that at the HL-LHC. Moreover, the $gg\to hh \to b\bar{b}\gamma\gamma$ process can be discovered with $\mathcal{L}=256~\rm fb^{-1}$ at the 100 TeV $pp$-collider. Accumulating more luminosity enables us to discover NP effects in the double Higgs productions through $b\bar{b}\gamma\gamma$ channel. 

\begin{figure}
 \centering
 \includegraphics[width=0.23\textwidth]{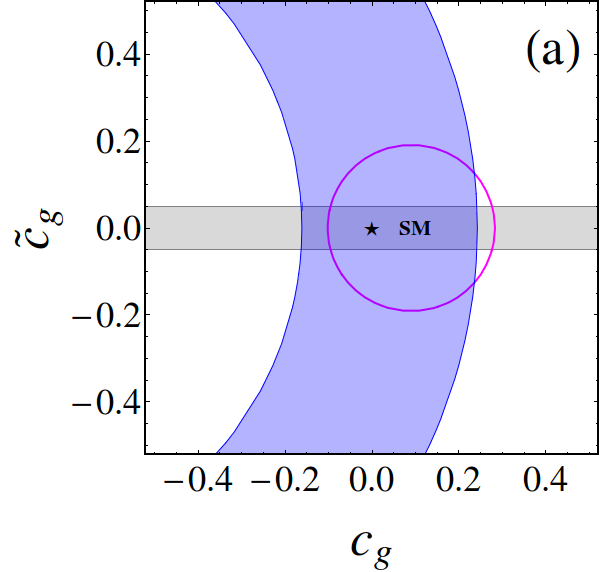}
 \includegraphics[width=0.23\textwidth]{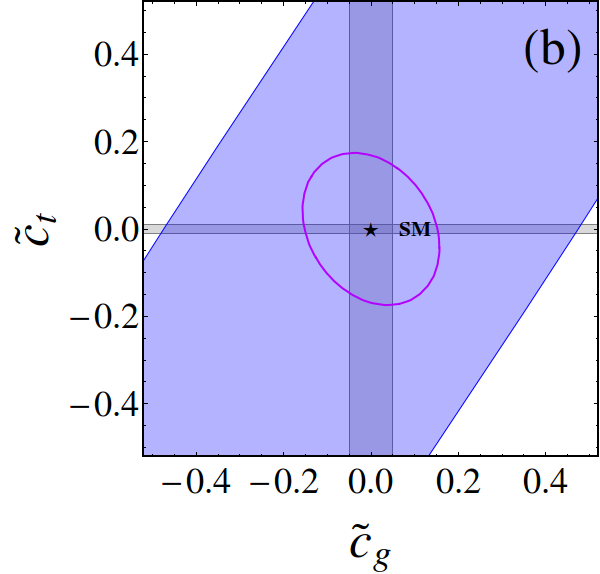}
 \includegraphics[width=0.23\textwidth]{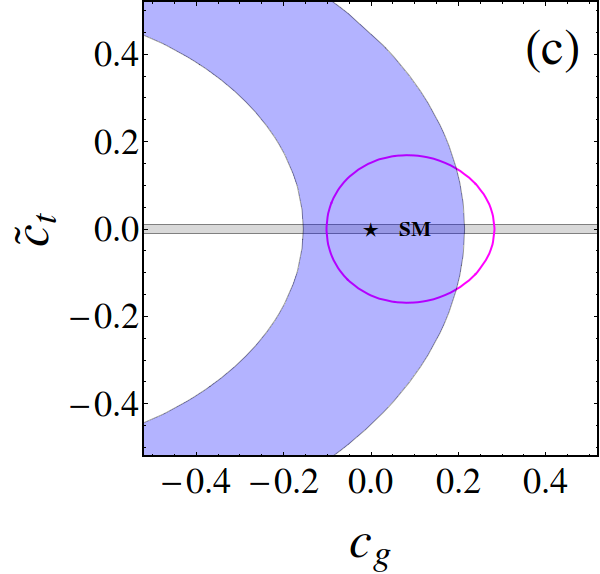}\\
 \includegraphics[width=0.23\textwidth]{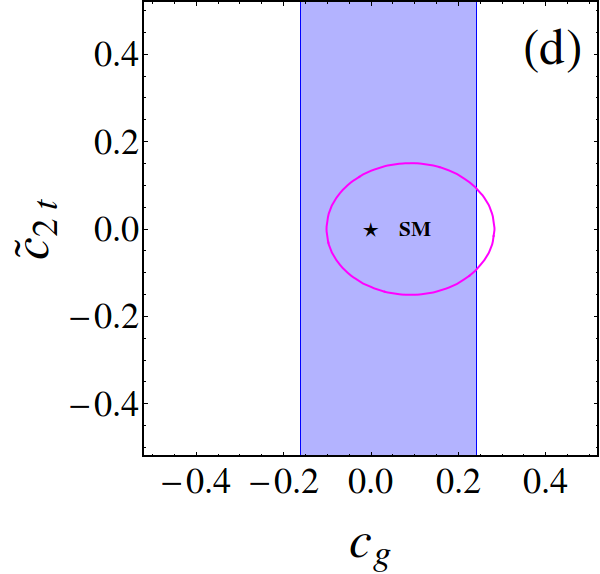}
 \includegraphics[width=0.23\textwidth]{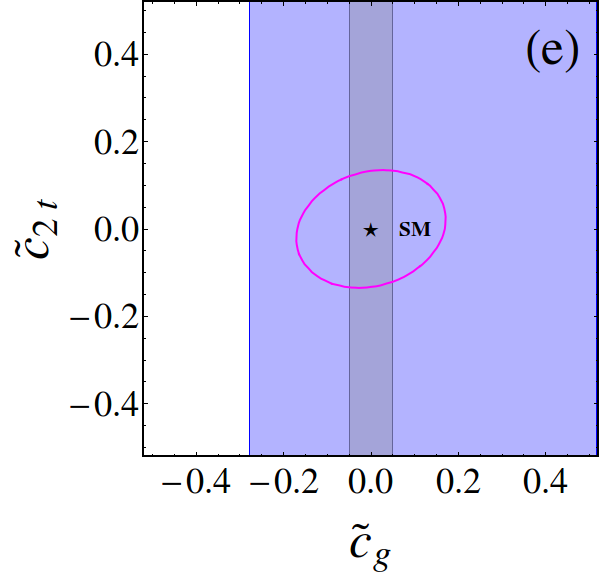}
 \includegraphics[width=0.23\textwidth]{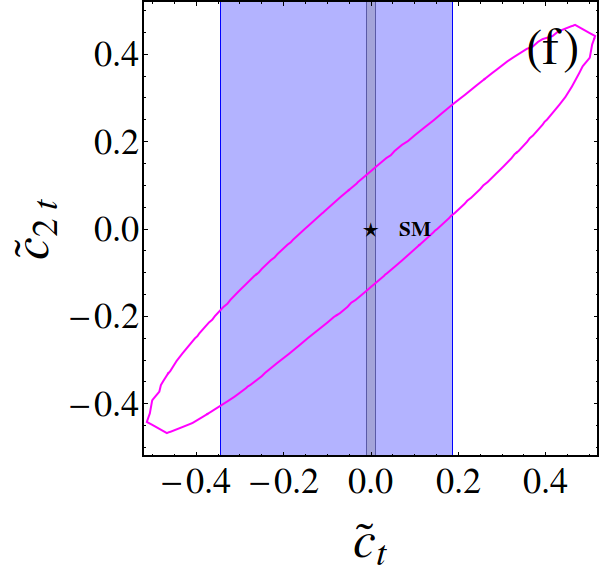}
  \caption{ $5\sigma$ discovery contours for different combinations of Higgs effective couplings at the 100 TeV $pp$-collider with the integrated luminosity $\mathcal{L}=30~\rm ab^{-1}$. The blue regions represent the $95\%$ CL constraints at the 7 and 8 TeV LHC.}
 \label{fig:100TeV_0}
 \end{figure}

\begin{figure}[h!]
\includegraphics[width=0.23\textwidth]{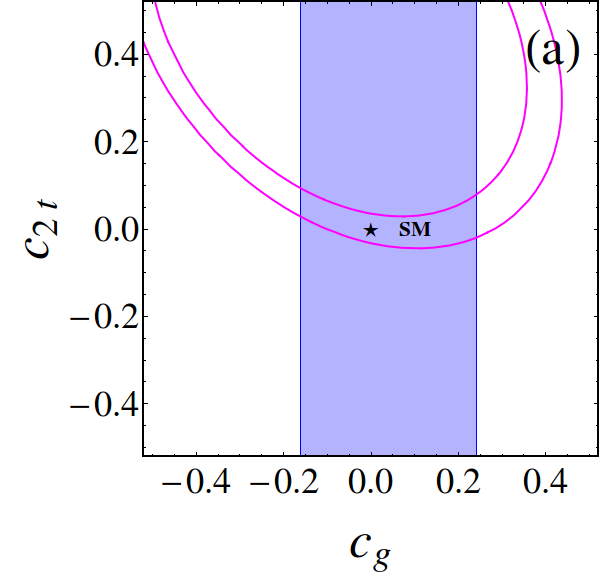}
\includegraphics[width=0.23\textwidth]{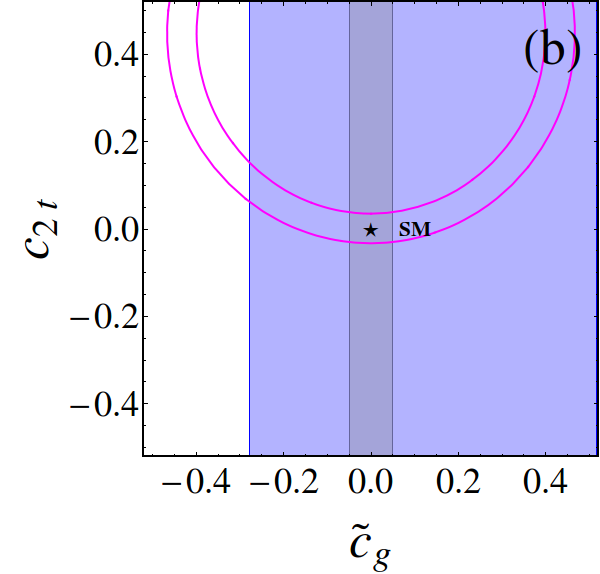}
\includegraphics[width=0.23\textwidth]{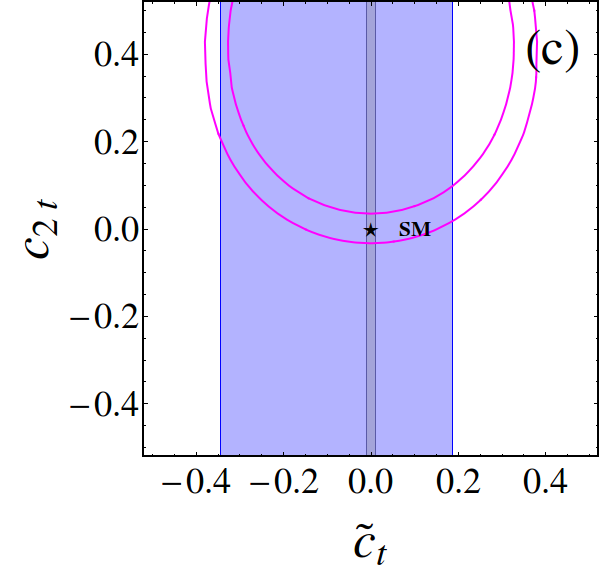}
\includegraphics[width=0.23\textwidth]{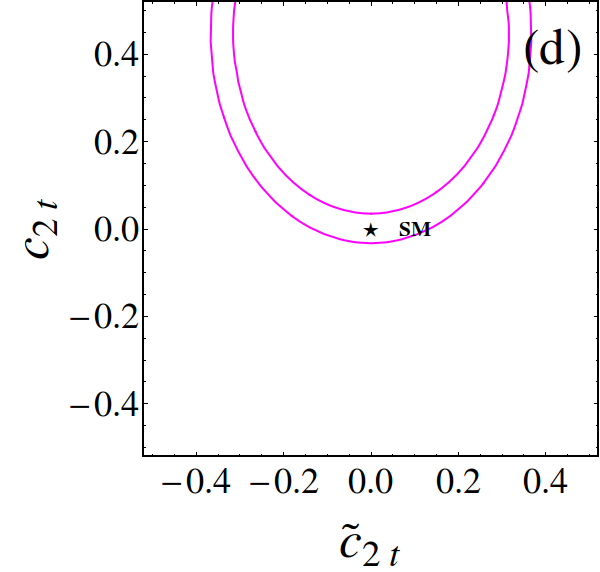} \\
\includegraphics[width=0.23\textwidth]{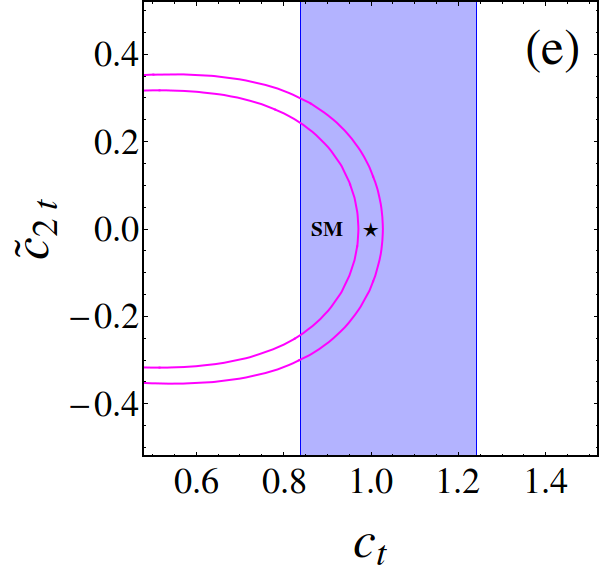}
\includegraphics[width=0.23\textwidth]{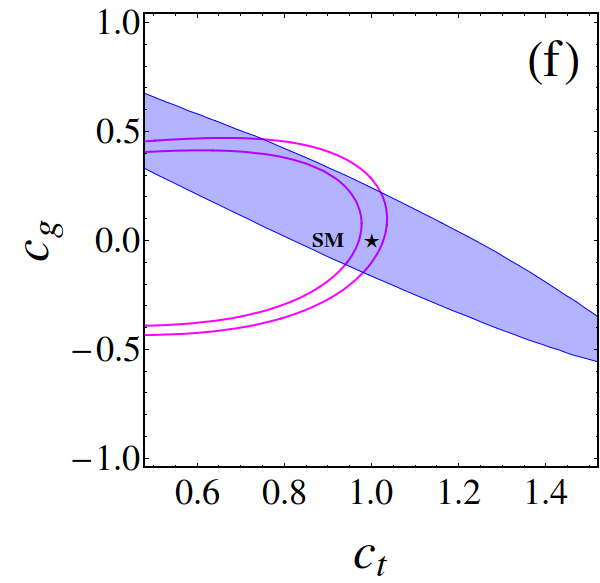}
\includegraphics[width=0.23\textwidth]{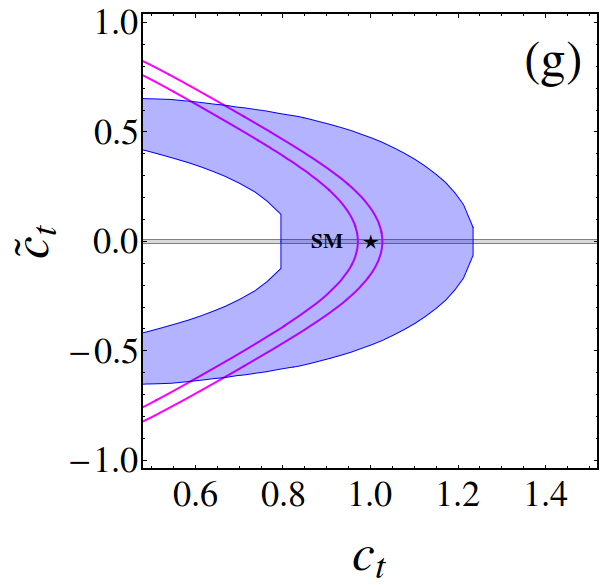}
\includegraphics[width=0.23\textwidth]{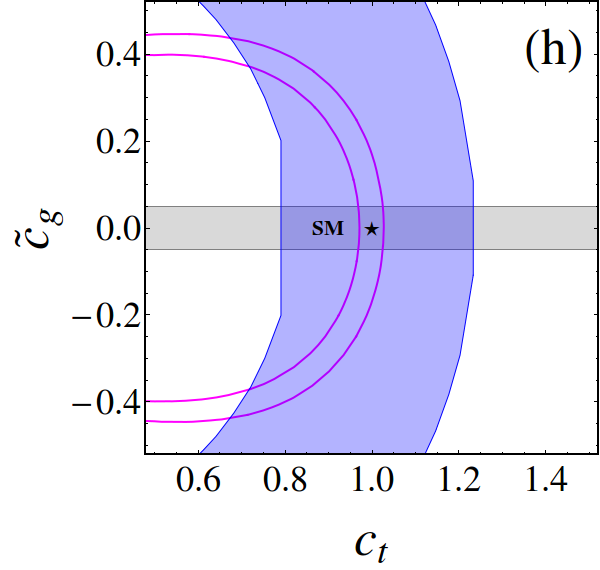}
\caption{ $5\sigma$ discovery contours for different combinations of Higgs effective couplings at the 100 TeV $pp$-collider with the integrated luminosity $\mathcal{L}=30~\rm ab^{-1}$. The blue regions represent the $95\%$ CL constraints at the 7 and 8 TeV LHC.}
 \label{fig:100TeV_1}
 \end{figure}

\begin{figure}[b!]
  \includegraphics[width=0.23\textwidth]{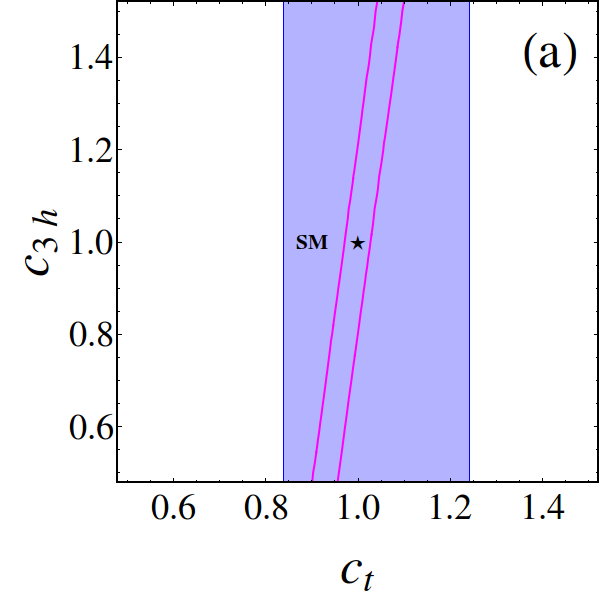}
  \includegraphics[width=0.23\textwidth]{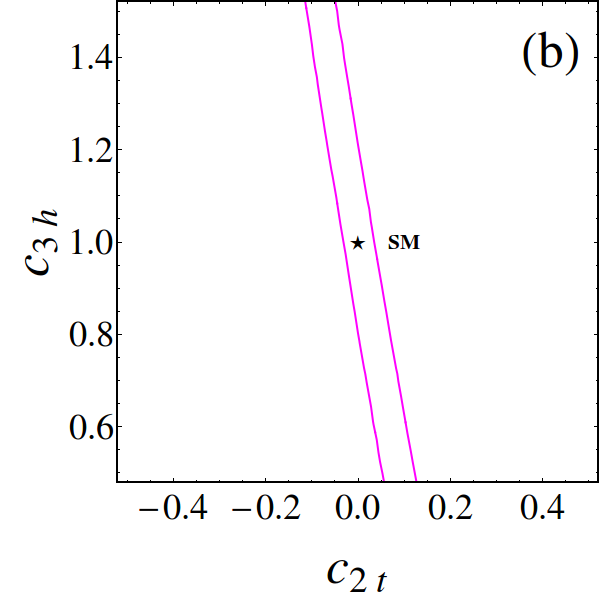}
 \includegraphics[width=0.23\textwidth]{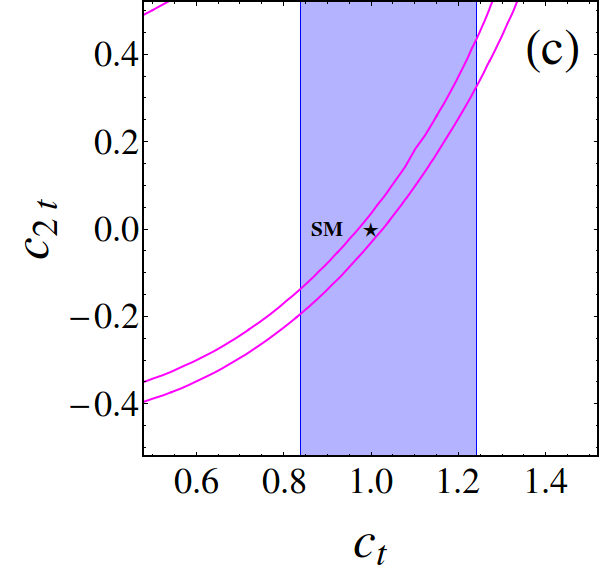}
  \caption{ $5\sigma$ discovery contours for different combinations of Higgs effective couplings at the 100 TeV $pp$-collider with the integrated luminosity $\mathcal{L}=30~\rm ab^{-1}$. The blue regions represent the $95\%$ CL constraints at the 7 and 8 TeV LHC.}
 \label{fig:100TeV_2}
 \end{figure}

\begin{figure}
 \includegraphics[width=0.23\textwidth]{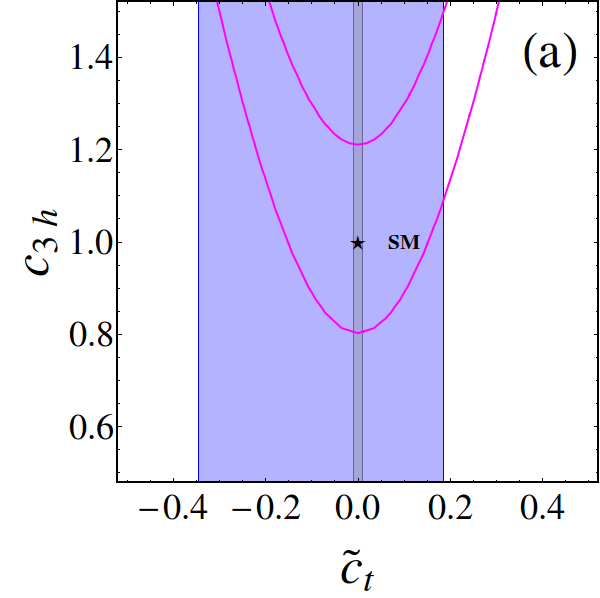}
 \includegraphics[width=0.23\textwidth]{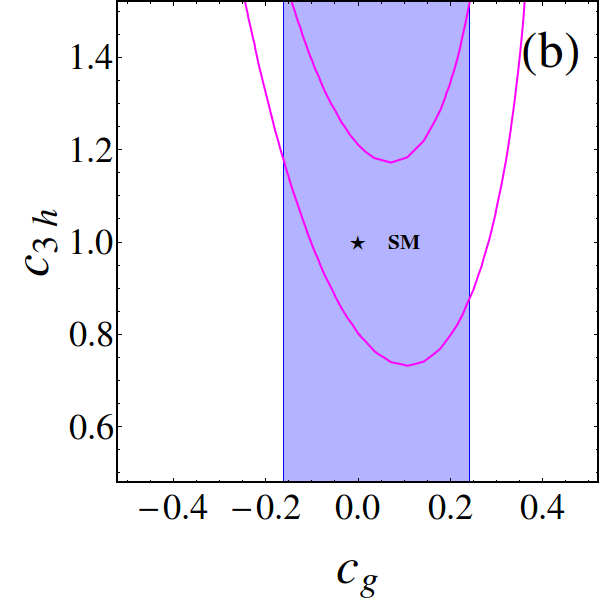}
 \includegraphics[width=0.23\textwidth]{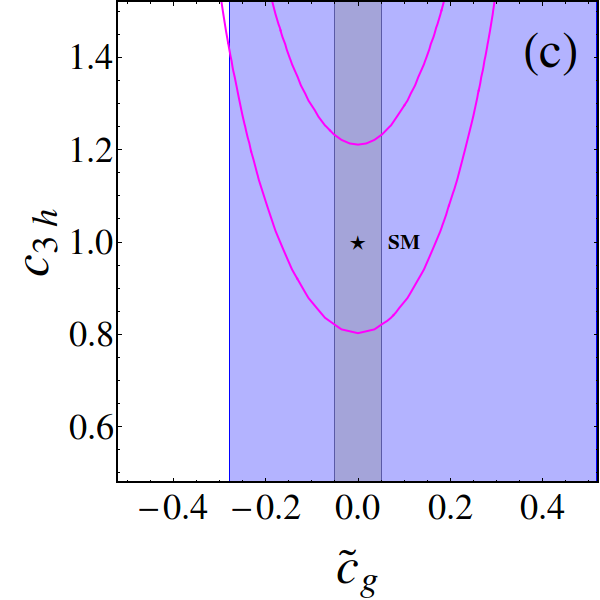}
 \includegraphics[width=0.23\textwidth]{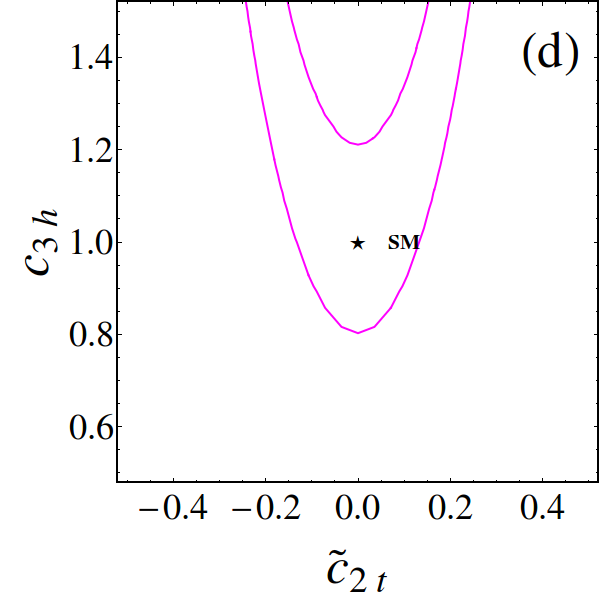}
 \caption{$5\sigma$ discovery contours for different combinations of Higgs effective couplings at the 100 TeV $pp$-collider with the integrated luminosity $\mathcal{L}=30~\rm ab^{-1}$. The blue regions represent the $95\%$ CL constraints at the 7 and 8 TeV LHC.}
 \label{fig:100TeV_3}
 \end{figure}

As it is guaranteed to observe the Higgs pair signal in the SM at the 100 TeV machine, we focus on the NP searches hereafter. Figures~\ref{fig:100TeV_0}-\ref{fig:100TeV_3} display the
$5\sigma$ contours of discovering NP with
an integrated luminosity of $30~\rm ab^{-1}$. The
SM process $gg\to hh \to b\bar{b}\gamma\gamma$  is recognized
as a background. The regions with the significance $Z_0<5$ are depicted
with magenta curves. Outside of those magenta regions, the NP is expected to be observed. Again the constraints from the current single Higgs
measurements are denoted in blue regions. We also include the EDM constraints on the CP-odd
couplings $\tilde{c}_t$ and $\tilde{c}_g$; see the grey bands. The EDM 
constraints are very strigent on $\tilde{c}_t$ or $\tilde{c}_g$.  The
double Higgs production provides an alternative way to check $\tilde{c}_t$ and $\tilde{c}_g$. If the Higgs pair signal in the NP model is discovered in the parameter space outside the EDM bound, then additional CP-violating interaction has to be included to respect the EDM constraint.

We classify those figures into four categories according to the
shapes of the boundary of $5\sigma$ discovery region.
All the discovery regions in Fig.~\ref{fig:100TeV_0} are in a shape of ellipse; see the magenta curve. The parameter outside those ellipses can be discovered at more than $5\sigma$ confidence level. In the parameter space that is close to the SM, the modification of the decay branching ratios $\mu_{\gamma\gamma}$ and $\mu_{b\bar{b}}$ can be ignored.  We obtain  analytic expressions corresponding to the $5\sigma$ discovery of NP effects as follows:
\begin{align}
&(c_g,\tilde{c}_g): ~~3.70 c_g^2-0.290 c_g+3.70 \tilde{c}_g^2 \geq 0.1~,\nn\\
&(\tilde{c}_g,\tilde{c}_t):~~3.70 \tilde{c}_g^2 + 1.40 \tilde{c}_g \tilde{c}_t + 0.806 \tilde{c}_t^4 + 4.54 \tilde{c}_t^2 \geq 0.1~,\nn\\
&(c_g,\tilde{c}_t):~~3.70 c_g^2+c_g (1.90 \tilde{c}_t^2-0.290)+0.806 \tilde{c}_t^4+4.54 \tilde{c}_t^2\geq 0.1~,\nn\\
&(c_g,\tilde{c}_{2t}):\ 5.68 \tilde{c}_{2t}^2+3.70 c_g^2-0.290 c_g\geq 0.1~,\nn\\
&(\tilde{c}_g,\tilde{c}_{2t}):\ 5.68 \tilde{c}_{2t}^2 - 1.46 \tilde{c}_{2t} \tilde{c}_g + 3.70 \tilde{c}_g^2\geq 0.1~,\nn\\
&(\tilde{c}_t,\tilde{c}_{2t}):\ 5.68 \tilde{c}_{2t}^2 - 9.73 \tilde{c}_{2t} \tilde{c}_t + 0.806 \tilde{c}_t^4 + 4.54 \tilde{c}_t^2\geq 0.1~.
\label{eq:analytic0}
\end{align}
The analytical expressions of one effective coupling can be derived from the above inequalities by setting the other coupling to be zero.  
The $5\sigma$ curve in $(\tilde{c}_{t},\tilde{c}_{2t})$ is stretched as a result of 
the significant interference effect between $\tilde{c}_t F_\Box^{(2)}$
and $\tilde{c}_{2t} F_\triangle^{(1)}$.

Figure~\ref{fig:100TeV_1} displays the correlation among effective couplings, of which the $5\sigma$ discovery boundary exhibits a ring type shape. 
Most of parameter space allowed by the single Higgs production can be covered by double Higgs production. The parameter outside the $5\sigma$ band produces more Higgs pair events,  while the parameter inside the band reduces Higgs pair events. The bands of discovery potential at a confidence level less than $5\sigma$ are listed as follows:
\begin{align}
&(c_g,c_{2t}):\ ~-0.1 <3.33 c_{2t}^2+c_{2t} (2.33 c_g-2.98)+3.70 c_g^2-0.290 c_g<0.1,\nn\\
&(\tilde{c}_g,c_{2t}):\ ~-0.1 <3.33 c_{2t}^2 - 2.98 c_{2t} + 3.70 \tilde{c}_g^2<0.1,\nn\\
&(\tilde{c}_t,c_{2t}):\ ~-0.1 <3.33 c_{2t}^2 + c_{2t} (1.92 \tilde{c}_t^2 - 2.98) + 0.806 \tilde{c}_t^4 + 4.54 \tilde{c}_t^2<0.1,\nn\\
&(\tilde{c}_{2t},c_{2t}):\ -0.1 <3.33 c_{2t}^2-2.98 c_{2t}+5.68 \tilde{c}_{2t}^2<0.1,\nn\\
&(c_t,\tilde{c}_{2t}):\ ~-0.1 <5.68 \tilde{c}_{2t}^2+1.58 c_t^4-0.671 c_t^3+0.0880 c_t^2-1.00<0.1,\nn\\
&(c_t,c_g):\ ~-0.1 <3.70 c_g^2+c_g (0.457-0.746 c_t) c_t+1.58 c_t^4-0.671 c_t^3+0.0880 c_t^2-1.00<0.1,\nn\\
&(c_t,\tilde{c}_t):\ ~-0.1 <1.58 c_t^4-0.671 c_t^3+c_t^2 (6.46 \tilde{c}_t^2+0.0880)-2.14 c_t \tilde{c}_t^2+0.806 \tilde{c}_t^4+0.222 \tilde{c}_t^2-1.00<0.1,\nn\\
&(c_t,\tilde{c}_g):\ ~-0.1 <3.70 \tilde{c}_g^2+1.58 c_t^4-0.671 c_t^3+0.0880 c_t^2-1.00<0.1~.
\label{eq:analytic1}
\end{align}
Couplings violating the above inequalities lead to a discovery of Higgs pair signal in the NP model.

Figure~\ref{fig:100TeV_2} displays the $5\sigma$ contour with a line shape. We notice that, owing to the insensitivity to $c_{3h}$, the $5\sigma$ discovery band in $(c_{t},c_{3h})$ and $(c_{2t},c_{3h})$ appears as a vertical line. The $5\sigma$ band in $(c_t,c_{2t})$ is determined by the cancellation among $c_t^2 F_\Box$ and $c_{2t} F_\triangle$ terms. The bands of discovery potential at a confidence level less than $5\sigma$ are 
\begin{align}
&(c_t,c_{3h}):\ ~-0.1 <0.0880 c_{3h}^2 c_t^2-0.671 c_{3h} c_t^3+1.58 c_t^4-1.00<0.1,\nn\\
&(c_{2t},c_{3h}):\ -0.1< 3.33 c_{2t}^2+c_{2t} (0.889 c_{3h}-3.87)+0.0880 c_{3h}^2-0.671 c_{3h}+0.583<0.1,\nn\\
&(c_{t},c_{2t}):\ ~ -0.1<3.33 c_{2t}^2+c_{2t} (0.889-3.87 c_t) c_t+1.58 c_t^4-0.671 c_t^3+0.0880 c_t^2-1.00<0.1~.
\label{eq:analytic2}
\end{align}

Finally, we plot in Fig.~\ref{fig:100TeV_3} the $5\sigma$ contour with a irregular shape. The bands of discovery potential at a confidence level less than $5\sigma$ are 
\begin{align}
&(\tilde{c}_t,c_{3h}):\ -0.1<c_{3h}^2 (0.222 \tilde{c}_t^2+0.0880)+c_{3h} (-2.14 \tilde{c}_t^2-0.671)+0.806 \tilde{c}_t^4+6.46 \tilde{c}_t^2+0.583<0.1~,\nn\\
&(c_g,c_{3h}):\ -0.1<c_{3h}^2 (0.0347 c_g^2+0.0846 c_g+0.0880)+c_{3h} (0.465 c_g^2+0.157 c_g-0.671)+3.20 c_g^2-0.531 c_g+0.583<0.1~,\nn\\
&(\tilde{c}_g,c_{3h}):\ -0.1<c_{3h}^2 (0.0347 \tilde{c}_g^2+0.0880)+c_{3h} (0.465 \tilde{c}_g^2-0.671)+3.20 \tilde{c}_g^2+0.583<0.1~,\nn\\
&(\tilde{c}_{2t},c_{3h}):\ -0.1<5.68 \tilde{c}_{2t}^2+0.0880 c_{3h}^2-0.671 c_{3h}+0.583<0.1~.
\label{eq:analytic3}
\end{align}

\section{Conclusions}\label{sec:5}
We considered effective Higgs boson couplings that affect the double Higgs production.
For generality we included both CP-even and CP-odd effective couplings.  Some of the effective couplings are loosely constrained by the single Higgs measurements at the 7 TeV and the 8 TeV LHC. The correlations of those effective couplings are different in single and double Higgs productions, therefore, one can probe those effective couplings by combining both the single and double Higgs productions. We examined the impact of the effective couplings on double Higgs production at the high luminosity LHC with an integrated luminosity of $3000~{\rm fb}^{-1}$ and at a future $pp$-collider operating at an energy of 100~TeV with an integrated luminosity of  $30~\rm{ab}^{-1}$.

The amplitude of the double Higgs production depends on several form factors. From partial wave analysis, we found that the double Higgs production is still dominated by the $s$-wave component even at the 100 $pp$-collider.
Making use of the $s$-wave dominant feature, we propose a universal cut efficiency function $\mathcal{A}({m}_{hh})$ to mimic the experimental cuts and detector effects. Convoluting inclusive distribution of the invariant mass of Higgs pair with the cut efficiency function gives rise to the signal events after experimental cuts.  We followed the analysis in Refs.\cite{Aad:2015,Contino:2016spe} to derive the cut efficiency functions at the 14 TeV LHC and the 100 TeV $pp$-collider. 
Using the cut efficiency functions, we obtain the differential cross sections of $m_{hh}$ and total cross sections of $gg\to hh \to b\bar{b}\gamma\gamma$ after kinematics cuts. From there we obtained the potential of probing those effective couplings at the 14 TeV HL-LHC and at the 100 TeV $pp$-collider. 

We varied two effective couplings at a time and fixed other couplings to be the SM values. With the tremendously high luminosity, the HL-LHC could cover a lot of parameter space, which could yield a $5\sigma$ discovery. Negative results of Higgs pair searches also exclude a vast amount of parameter spaces. There are two islands in the parameter space of $(c_t, c_g)$, $(c_g, c_{3h})$, $(c_g, c_{2t})$ and $(c_g, \tilde{c}_{2t})$, which cannot be resolved by the single Higgs production. The double Higgs production could exclude the island that does not consist of the SM. We also presented the analytical expressions of those $2\sigma$ exclusion limits in the parameter space.

We found that the double Higgs production can be discovered in the process $gg\to hh\to b\bar{b}\gamma\gamma$ at the 100 TeV $pp$-collider with an integrated luminosity of $256~{\rm  fb}^{-1}$. We thus focused on searching for Higgs effective couplings at the 100 TeV machine 
with $\mathcal{L}=30~\rm{ab}^{-1}$ and treat the SM double Higgs production as a background. 
Thanks to the large center of mass energy, the 100~TeV $pp$-collider could cover almost entire parameter space of effective couplings, except $c_{3h}$ which is not sensitive to the Higgs pair production. Finally, we listed analytical expressions of the $5\sigma$ discovery bands which, together with the analytical expressions of the $2\sigma$ exclusion limits at the HL-LHC, is useful to probe new physics models. 

\begin{acknowledgments}
The work is supported in part by the National Science Foundation of China under Grand No. 11275009, 11635001, 11135003 and 11375014. 
\end{acknowledgments}

\appendix
\section{The expressions of form factors }
\label{formfactors}
In this appendix, we collect the explicit expressions of the form factors in the single and double Higgs productions,
\bea
F_{\Box}&=&\frac{2m_t^2}{\hat{s}}\{m_t^2 (8m_t^2-\hat{s}-2m_h^2)
(D_0^t+D_0^u+D_0^{tu})+p_T^2(4m_t^2-m_h^2)D_0^{tu}\nn\\
&&+2+4m_t^2 C_0^s+\frac{2}{\hat{s}}(m_h^2-4m_t^2)[(\hat{t}
-m_h^2)C_0^t+(\hat{u}-m_h^2)C_0^u] \},\\
G_{\Box}&=&\frac{m_t^2}{\hat{s}}\{2(8m_t^2+\hat{s}-2m_h^2)
[m_t^2(D_0^t+D_0^u+D_0^{tu})-C_0^{sm}]-2[\hat{s}C_0^s+(\hat{t}
-m_h^2)C_0^t+(\hat{u}-m_h^2)C_0^u]\nn\\
&&+\frac{1}{\hat{s}p_T^2}[\hat{s}\hat{u}(8\hat{u}m_t^2-\hat{u}^2
-m_h^4)D_0^u+\hat{s}\hat{t}(8\hat{t}m_t^2-\hat{t}^2-m_h^4)D_0^t
+(8m_t^2+\hat{s}-m_h^2)\nn\\
&&[\hat{s}(\hat{s}-2m_h^2)C_0^s+\hat{s}(\hat{s}-4m_h^2)C_0^{sm}
+2\hat{t}(m_h^2-\hat{t})C_0^t+2\hat{u}(m_h^2-\hat{u})C_0^{u}]]\},\\
F_{\Box}^{(1)}&=&\frac{2m_t^2}{\hat{s}^2}\{m_h^2(2\hat{t}C_0^t
+2\hat{u}C_0^u-\hat{t}\hat{u}D_0^{tu})
-2m_h^4(C_0^t+C_0^u)+m_h^6D_0^{tu}\nn\\
&&+\hat{s}[2+m_t^2[4C_0^s-(D_0^t+D_0^u+D_0^{tu})(\hat{t}+\hat{u})]]\}\\
G_{\Box}^{(1)}&=&\frac{m_t^2}{2\hat{s}}\{\frac{2}{m_h^4
-\hat{t}\hat{u}}[-\hat{s}(2m_h^4+\hat{t}^2+\hat{u}^2)C_0^s
+2(m_h^2-\hat{t})(m_h^4+\hat{t}^2)C_0^t+2(m_h^2
-\hat{u})(m_h^4+\hat{u}^2)C_0^u\nn\\
&&-(\hat{t}+\hat{u})(2m_h^4-\hat{t}^2-\hat{u}^2)C_0^{sm}
+\hat{s}\hat{t}(\hat{t}^2+m_h^4)D_0^t+\hat{s}\hat{u}
(\hat{u}^2+m_h^4)D_0^u]\nn\\
&&-4m_t^2(\hat{t}+\hat{u})(D_0^t+D_0^u+D_0^{tu})\},\\
F_{\Box}^{(2)}&=&4m_t^{4}(D_0^t+D_0^u+D_0^{tu}),\\
F_{\triangle}&=&\frac{2m_t^2}{\hat{s}}[2+(4m_t^2-s)C_0^s],\\
F_\triangle^{(1)}&=&2m_t^2 C_0^s.
\eea
In the above we have the conventions~\cite{Shao:2013bz}
\bea
C_0^s&=&C_0(0,0,\hat{s},m_t^2,m_t^2,m_t^2),\ C_0^t=
C_0(0,\hat{t},m_h^2,m_t^2,m_t^2,m_t^2),\nn\\
C_0^u&=&C_0(0,\hat{u},m_h^2,m_t^2,m_t^2,m_t^2),\
C_0^{sm}=C_0(m_h^2,\hat{s},m_h^2,m_t^2,m_t^2,m_t^2)\nn\\
D_0^t&=&D_0(m_h^2,0,0,m_h^2,\hat{t},\hat{s},m_t^2,m_t^2,m_t^2,m_t^2),\nn\\
D_0^u&=&D_0(m_h^2,0,0,m_h^2,\hat{u},\hat{s},m_t^2,m_t^2,m_t^2,m_t^2),\nn\\
D_0^{tu}&=&D_0(m_h^2,0,m_h^2,0,\hat{t},\hat{u},m_t^2,m_t^2,m_t^2,m_t^2),
\eea
and the definitions of the scalar Passarino-Veltman
functions are as follows~\cite{Denner:1991kt}
\bea
&&C_0(p_1^2,p_2^2,(p_1+p_2)^2,m_1^2,m_2^2,m_3^2)\nn\\
&=&\frac{(2\pi\mu)^{4-D}}{i\pi^2}\int d^Dq\frac{1}{(q^2-m_1^2)
[(q+p_1)^2-m_2^2][(q+p_1+p_2)^2-m_3^2]},\nn\\
&&D_0(p_1^2,p_2^2,p_3^2,p_4^2,(p_1+p_2)^2,(p_2+p_3)^2,
m_1^2,m_2^2,m_3^2,m_4^2)\nn\\
&=&\frac{(2\pi\mu)^{4-D}}{i\pi^2}\int d^Dq\frac{1}{(q^2-m_1^2)
[(q+p_1)^2-m_2^2][(q+p_1+p_2)^2-m_3^2][(q+p_1+p_2+p_3)^2-m_4^2]},
\eea
where $\mu$ is the renormalization scale and $D$ is the space-time dimension.

\bibliographystyle{apsrev}
\bibliography{reference}

\end{document}